\documentclass[12pt,twoside]{article}
\usepackage{amssymb}
\usepackage{amsmath}
\usepackage{mathabx}
\usepackage{latexsym}
\usepackage{longtable}
\usepackage{epsfig}
\usepackage{graphicx,bbm,psfrag}
\graphicspath{{images/}}
\usepackage{cite}

\DeclareMathAlphabet{\mathsfit}{T1}{\sfdefault}{\mddefault}{\sldefault}
\SetMathAlphabet{\mathsfit}{bold}{T1}{\sfdefault}{\bfdefault}{\sldefault}

\begin{document}
\begin{titlepage}
\vspace*{5cm}
\begin{center}
\Large{\textbf{Hydrodynamic Equations  for the Toda Lattice}}\bigskip\bigskip\bigskip
\end{center}
\begin{center} 
{\large{Herbert Spohn}}\bigskip\bigskip\\
Departments of Mathematics and Physics, Technical University Munich,\smallskip\\
Boltzmannstr. 3, 85747 Garching, Germany
\end{center}
\vspace{3cm}
\begin{flushright}
January 15, 2021
\end{flushright}
\vspace{6cm}
\begin{center}
\textit{Dedicated to my grandson Lio Spohn, who  step by step closely followed my enterprise.}
\end{center}
\end{titlepage}

\tableofcontents
\newpage
\noindent
{\large\textbf{Preface}}\\\\
In August 1974, for my first time I attended  a summer school, which happened to be number three in the series on
``Fundamental Problems in Statistical Mechanics". The school took place at the  Agricultural University of Wageningen.
We were approximately 70 participants from 25 countries. The school lasted nearly three weeks with 11 lecture courses, each four hours long, and various more specialized seminars. As to be expected, present were the big shots, as Eddy Cohen who with moderate success tried to slow down the speaker through posing questions. Besides lectures, the truly exciting part of the school was to meet fellow youngsters who had similar interests and struggled more or less  with the  same difficulties. 
    
At the time, critical phenomena and the just invented RG methods were the overwhelming topic. Fortunately the Dutch physics community has a long tradition in Statistical Mechanics and thus a wide range of topics were covered. I vividly recall the lectures by Nico van Kampen
on ``Stochastic differential equations''. Joe Ford lectured on ``The statistical mechanics of classical analytic dynamics", the early days of deterministic chaos. In fact, at the end of his lectures Joe mentioned the Toda lattice describing the just confirmed integrability.
Perhaps I should have had listened with more care. For me the strongest impact had the lectures presented by  Piet Kasteleyn
on ``Exactly solvable lattice models", explaining the fascinating link between equilibrium statistical mechanics and models from quantum many-body physics.

My second encounter with integrable systems is related to the study of Dyson Brownian motion, which is an integrable stochastic particle system. At the time,  the model  was  to me an intriguing example for the hydrodynamics of a many-particle system with long range forces. The third encounter was triggered by the KPZ revolution, which brought me in contact with other corners of integrable systems.
Around 2016 I first learned about the activities investigating the hydrodynamic equations for integrable many-body quantum systems. I could not resist.
Of course, major insights had been accomplished already. But, apparently, the classical Toda lattice was in a state of dormancy. This is how my project got started.

During the ongoing project, I had many insightful comments and good advice. Gratefully acknowledged are   Vir Bulchandani, Xiangyu Cao,   Percy Deift,  
Jacopo De Nardis, Atharv Deokule, Kedar Damle, Avijit Das, Abhishek Dhar, Maurizio Fagotti, Patrik Ferrari, Pablo Ferrari,
Alice Guionnet,  David Huse, Thomas Kappeler, Manas Kulkarni,   Christian Mendl, Joel Moore, Fumihiko Nakano, Neil O'Connell,
Stefano Olla, Lorenzo Piroli, Bal\'{a}zs Pozsgay, 
Michael Pr\"{a}hofer, Tomaz Prosen, Keiji Saito, Makiko Sasada,Tomohiro Sasamoto, J\"{o}rg Teschner,  Khan Duy Trinh,
Simone Warzel, and Takato Yoshimura.  Special thanks are due to Benjamin Doyon. Our encounter at Pont-\`{a}-Mousson is well remembered.\\


\newpage
\section{Introduction}
\label{sec1}
Hydrodynamics is based on the observation that the motion of a large assembly of strongly interacting particles is constrained by local conservation laws.
As a result, local equilibrium is established over an initial time span and followed by a much longer time window when the local equilibrium parameters are governed by the hydrodynamic evolution equations. The initial time span could shrink to microscopic times when the 
system starts out already in local equilibrium. It is a matter of fact that a vast amount of interesting physics is covered by the hydrodynamic approach. 
Historically the best known example are simple fluids for which hydrodynamics is synonymous with fluid dynamics. Even for fluids the hydrodynamic approach already carries the seed for further extensions, since the equilibrium phase diagram is richly structured. Most commonly there is a liquid-gas
phase transition. This discrete order parameter has now to be included as part of local equilibrium.  For example, gas and fluid phase may spatially coexist and
the respective interface is then an additional slow degree of motion, to be included in the macroscopic dynamics.   Close to the critical point, conventional
hydrodynamics has to be augmented by more refined theories. At lower temperatures, there will be a solid 
phase. Because of slow relaxation of solids, for the dynamics of the solid-gas interface mostly non-hydrodynamic modelling is used.
Bosonic particles at low temperatures will form a condensate. One then employs a hydrodynamic two-fluid model, which governs the 
superfluid interacting with the normal fluid. Going beyond short-range interactions, magneto-hydrodynamics describes the motion of fluids made up of charged particles,
also including the Maxwell field as additional dynamical degrees of freedom. Relativistic hydrodynamics becomes relevant for extreme events such as
 the formation of superdense neutron stars, relativistic jets, and Gamma ray bursts. Each mentioned topic is part of ongoing research.
 
Approximately  five years ago a novel item was added to our list under the name of \textit{generalized hydrodynamics}. Perhaps not as far reaching as other examples mentioned, generalized hydrodynamics relies on an amazing novel twist. The added topic is concerned with integrable systems
 for which the number of conserved fields is proportional to system size, in sharp contrast to the items listed above which have 
 only a few conserved fields, as number, momentum, energy, plus broken symmetries. At first glance the mere idea of a hydrodynamic description of the time evolution of an integrable system sounds like an intrinsic contradiction.  After all, establishing local equilibrium relies on chaotic dynamics which is just the opposite of integrability.
 But the large number of degrees of freedom helps. A convincing and easily accessible example is the classical fluid of hard rods in one space dimension, which will be discussed in Section \ref{sec5}. Since integrable systems have  such a large number of  local conservation laws,
 local equilibrium must now be characterized by a correspondingly large number of chemical potentials. It is this feature which is called ``generalized". In the limit of infinite system size, the hydrodynamic 
 fields are labelled by a parameter taking a continuum of values, or possibly by more complicated schemes. As a consequence,
 even writing down the coupled set of hyperbolic conservation laws is a major obstacle. As the title of our treatise indicates, the main focus will be to figure out the   structure of the hydrodynamic equations. While a few predictions will be discussed, for numerical confirmations and experimental results the reader is asked to consult the original literature.
 
 Integrable systems fall into two classes, interacting and non-interacting. In the latter case, particles, or quasiparticles, just pass through each other.
 The system of hyperbolic conservation laws decouples and each mode separately is governed by a linear equation. In contrast, interacting models have a non-vanishing two-particle scattering shift, which is the mechanism behind the nonlinearity 
 in the equations of generalized hydrodynamics. While a focussed research started only say roughly six years ago, there have been precursors. In the early 1980ies Roland Dobrushin analyzed in great detail the system of hard rods. He well understood the hydrodynamic 
 perspective. But at the time no tools for handling more intricate models were available. In retrospect, the true simplification of hard rods is a two-particle scattering shift which is independent of the  incoming quasiparticle velocities. Another early work concerns the Korteweg-de-Vries equation which is an integrable nonlinear wave equation. In the mid 1990ies Vladimir E. Zhakarov obtained the kinetic equation for a low density gas of solitons. The hydrodynamic equation, claimed  to be valid at any density,  has been derived by Gennady El in 2005.
     
 The advance in 2016 established a general rule of how to write down the average currents, thereby covering classical field theories, as the sinh-Gordon model,
 and many-body quantum systems, as the Lieb-Liniger $\delta$-Bose gas, the XXZ spin chain, and the one-dimensional Fermi-Hubbard model.  Only with such an input the equations of generalized hydrodynamics could be written, herewith opening the door to applications of physical interest. As a result of such detailed studies, step by step one arrived at the conclusion that all integrable interacting models have a structurally identical  generalized hydrodynamics.
 
My exposition is \textit{not} a review, even though much of the relevant literature will be cited. I will focus exclusively on the Toda lattice, both classical and quantum,  with 
random initial data. The larger part of the article will deal with the classical chain, simply because our understanding is more complete. The two-particle scattering shift of the classical Toda lattice equals $2\log|v-v'|$, hence the model is interacting. Since the classical Toda lattice is a Hamiltonian system of interacting particles, rather modest prerequisites are required. My goal is to have a text, where the reader can follow each argument
without having to consult the original literature. As to be anticipated, this rule can be followed only partially, and additional inserts on more advanced topics
 are added. 
I allowed myself one further exception. In Section 
\ref{sec4} mean-field Dyson Brownian motion will be explained, since its stationary measure in essence coincides with a generalized Gibbs ensemble of the Toda lattice.
The quantum Toda chain will be taken up only in Section \ref{sec11}. While somewhat hidden, its structure has surprising similarities with the Lieb-Liniger $\delta$-Bose gas.  As preparation, we will thus first discuss at some length the hydrodynamic equations  of the $\delta$-Bose gas. 

Achieving familiarity with one very particular model  may help, I hope, to  understand  as well the structure of hydrodynamics for other integrable models.  
 \bigskip\\\\
 \textbf{\large{Notes and references}}
  \bigskip\\
\textbf{ad Preface}: The Proceedings of the summer school have been edited by E.D.G. Cohen \cite{C75}. My work on Dyson Brownian motion is published in \cite{S87}. The KPZ revolution is covered in many articles from which only \cite{C12,QS15,S17} are quoted.\medskip\\      
\textbf{ad 1}: The upswing of generalized hydrodynamics can be traced back to the two independent seminal contributions \cite{CDY16,BCDF16}, in which a general prescription for the GGE averaged currents is presented, GGE being the standard acronym for ``generalized Gibbs ensemble''. Thus, for the first time the Euler type equations could be written down with confidence. Boldrighini, Dobrushin, and Suhov \cite{BDS83}
 prove, under fairly general assumptions on the initial probability measure, the validity of the Euler equation starting from hard rods under ballistic scaling.
 They also established that the hydrodynamic equation has smooth solutions. Dobrushin's vision is outlined in \cite{D89}. Zakharov \cite{Z71} assumed a low density gas of solitons with Poisson distributed centers and  a general velocity distribution. He argued for a Boltzmann type kinetic equation. The hydrodynamic extension to a dense soliton fluid has been accomplished in \cite{EK05}, see also
 \cite{E05,EKPZ11,CDE16}. Interestingly enough,  their proposal agrees with the general scheme developed later in \cite{CDY16,BCDF16} upon implementing the appropriate adaptions. 
 
 A useful account on the developments prior to generalized hydrodynamics can be found in a special volume on  ``Quantum Integrability in Out-of-Equilibrium Systems'' \cite{CEM16}. Its central theme are quantum quenches starting from a spatially homogeneous initial state
 and their long time equilibration.
 
The notion ``scattering shift" is convenient but not so widely used. It refers to the fact that, when two particles undergo a scattering,  each trajectory is asymptotically of the form $v_j t +\kappa_j$, $j=1,2$, i.e. free motion linear in time and on top a constant displacement as first order correction, which by definition is the scattering shift. In quantum mechanical two-body scattering,  the wave function for the relative motion has asymptotically the form $\exp\big[\mathrm{i}(-kx +\tfrac{1}{2}k^2t + \theta(k))\big]$. Then a narrow wave packet, centered at  
$k_0$, travels with velocity   $k_0$ and is displaced by $\theta'(k_0)$. $\theta$ is the phase shift, while its derivative,  $\theta'$, is the scattering shift.

 \section{Dynamics of the classical Toda lattice}
\label{sec2}
\setcounter{equation}{0} 
As well known at the time when Morikazu Toda started his studies on the lattice with exponential interactions, shallow water waves in long channels have peculiar dynamical properties. One observes solitary waves and soliton collisions. In the latter, two incoming solitons  emerge with  their original shape after an intricate dynamical process. The Korteweg-de Vries (KdV) equation, a one-dimensional nonlinear wave equation, provides an accurate theoretical description of these phenomena. In 1967 Toda investigated whether also discrete wave equations might have solitary type dynamics. A point in case is the wave equation and  its discretization the harmonic lattice, both of them linear equations. With an ingenious insight Toda discovered that a lattice with exponential interactions, later known as Toda lattice, exhibits the same dynamical features as the KdV equation. It took another seven years until it had been firmly established that the $N$-particle Toda lattice is indeed integrable with $N+1$ conservation laws.

In the usual dimensionless form, the hamiltonian of the Toda lattice reads 
\begin{equation}\label{2.1}
H_\mathrm{toda} = \sum_{j \in \mathbb{Z}}\big( \tfrac{1}{2}p_j^2 + \mathrm{e}^{-(q_{j+1} - q_j)}\big),
\end{equation}
where $j \in \mathbb{Z}$ is the particle label and $(q_j,p_j)$ are position and momentum of the $j$-th particle. 
One could introduce a particle mass, coupling strength, and decay rate for the potential. But through rescaling spacetime 
the standard form \eqref{2.1} is recovered. The model has no free parameter. The equations of motion are
\begin{equation}\label{2.2}
\frac{d^2}{dt^2}q_j =   \mathrm{e}^{-(q_{j} - q_{j-1})} -\mathrm{e}^{-(q_{j+1} - q_j)}, \quad j \in \mathbb{Z}.
\end{equation}
Physically this equation can be naturally viewed in two ways. (i) The displacements $q_j(t), j \in \mathbb{Z}$, are regarded as the lattice discretization
of a continuum wave field $q(x,t)$, $x \in \mathbb{R}$. We call this the lattice (field theory) picture. (ii) The fluid picture is to literally
regard $q_j(t), j = 1,...,N$, as positions of particles moving on the real line. However they do not interact  pairwise as would be the case for a real fluid. Toda was mostly thinking of a lattice discretization. But the fluid picture is easier to visualize. Of course, there is only a single set of equations of motion and 
one can switch back and forth between the two options.

At this stage, a review of the vast research on the Toda lattice can neither be supplied nor is intended. I have to refer to monographs, reviews, and textbooks. 
However, it can be safely summarized that almost exclusively problems have been studied for which physically the chain is at zero temperature.  Examples are multi-soliton solutions and the spatial spreading of local perturbations of an initially  periodic particle configuration. In contrast, our focus are random initial 
data with an energy proportional to the system size and away from a ground state energy. A paradigmatic set-up would be the thermal state at
some non-zero temperature. For such an enterprise novel techniques are required. In particular, the issue of large system size has to be properly understood.
\subsection{Locally conserved fields and their currents}\label{sec2.1}
Our first task is to elucidate the integrable structure of the Toda lattice. For this purpose we introduce the \textit{stretch}
\begin{equation}\label{2.3}
r_j = q_{j+1} - q_j,
\end{equation}
also the free distance between particles $j$ and $j+1$, and the \textit{Flaschka variables}
\begin{equation}\label{2.4}
a_j = \mathrm{e}^{-r_j/2},\quad b_j =  p_j.
\end{equation}
The stretch can have either sign, while $a_j >0$. The $a$'s and $b$'s are a conventional notation, but we will avoid the duplication of symbols  by using only the momentum $p_j$. In principle, we could have set $a_j = \kappa  \mathrm{e}^{-r_j/2}$ and $b_j = \kappa p_j $, which amounts to a mere time change. Flaschka picked $\kappa = \tfrac{1}{2}$.
Below we will explain why $\kappa = 1$ is singled out for our purposes. In terms of the variables in \eqref{2.4} the equations of motion read
\begin{equation}\label{2.5}
\frac{d}{dt} a_j = \tfrac{1}{2}a_j(p_j - p_{j+1}),\quad \frac{d}{dt} p_j = a_{j-1}^2 - a_{j}^2.
\end{equation}
Hence $a_j$ couples to the right neighbor and $p_j$ to the left one. We consider the finite volume $[1,...,N]$ and impose periodic boundary conditions as
\begin{equation}\label{2.6}
a_{-1} = a_N,\quad  p_{N+1} = p_1.
\end{equation}
This choice is also called the \textit{closed} or \textit{periodic} chain. Note that
\begin{equation}\label{2.7} 
\frac{d}{dt}\sum_{j=1}^N r_j = 0,
\end{equation}
implying 
\begin{equation}\label{2.8} 
\sum_{j=1}^N r_j(t) = \ell
\end{equation}
with some constant $\ell$, which can have either sign. An equivalent description is to consider the infinite Toda lattice and to impose the initial condition
\begin{equation}
\label{2.9} 
q_{j+N}= q_j +\ell, \quad p_{j+N} = p_j
\end{equation}
 for all $j \in \mathbb{Z}$, which then holds at any time. Cutting the lattice into cells, each of size $N$,
in every cell there is the same dynamics of stretches. In the literature, periodic boundary conditions are often stated as
$q_{N+1} = q_1$, which corresponds to the special case $\ell = 0$. Physically the parameter $\ell$ is of crucial importance because through it the stretch per particle, $\ell/N$, is controlled.  In the fluid picture
the unit cell has length $\ell$ and contains $N$ particles. The physical particle density is $N/|\ell|$. But often it is more natural to work with the signed particle density
$N/\ell$, which could be negative.

Out of the Flaschka variables one forms the tridiagonal \textit{Lax matrix}, $L_N$, $N\geq 3$, and its companion matrix $B_N$ as
\begin{equation}\label{2.10} 
L_N = 
\begin{pmatrix}
p_1 & a_1&0  &\cdots&a_N\\
a_1 & p_2 & a_2 & \ddots&0\\
0&a_2& p_3&\ddots&\vdots\\
\vdots&\ddots &\ddots&\ddots&a_{N-1}\\
a_N&0&\cdots &a_{N-1}&p_N\\
\end{pmatrix}, 
\qquad B_N = \frac{1}{2}
\begin{pmatrix}
0 & -a_1& 0 &\cdots& a_N\\
a_1 & 0 &-a_2&\ddots  &0\\
0&a_2 & 0&\ddots&\vdots\\
\vdots&\ddots&\ddots& \ddots&-a_{N-1}\\
- a_N&0&\cdots &a_{N-1}&0\\
\end{pmatrix}.
\end{equation}
Clearly, $L_N$ is symmetric and $B_N$  skew symmetric, $(L_N)^\mathrm{T} = L_N$, $(B_N)^\mathrm{T} = - B_N$, with $^\mathrm{T}$ denoting the transpose of a matrix.  Later on for the adjoint of an operator also the more common $^*$ will be used.
From the equations of motion \eqref{2.5}, one verifies that
\begin{equation}\label{2.11} 
\frac{d}{dt} L_N = [B_N,L_N]
\end{equation}
with $[\cdot,\cdot]$ denoting the commutator, $[B_N,L_N] = B_NL_N -L_NB_N$. Since $B_N$ is skew symmetric, $L_N(t)$ is isospectral to $ L_N =  L_N(0)$. Hence the eigenvalues of $L_N$ are conserved.\bigskip\\
$\blackdiamond\hspace{-1pt}\blackdiamond$~\textit{Vector notation}.\hspace{1pt} In our text various $N$-vectors will appear. The standard notation $x\in \mathbb{R}^N$, $x = (x_1,...,x_N)$, is adopted. The  $N$-dimensional volume element is denoted by $\mathrm{d}x$. Also $x\in \mathbb{R}$ will be used. The distinction should be obvious from the context. \hfill$\blackdiamond\hspace{-1pt}\blackdiamond$\\

Let us write the eigenvalue problem for $L_N$ as 
\begin{equation}\label{2.12} 
L_N \psi_\alpha = \lambda_\alpha \psi_\alpha, \quad\alpha = 1,\dots,N.
\end{equation}
Then 
$\lambda_\alpha(a,p)$ is some function on phase space, which  does not change under the Toda time evolution. 
However $\lambda_\alpha$ is a highly nonlocal function, in general. For example, considering  the dependence of $\lambda_\alpha$
only on $p_1$
and $p_{N/2}$, this will not split into a sum as $g(p_1) + \tilde{g}(p_{N/2})$ even approximately. Physically more relevant are \textit{local 
conservation laws}. For the Toda lattice they are easily obtained through forming the trace as
\begin{equation}\label{2.13} 
Q^{[n],N} = \mathrm{tr}\big[(L_N)^n\big] = \sum_{j=1}^N  (\lambda_j)^n = \sum_{j=1}^N ((L_N)^n)_{j,j}
= \sum_{j=1}^N Q^{[n],N}_j.
\end{equation}
The second identity confirms that $Q^{[n],N}$ is conserved and the third identity that $Q^{[n],N}_j$ is local. Indeed, taking $n < N/2 $
and expanding as
\begin{equation}\label{2.14}  
Q^{[n],N}_j = \sum_{j_1=1}^N\dots \sum_{j_{n-1}=1}^N (L_N)_{j,j_1}(L_N)_{j_1,j_2}\dots (L_N)_{j_{n-1},j},
\end{equation}
the density $Q^{[n],N}_j$ depends only on the variables $\{a_{j-i},p_{j-i},...,a_{j+i},p_{j+i},i = 0,...,n-1\}$, modulo $N$. 

The two-sided infinite volume limit of $L_N$ is the tridiagonal Lax matrix $L$, and correspondingly its companion matrix $B$, which are operators acting on the Hilbert space $\ell_2(\mathbb{Z})$ of square-integrable two-sided sequences over the lattice $\mathbb{Z}$. They still satisfy
\begin{equation}\label{2.15} 
\frac{d}{dt} L = [B,L].
\end{equation}
Of course $\mathrm{tr}[L^n]$ makes no sense, literally. But the infinite volume density  
\begin{equation}\label{2.16} 
Q^{[n]}_j = L_{j,j} 
\end{equation}
is well defined.  $Q^{[n]}_j$ is a finite polynomial in the variables $\{a_i,p_i, |i - j| \leq n-1\}$, which can more easily grasped by using the random walk expansion derived from the $N=\infty$ version of \eqref{2.14}. The walk is on $\mathbb{Z}$ with $n$ steps, starting and ending at $j$, of step size $0,\pm1$. A step from $i $ to $i$ carries the variable $p_i$ and the step from either $i$ to $i+1$ or from $i+1$ to $i$ carries the variable $a_i$. For a given walk one forms the product along the path, which is a monomial of degree $n$. $Q^{[n]}_j$ is then obtained by summing over all admissible walks.
By translation invariance of the model, $Q^{[n]}_j$ and $Q^{[n]}_{j+i}$ are identical polynomials, except that the particle label is shifted by $i$.  Just as for the hamiltonian \eqref{2.1}, formally we still write
\begin{equation}\label{2.17} 
Q^{[n]} =\sum_{j\in\mathbb{Z}}Q^{[n]}_j,  
\end{equation}
 where the index $n$ ranges over the positive integers. 
 
 As noted already before, 
in addition the stretch is locally conserved, which carries the label $0$, 
 \begin{equation}\label{2.18} 
Q^{[0]}_j = r_j.
\end{equation}
The three lowest order fields have an immediate physical interpretation as stretch, momentum, and energy density,
 \begin{equation}\label{2.19} 
Q^{[0]}_j = r_j, \quad Q^{[1]}_j = p_j, \quad \tfrac{1}{2} Q^{[2]}_j = \tfrac{1}{2}\big(p_j^2 + a_j^2 + a_{j-1}^2\big).
\end{equation}
 Obviously, there cannot be physical names for the higher orders. In the community of quantum integrable systems, $Q^{[n]}$ is called the 
 $n$-th conserved charge or merely the $n$-th charge, which serves as a concise notion but carries no specific physical meaning.
 Notwithstanding we stick to the more bulky ``conserved field'' in the tradition of the theory of fluids.
 
 Since the Toda hamiltonian has a local energy density, any locally conserved field must satisfy a continuity equation,
 in other words the lattice version of a local conservation law. We consider the infinite lattice and compute
 \begin{equation}\label{2.20}
\frac{d}{dt} Q_{j}^{[n]} = (BL^n - L^nB)_{j,j} =  a_{j-1}(L^n)_{j,j-1} - a_j (L^n)_{j+1,j} = J_{j}^{[n]} - J_{j+1}^{[n]}.
\end{equation} 
Hence $ J_{j+1}^{[n]}$ is the current of the $n$-th conserved field from $j$ to  $j+1$ and $ J_{j}^{[n]}$ the current from 
$j-1$ to  $j$. Defining the lower triangular matrix $L^{\scriptscriptstyle \downarrow}$ by $(L^{\scriptscriptstyle \downarrow})_{j+1,j}=a_j$
and $(L^{\scriptscriptstyle \downarrow})_{i,j} = 0$ otherwise, a more concise expression is
\begin{equation}\label{2.21}
J_{j}^{[n]}  = (L^nL^{\scriptscriptstyle \downarrow})_{j,j}, \quad n=1,2,...\,.
\end{equation} 
For the stretch
\begin{equation}\label{2.22}
\frac{d}{dt} Q_{j}^{[0]}  = -p_j +p_{j+1}, \quad J_{j}^{[0]}  = - p_j,
\end{equation} 
and hence
\begin{equation}\label{2.23}
J_{j}^{[0]}  = - Q_{j}^{[1]}.
\end{equation} 
This innocent looking equation will have surprising consequences.\bigskip\\
$\blackdiamond\hspace{-1pt}\blackdiamond$~\textit{Ambiguity of densities}.\hspace{1pt} As presented, the densities, both for field and current, seem to be unique. This is not the case, however. We have made a particular choice which will be useful when investigating the hydrodynamics of the Toda lattice. 
A further common scheme is to require the density to depend on a minimal number of lattice sites.  In terms of the random walk representation described above, this amounts to a density  $\tilde{Q}_j^{[n],N}$ defined by  summing over all closed  admissible paths with minimum equal to $j$.
As an example,  for $n =4$ the minimal version of the density is given by
\begin{equation}\label{2.24}
\tilde{Q}_{j}^{[4]} = p_j^4 + 4 (p_j^2 +p_jp_{j+1} +p_{j+1}^2)a_{j}^2  +2 a_j^4 +4a_j^2a_{j+1}^2.
\end{equation} 

Rather than trying to dwell on generalities, we illustrate the issue by considering the energy density, $n=2$. From \eqref{2.16} we have
 \begin{equation}\label{2.25}
Q_{j}^{[2]} = p_j^2 +a_{j-1}^2+ a_j^2
\end{equation} 
with the current density
\begin{equation}\label{2.26}
J_{j}^{[2]}  = a_{j-1}^2(p_{j-1} +p_{j}).
\end{equation} 
The minimal version of the energy density would be
\begin{equation}\label{2.27}
\tilde{Q}_{j}^{[2]} = p_j^2 + 2 a_j^2,
\end{equation} 
having the current density
\begin{equation}\label{2.28}
\tilde{J}_{j}^{[2]}  = 2 a_{j-1}^2 p_{j}.
\end{equation} 
At infinite volume, we consider the spatial sum $\sum_{j=1}^N$, denoted by $Q^{[2],N\diamond}$ and $\tilde{Q}^{[2],N\diamond}$. They  differ only by a boundary term and  hence the spatial average
$N^{-1}\big( Q^{[2],N\diamond}- \tilde{Q}^{[2],N\diamond}\big) = \mathcal{O}(1/N)$. On the other hand, the corresponding total currents, $J^{[2],N\diamond}$ and $\tilde{J}^{[2],N\diamond}$, differ by $\mathcal{O}(N)$, but
\begin{equation}\label{2.29}
\big(J_{j}^{[2],N\diamond} - \tilde{J}_{j}^{[2],N\diamond}\big) = \frac{d}{dt}(a_{j-1})^2.
\end{equation} 
The difference is a total time derivative and thus vanishes when averaged over a time-stationary probability measure,
e.g. thermal average. As conclusion, while there is ambiguity on the microscopic scale, upon averaging over
large spacetime cells this amounts to only small correction terms. In particular, the hydrodynamic equations for the Toda lattice  do not depend on the particular choice of microscopic densities.  \hfill$\blackdiamond\hspace{-1pt}\blackdiamond$
\subsection{Action-angle variables, notions of integrability}\label{sec2.2}
A  cornerstone of hamiltonian dynamics is the abstract characterization of integrable systems. 
Just to recall, given is some hamiltonian $H$ on a phase space $\Gamma_n$ of dimension $2n$. The dynamics generated by $H$ is called integrable, 
if there are $n$ differentiable functions, $I_1,...,I_n$, the action variables, on phase space which have the following properties: (i) They span an $n$-dimensional hypersurface in $\Gamma_n$, 
(ii) they are conserved, which means that the Poisson brackets $\{I_j,H\} = 0$, (iii) they are in involution, i.e. $\{I_i,I_j\} =0$ for all $i,j=1,...,n$. There exists then a canonical transformation to the action variables $(I_1,...,I_n)$ and the canonically conjugate angle variables 
$(\vartheta_1,...,\vartheta_n)\in (S^1)^n$ such that the transformed hamiltonian $\tilde{H}$ depends only on $I$.
In these variables the dynamics trivializes as
\begin{equation}\label{2.30}
\frac{d}{dt}\vartheta_j = \omega_j, \qquad \omega_j = \partial_{I_j} \tilde{H}(I), \quad j = 1,...,n.
\end{equation} 

This characterization applies also to the Toda chain. 
We consider a lattice of $N$ sites, the phase space $\Gamma_N = (\mathbb{R}_+ \times \mathbb{R})^N$, and  the evolution \eqref{2.5} in terms of the Flaschka variables $a= (a_1,...,a_N)$ and $p =(p_1,...,p_N)$.
These are not canonical variables. However, instead of the usual Poisson bracket, one can introduce a nonstandard Poisson bracket 
by first defining the $N\times N$ matrix 
\begin{equation}\label{2.31}
\qquad A_N = \frac{1}{2}
\begin{pmatrix}
-a_1& 0&\cdots  &0&a_N\\
a_1 & -a_2 &0  & \ddots&0\\
0 &a_2& -a_3&\ddots& \vdots\\
\vdots&\ddots&\ddots &\ddots&0\\
0 &\cdots&0 &a_{N-1}&-a_N
\end{pmatrix}
\end{equation}
and adjusting  the Poisson bracket to
\begin{equation}\label{2.32}
\{f,g\} = \langle \nabla _p f, A_N \nabla_a g\rangle  - \langle \nabla _a f, (A_N)^\mathrm{T} \nabla_p g\rangle
\end{equation} 
with $\langle\cdot,\cdot \rangle$ denoting the inner product in $\mathbb{R}^N$. For the usual Poisson bracket, $A_N$ would 
be the identity matrix. In the Flaschka variables
\begin{equation}\label{2.33}
H_{\mathrm{toda},N} = \sum_{j =1}^N\big( \tfrac{1}{2}p_j^2 + a_j^2\big)
\end{equation}
and the equations of motion  \eqref{2.5} can be written in hamiltonian form as 
\begin{equation}\label{2.34}
\frac{d}{dt} p_j = \{p_j,H_{\mathrm{toda},N} \}, \qquad \frac{d}{dt} a_j = \{a_j,H_{\mathrm{toda},N} \}.
\end{equation}

The matrix $A_N$ is of rank $N-1$, the oblique projection for the eigenvalue $0$ being
\begin{equation}\label{2.35}
 |a_1^{-1},a_2^{-1},...,a_N^{-1}\rangle \langle 1,1,...,1|,
\end{equation}
i.e. the corresponding left eigenvector of $A_N$ equals  $\nabla_a Q^{[0],N}$ and the right one $\nabla_a Q^{[1],N}$. 
 The Poisson structure \eqref{2.32} is degenerate. But it can be turned  non-degenerate  simply by fixing
the two conservation laws as $Q^{[0],N} = c_0$, $Q^{[1],N} = c_1$ with an arbitrary choice of the real parameters $c_0,c_1$. The new phase space becomes $\Gamma_{N-1}$ and the dynamical evolution equations  involve only  the variables
$(a_1,...,a_{N-1}, p_1,...,p_{N-1})$, which are a hamiltonian system with a non-degenerate Poisson bracket structure. 
The phase space for the action-angle variables is $(\mathbb{R}^2)^{N-1}= \Gamma_\mathrm{aa}$ with coordinates $(x_1,...,x_{N-1},y_1,...,y_{N-1})$ and action-angle variables as
\begin{equation}\label{2.36}
 x_j = \sqrt{I_j} \cos \vartheta_j,\quad   y_j = \sqrt{I_j} \sin \vartheta_j ,\quad j = 1,...,N-1,
\end{equation}
which are known as global Birkhoff coordinates.
 As proved in 2008 by A. Henrici and T. Kappeler, there is a canonical transformation $\Phi: \Gamma_{N-1} \to  \Gamma_\mathrm{aa}$ such that the transformed hamiltonian,  $H_\mathrm{aa}$, depends only on the action variables, $H_\mathrm{aa} = H_\mathrm{aa}(I_1,...,I_{N-1})$. 
In fact, the for us crucial property, $H_\mathrm{aa}$ is a strictly convex, real-analytic function. This means that the phases $\omega_j = \partial_{I_{j}}H_\mathrm{aa}$ are incommensurate Lebesgue almost surely. In other words, $H_\mathrm{aa}$ has no linear pieces, as would be the case for a system of harmonic oscillators. The Toda lattice is 
 phase-mixing:  starting with a probability measure on $\Gamma_{N-1}$ with  a continuous density function,  in the long time limit the density
 will become uniform on each torus of dimension $N-1$ with an amplitude computed from the initial density. 
 
The Toda lattice is a very peculiar dynamical system in the sense that it is integrable for every system size $N$, which we call  \textit{many-body
integrable}. Now the large $N$ limit is in focus and from a physics perspective the conventional definition might have to be reconsidered. In
particular one would like to have a notion which refers directly to the infinite lattice. From a hydrodynamic perspective the central building block are local conservation laws. Because of locality, for $N$ larger than the range of the density, the notion of local conservation law becomes independent of $N$ and refers 
only to phase spaces of finite dimensions. Note that local conservation laws have a linear structure, in the sense that the sum of two local conservation laws are again a local conservation law. We thus propose to call an infinitely extended system \textit{nonintegrable}, if it admits only for a few strictly local conservation laws. The system is called \textit{integrable}, if it possesses an infinite number of linearly independent local conservation laws. 
The condition of being in involution is dropped, although it seems to hold in most examples. A concrete example for which the involution property fails is the Landau-Lifshitz chain of classical
spins with isotropic nearest neighbor interactions. For a particular choice of the interaction potential, the chain is integrable in both the conventional and our sense.
The three components of spin are locally conserved, but they are not in involution.\bigskip\\
$\blackdiamond\hspace{-1pt}\blackdiamond$~\textit{Local conservation laws}.\hspace{1pt} The Toda chain is integrable in our sense with the densities of the locally conserved fields stated in \eqref{2.16}. But, as a stronger property, one would like to establish that there are no further local conservation laws.  Currently this is a conjecture and more studies would be needed. Still, a precise formulation is worthwhile. As before, periodic boundary conditions are understood.  We assume some general density function $f$ 
of support  $\kappa$, in other words $f(p_1,a_1,...,p_\kappa,a_\kappa)$.  Then the shifted densities are $f_j(a,p)= f(p_j,a_j,...,p_{j +\kappa -1}, a_{j+\kappa-1})$ and the 
conditions for being a local conservation law read
\begin{equation}\label{2.37}
Q^{(f),N} = \sum_{j=1}^N f_j, \qquad   \sum_{j=1}^N\{ f_j, H_{\mathrm{toda},N}\} =0, \qquad N > 2\kappa. 
\end{equation}
If so, each of the Poisson brackets is a local function. By translation invariance the sum has to be telescoping and thus necessarily there exists a strictly local current function, $J_j$, such that
$\{ f_j, H_{\mathrm{toda},N}\} = J_j - J_{j+1}$. 
According to the already proven conventional integrability there exists some function, $G$, such that 
\begin{equation}\label{2.38}
Q^{(f),N} = G\big( Q^{[0],N} ,..., Q^{[N],N} \big).
\end{equation}
Our conjecture claims that $G$ is necessarily  \medskip  linear.\\
\textit{Conjecture}: For fixed $\kappa$ and $N$ sufficiently large, there exists coefficients $c_0,c_1,...,c_\kappa$ such that 
\begin{equation}\label{2.39}
Q^{(f),N} =  \sum_{m=0}^\kappa c_m Q^{[m],N} . 
\end{equation}
One argument in favor of the conjecture comes from a simple observation. Consider some locally conserved field, $Q^{[n],N}$.
Then $(Q^{[n],N})^2$ is also conserved, but no longer local. The condition of locality should be strong enough to force a linear function in \eqref{2.38}. 

The reader might find the evidence for linking integrability with conservation laws not so convincing. Agreed, but it should be noted that our definition translates one-to-one to quantum spin chains, and also continuum quantum models. In the former case there are two concrete results,
with their wider validity to be expected. Considered is the XYZ spin chain with external magnetic field, $h$, pointing in the $z$-direction. In terms of 
the Pauli spin-$\tfrac{1}{2}$ matrices, $\sigma^x,\sigma^y, \sigma^z$, the hamiltonian reads
\begin{equation}\label{2.40}
H_\mathrm{XYZ} = \sum_{j \in \mathbb{Z}}\big( J_x  \sigma_j^x \sigma_{j+1}^x +  J_y\sigma^y_j \sigma_{j+1}^y + J_z\sigma^z_j \sigma^z_{j+1}  - h \sigma_j^z\big).
\end{equation}
Now, in case of non-zero coupling constants and $h\neq 0$, $J_x \neq J_y$, it is proved that there is only one local conservation law, 
namely the hamiltonian $H_\mathrm{XYZ}$ itself. This is, so to speak, the fully chaotic case. Only energy is transported and
the chain thermalizes, in the sense that expectations of local observables converge to the thermal average in the long time limit.
On the other hand 
for $h = 0$ the model is integrable. The strictly local conserved quantities can be computed from the transfer matrix, compare with  
Section \ref{sec11.2}, or through the boost operator. The second result states that, if the  coupling constants are non-vanishing, then any local conservation law is a 
finite linear combination of the already known conservation laws. This is the precise analogue of our conjecture.
 \bigskip \hfill$\blackdiamond\hspace{-1pt}\blackdiamond$ 

Our discussion raises some difficult issues. From the available evidence, there is a dichotomy, either a few conservation laws or 
infinitely many. One does not know whether there is a deep reason behind or merely reflects the limited class of models studied.
Personally I believe in the first option. In this context, 
particularly intriguing are nearly integrable systems. For finite $N$, the KAM theorem provides information on the stability of the invariant tori, confirming the coexistence of chaotic and integrable regions in phase space. But in our definition the limit $N \to \infty$ is taken first. 
Another point, strict locality is used here in an essential way. From a physics perspective also exponential tails should be admissible.
For sure, they would have to be included in the hydrodynamic equations.  
For the Toda lattice there is no indication
of additional conservation laws. On the other hand, it seems difficult to exclude such quasilocal conservation laws on abstract grounds.
A famous, well-understood case, is the XXZ model for which only the strictly local conservation laws were listed originally. Through numerical and theoretical studies it  became clear that additional conservation laws had been missed, which are now included
in the correct Euler scale hydrodynamics. 
 \subsection{Scattering theory}\label{sec2.3}
In hydrodynamics the system is confined and  thus interactions  persist without interruption, which in the long time limit then leads to
some sort of statistical equilibrium.  A dynamically less intricate  set-up is scattering: in the distant past particles are in the incoming configuration, for which they are far apart and do not interact. A time span of multiple collision processes follows. In the far future the outgoing particles move 
freely  again. To illustrate the special features of scattering for a many-body integrable systems, we first discuss a fluid consisting of hard rods, which will serve as an
instructive example also later on.  \bigskip\\
\textbf{Hard rod fluid}.\hspace{1pt} We consider $N$ hard rods, rod length $\mathsfit{a}$, moving on the real line. The hamiltonian reads
\begin{equation}\label{2.41}
H_\mathrm{hr} = \sum_{j = 1}^N  \tfrac{1}{2}p_j^2 + \sum_{j = 1}^{N-1} V_\mathrm{hr}(q_{j+1} - q_j),
\end{equation}
with the hard rod potential $V_\mathrm{hr}(x) = \infty$ for $|x| < \mathsfit{a}/2$ and $V_\mathrm{hr}(x) = 0$ for $|x| \geq \mathsfit{a}/2$.
Hard rods collide elastically with their two neighbors. Obviously, the system is integrable with one-particle sum functions, $\sum_{j=1}^N \phi(p_j)$, being conserved.

For scattering it is customary  to order the incoming momenta as $p_1 <...<p_N$, $p_j= p_j(-\infty)$. Then in the distant past
$q_j(t) = p_jt$ and $q_N(-\infty) < ...< q_1(-\infty)$. In the course of time there are exactly $N(N-1)/2$ collisions leading to the outgoing 
configuration $p_j(\infty) = p_{N-j+1}(-\infty)$, in particular $p_1(\infty) >...> p_N(\infty)$. Because of collisions, on top of the free motion
the asymptotic positions are shifted by a distance of order $1$, which we call the  \textit{scattering shift}. For the $j$-th particle it is defined by 
\begin{equation}\label{2.42}
\lim_{t \to \infty } \frac{1}{t}\big((q_{N-j +1}(t) -p_{N-j +1}t)- (q_{j}(-t) +p_{j}t)\big) =  \kappa_{j}. 
\end{equation} 
For $N=2$, particle $1$ is shifted by $ -\mathsfit{a}$
 and particle $2$  by $\mathsfit{a}$. For $N$ particles it suffices to study the intersections of $N$ freely moving particles to obtain
 \begin{equation}\label{2.43}
\kappa_j = \sum_{1 \leq i \leq N, i\neq j} \mathrm{sgn}(i - j)\phi_{ij} 
\end{equation} 
with $ \phi_{ij} = -\mathsfit{a}$ independent of the index. The scattering shifts of the $N$-particle system are the sum
of two-particle scattering shifts. Surprisingly, this property also holds for the Toda lattice despite much more intricate many-body
collisions. In fact, the additivity of scattering shifts seems to be a general property of  many-body integrable systems, both classical and quantum. However, except for hard rods, the scattering shift will depend on the incoming momenta.\bigskip\\
\textbf{Two-particle Toda lattice}.\hspace{1pt} The equations of motion are
\begin{equation}\label{2.44}
\ddot{q}_1(t) = -\mathrm{e}^{q_1-q_2}, \quad \ddot{q}_2(t) = \mathrm{e}^{q_1-q_2}
\end{equation} 
with the asymptotic condition $p_1(-\infty) = p_1$,  $p_2(-\infty) = p_2$, and $p_1 < p_2$. The solution is still explicit,
\begin{eqnarray}\label{2.45}
&&q_1(t) = \tfrac{1}{2}\big((p_1+p_2)t -\log (2\gamma)^2 + \log (\mathrm{e}^{\gamma t} + \mathrm{e}^{-\gamma t})\big), \nonumber\\[1ex]
&& q_2(t) = \tfrac{1}{2}\big((p_1+p_2)t +\log (2\gamma)^2 - \log (\mathrm{e}^{\gamma t} + \mathrm{e}^{-\gamma t})\big),
\end{eqnarray} 
with $\gamma = \tfrac{1}{2}(p_1 - p_2)<0$. Hence the large time asymptotics is given by 
\begin{equation}\label{2.46}
q_1(t) = 
\begin{cases}p_1t - \log|p_1 - p_2|,\\[1ex]
p_{2}t -  \log|p_1 - p_2|,
\end{cases} 
q_2(t) = 
\begin{cases}p_2t + \log|p_1 - p_2|, &\quad t \to -\infty,\\[1ex]
p_{1}t + \ \log|p_1 - p_2|, &\quad t \to \infty.
\end{cases} 
\end{equation}
We conclude that the Toda two-particle scattering shift is given by
\begin{equation}\label{2.47}
\phi_{ij} = 2 \log |p_i - p_j|.
\end{equation} 
 The scattering shift has no definite sign. For $|p_1 - p_2| = 1$ the scattering shift vanishes. 
 For  $|p_1 - p_2| <1$ the scattering shift is negative, just as for hard rods. The trajectories of the two Toda particles look similar to the ones
 of hard rods, 
 but the hard rod zero collision time  is smeared to an exponential with rate $\gamma = |p_2 - p_1|/2$. For $|p_1 - p_2| >1$ the 
 scattering shift is positive. The trajectories cross each other, still approaching their asymptotic motion exponentially fast. \bigskip\\
 $\boldsymbol{N}$\textbf{-particle Toda lattice}.\hspace{1pt} For free boundary conditions, also called the \textit{open} chain, the $N$-particle hamiltonian reads
\begin{equation}\label{2.48}
 H^\diamond = \sum_{j=1}^N \tfrac{1}{2}p_j^2 +  \sum_{j=1}^{N-1}\mathrm{e}^{-(q_{j+1} - q_j)},
\end{equation} 
 where for this section only we omit the index $N$ and put the superscript $^\diamond$ to avoid any potential ambiguities.
 In the Flaschka variables the equations of motion become 
 \begin{eqnarray}\label{2.49}
&&\hspace{0pt}\frac{d}{dt} a_j = \tfrac{1}{2}a_j(p_j - p_{j+1}),\quad j = 1,...,N-1,\nonumber\\[1ex]
&&\hspace{0pt} \frac{d}{dt} p_j = a_{j-1}^2 - a_{j}^2, \quad j = 1,...,N, 
\end{eqnarray}
with the boundary conditions $a_0 = 0, a_N = 0$. The Lax matrix $L^\diamond$ equals $L_N$ except for $a_N = 0$, correspondingly
for the companion matrix $B_N$. The time evolution is still encoded as
\begin{equation}\label{2.50} 
\frac{d}{dt} L^\diamond = [B^\diamond,L^\diamond].
\end{equation}
Thus the dynamics of the open chain is also integrable which is a special feature of the Toda lattice. In general, modifying boundary conditions would
break integrability.

As before, the eigenvalue problem is $L^\diamond \psi_j = \lambda_j \psi_j$, $j=1,...,N$. Since the eigenvalues are time-independent,
the eigenvectors are governed by
\begin{equation}\label{2.51} 
\frac{d}{dt} \psi_j(t) = B^\diamond(t)\psi_j(t).
\end{equation}
Particularly simple are the equations for the boundary values $\psi_j(1), \psi_j(N)$. We choose the first entry with the result
\begin{equation}\label{2.52} 
a_1 \psi_j(2)+ p_1\psi_j(1) = \lambda_j \psi_j(1),\qquad \frac{d}{dt}\psi_j(1) = -\tfrac{1}{2}a_1\psi_j(2).
\end{equation}
By the completeness of eigenvectors
\begin{equation}\label{2.53} 
\sum_{j=1}^N\psi_j(m)\psi_j(m') = \delta_{m,m'},
 \end{equation}
which by \eqref{2.52}, left hand side, implies 
\begin{equation}\label{2.54} 
\sum_{j=1}^N\lambda_j \psi_j(1)^2 = p_1. 
 \end{equation}
 Thus rewriting  \eqref{2.52},  right hand side, one obtains
\begin{equation}\label{2.55} 
\frac{d}{dt}\psi_j(1,t) = - \frac{1}{2} \Big( \lambda_j - \sum_{j'=1}^N\lambda_{j'}\psi_{j'}(1,t)^2\Big) \psi_j(1,t),
 \end{equation}
whose solution preserves $\sum_{j=1}^N|\psi_j(1,t)|^2$  as required by \eqref{2.53}. It is convenient to simplify our notation by setting
$\psi_j(1,t) = \eta_j(t)$. Then \eqref{2.55} is solved through
\begin{equation}\label{2.56} 
\eta_j(t)^2 = \frac{\eta_j^2 \mathrm{e}^{- \lambda_jt}}{ \sum_{j'=1}^N\eta_{j'}^2\mathrm{e}^{- \lambda_{j'}t}},
\quad \eta_j = \eta_j(0).
\end{equation}

To step back, in the original problem the $p$'s and the $a$'s build the phase space $\Gamma^\diamond_N = \mathbb{R}^{N} \times  \mathbb{R}_+^{N-1}$. But we now introduced the new phase space
\begin{equation}\label{2.57} 
\Lambda = \big\{\lambda_1, ....\lambda_N, \eta_1,...,\eta_{N}\,\,\mathrm{with}\,\,
\lambda_1 <...<\lambda_N, \eta_j > 0, \sum_{j=1}^N\eta_j^2 = 1\big\}.
\end{equation}
It is a general fact about symmetric Jacobi matrices with strictly positive off-digonal matrix elements, that the map $\Phi:  \Gamma^\diamond_N  \to \Lambda$ is one-to-one and onto.
In our context the map is smooth.  In the old variables the dynamics is complicated, but the observables of interest are 
simple. The new variables trivialize the dynamics, see \eqref{2.56}, but the functions of interest might have become convoluted.
This is a familiar theme in dynamical systems. An obvious example are action-angle variables. Their dynamics is simple,
but the invariant tori might have a complicated shape in the physical phase space. The scattering map $\Phi$ is rather analogous to action-angle variables.
J. Moser provides continued fraction type formulas for $\Phi$ and manages to extract the scattering shift for general $N$. More specifically,  
for the $N$-particle Toda system with open boundary conditions he proves the asymptotics \eqref{2.42} with exponentially small errors.
The scattering shifts $\kappa_j$ are still given by Eq. \eqref{2.43} upon inserting  the Toda scattering shift $\phi_{ij} = 
2 \log|p_i- p_j|$. 

This result leaves us with a puzzle. In generalized hydrodynamics the two-particle scattering shift plays a central role and, in the context  quantum models, is tightly 
linked to the Bethe ansatz for the eigenvectors of the hamiltonian. So how does the logarithmic scattering shift show up in static properties of  the Toda lattice?
\bigskip\\
 \textbf{\large{Notes and references}}
  \bigskip\\ 
  \textbf{ad 2.0}: The discovery by Toda goes back to \cite{T67a,T67b}. The second edition of Toda's book  ``Theory of Nonlinear Lattices'' \cite{T89} is still the most complete account up to 1989.  A standard reference on classical integrable systems is \cite{FT07}. More specifically on the Toda lattice is the review \cite{KT09}. Closer to hydrodynamics is the study of a 
  particular shock problem \cite{VDO91}.  The anniversary volume \cite{BDKT18} provides a glimpse on research in vastly different directions.
  
Toda's work also motivated the search for integrable super-discrete wave equations. A much studied model is called box-ball \cite{TS90},
referring to the fact that lattice sites can be either empty or occupied. The dynamics is given through a deterministic map iterated
many times. The hydrodynamics of the box-ball system has been studied in considerable detail \cite{F18,CS20, KMP20,KMP20a}. 

Early molecular dynamics simulations of the Toda lattice are \cite{SS80,S83}.
  \medskip \\ 
 \textbf{ad 2.1}: Based on explicit soliton solutions, Toda conjectured integrability. For the case of three particles, Ford \cite{F73} obtained very supporting Poincar\'{e} plots.
 The integrals of motion in full generality were obtained by H\'{e}non \cite{H74} by a tricky enumerative argument. H\'{e}non was worried about locality. 
 Flaschka \cite{F74} had the advantage of working at the Courant Institute. Once the Lax matrix had been discovered, locally conserved fields are easily constructed. Independently, the Lax matrix for the Toda lattice has been reported in \cite{M74}. Apparently, at the time and later on, currents were hardly in focus, one exception being \cite{SY10}, which studies  the energy Drude weight in thermal equilibrium. 
  \medskip\\ 
\textbf{ad 2.2}:  To find out the canonically conjugate angles is a much more technical enterprise, which was  accomplished in a series of papers by Henrici and Kappeler \cite{HK08a,HK08b,HK08c}, starting from an early proposal by Flaschka and McLaughlin \cite{FL76},
see also \cite{FFL82} for a complimentary aspects. For classical Hamiltonian systems with a few degrees of freedom, the conventional definition of integrability works fine. 
For an extensive system one simply requires the conventional definition to hold for every $N$. The Landau-Lifshitz chain consists of classical three-component spins, $\vec{S}_j$,
of unit length, $|\vec{S}_j| = 1$. The integrable version of the isotropic chain is governed by the hamiltonian
\begin{equation}\label{2.58} 
H _\mathrm{LL} = -\sum_j \log\big(1+ \vec{S}_j\cdot\vec{S}_{j+1}\big). 
 \end{equation}
As canonical variables one chooses amplitude and phase of the spin components 1 and 2. By isotropy all three components are conserved, but their mutual Poisson brackets do not vanish, see \cite{DDH20,DKSD19} for more details.

For quantum systems, the notion of integrability is controversial, since the naive transcription of the classical notion would mean that every eigenprojection of the Hamiltonian is conserved, in itself not such a helpful observation. I refer to Caux and Mossel \cite{CM11} for an exhaustive discussion. The link between integrability and local conservation laws has been mostly pushed by the quantum community, see  \cite{GM94,GM95} for early work.  For classical systems, in general, this particular avenue still needs to be further developed. 
The mentioned results for the XYZ chain are prototypical for what one would like to achieve. The nonintegrable case is a result of N. Shiraishi \cite{Sh19}. The integrable case,
$h =0$,  has been studied already in   \cite{GM94,GM95} with recent progress \cite{NF20}.
Using the strictly local charges of the XXZ spin chain, one computes  the spin Drude weight at zero magnetization,
i.e. the persistent spin current, by using the Mazur formula. By spin inversion symmetry this weight turns out to be identically $0$. On the other hand for $0 \leq \Delta < 1$, numerical evidence and exact steady state results for the boundary driven chain indicate that the Drude weight does not vanish.
The puzzle is resolved by the construction of quasilocal conserved charges \cite{MPP15}.  \medskip   \\
\textbf{ad 2.3}: In a beautiful piece of analysis J\"{u}rgen Moser \cite{M75} proves the scattering shift for the $N$-particle Toda lattice. An account of his work can be found in \cite{T89}. 
From a different perspective, a more recent discussion are the notes of Percy Deift \cite{De19} from his course  at the Courant Institute in Spring 2019.

 \section{Static properties}
\label{sec3}
\setcounter{equation}{0}
Considering the Toda lattice with periodic boundary conditions and a highly excited initial condition $q,p$, one would expect that in the long time limit a statistically stationary state is reached. For a simple fluid, this state would be thermal equilibrium. But the motion of the Toda lattice is highly constrained through the conservation laws. Still, there are lots and lots of random like collisions. Following Boltzmann, a natural guess for the statistically stationary  state is a generalized microcanonical ensemble, namely the uniform measure on the $(N-1)$-dimensional torus defined by $\ell$ and the Lax eigenvalues $\lambda _1,...,\lambda_N$, see the discussion in Section \ref{2.2}.   The corresponding thermodynamics thus depends on $N+1$ extensive parameters. 
As for a simple fluid, the first step towards hydrodynamics is a study of such generalized thermodynamics. 
\subsection{Generalized Gibbs ensembles}
\label{sec3.1}
For fixed number of lattice sites, the phase space is $(r, p) \in (\mathbb{R} \times \mathbb{R})^N= \tilde{\Gamma}_N$ with the a priori weight
\begin{equation}\label{3.1} 
\prod_{j=1}^N\mathrm{d}r_j \mathrm{d}p_j \delta\Big( \sum_{j=1}^N r_j- \ell\Big), \qquad \ell \in \mathbb{R},
\end{equation} 
where we included already the microcanonical constraint resulting from the boundary conditions \eqref{2.9}. This measure is invariant under the flow generated by Eq. \eqref{2.5}.  Since particles are distinguishable, there is no factor of $1/N!$.  The other conserved fields are taken into account  
through the grand canonical type Boltzmann weight
\begin{equation}\label{3.2} 
\exp\!\Big(  - \sum_{n=1}^N \mu_n Q^{[n],N} \Big).
\end{equation}
Here $\mu_1,...,\mu_N$ are the intensive parameters. Only the low order ones have a physical interpretation, specifically
$\mu_2 = \tfrac{1}{2} \beta$ with $\beta$ the inverse temperature and $\mu_1$ as control parameter for the average total momentum.
As common in Statistical Mechanics we invoke the equivalence of ensembles to lift the delta constraint by the substitution
\begin{equation}\label{3.3} 
\delta\big( Q^{[0],N} - \ell\big)\quad \Rightarrow \quad \exp\!\big( - P Q^{[0],N}\big).
\end{equation}
 This equivalence has been extensively studied in rigorous statistical mechanics and presumably
some of the techniques can be used also for the Toda lattice. Along with other items, we have to leave this problem for future studies.
For a general anharmonic chain the physical pressure, $P_\mathrm{phys}$, is defined as the average force between neighboring particles in thermal 
equilibrium. Using a simple integration by parts, one obtains the relation $ P_\mathrm{phys} = \beta^{-1}P$. We still refer to  $P$ as pressure, since it is the thermodynamic dual of the stretch. To have an integrable Boltzmann weight, $P> 0$ is required. In combination  the \textit{generalized Gibbs ensemble} (GGE) is defined through  
\begin{equation}\label{3.4} 
\prod_{j=1}^N\mathrm{d}r_j \mathrm{d}p_j \exp\!\Big( - P Q^{[0],N} - \sum_{n=1}^N \mu_n Q^{[n],N} \Big),
\end{equation}
which still has to be normalized. The GGE is invariant under our dynamical flow.

As for the dynamical evolution, a priori measure and Boltzmann weight are transformed to Flaschka variables. But before, our notation is improved by introducing the \textit{confining potential}
\begin{equation}\label{3.5}
V(w) = \sum_{n=1}^\infty\mu_n w^n,
\end{equation}
which is required to be independent of $N$. We assume that $V$ is smooth and has a lower bound as $V(w) \geq c_0 +c_1|w|$ with $c_1 >0$, hence the label ``confining". Instead of the chemical potentials $\mu_n$, the primary object will be the confining potential. The 
quadratic confining potential, $V(w) = \tfrac{1}{2}\beta w^2$, corresponds to thermal
equilibrium.  Using \eqref{3.5} and the representation \eqref{2.13} for the conserved fields, the Boltzmann weight transforms to
\begin{equation}\label{3.6} 
\exp\!\big(-\mathrm{tr}[V(L_N)]\big) \prod_{j=1}^{N}  \mathrm{d}p_j  \mathrm{d}a_j \frac{2}{a_j}
(a_j)^{2P},
\end{equation}
which is defined on the phase space of the Flaschka variables, namely $ \Gamma_N = (\mathbb{R}_+\times  \mathbb{R})^N$. \bigskip\\
$\blackdiamond\hspace{-1pt}\blackdiamond$~\textit{Infinite volume limit, exponential mixing}.\hspace{1pt} As for other Gibbs measures, one might want to know about the existence of the infinite volume limit for the sequence of measures in \eqref{3.6}, the limit measure being independent of boundary conditions, and the qualitative decay of correlations. If the confining potential is a
finite polynomial, bounded from below, such questions can be answered by using transfer matrix techniques.  In the language of statistical mechanics, $Q^{[n]}$ has a range of size $n$. By assumption the leading coefficient of $V$ is even, say $2\kappa$, and  strictly positive. 
One cuts $[1,...,N]$ in blocks of size $2\kappa$. The density in \eqref{3.6} can then be written as  an $(N/2\kappa)$-fold power of the transfer matrix. This is just like the familiar case of the one-dimensional Ising model, in which case the transfer matrix is a $2\times 2$ matrix. For the Toda lattice the transfer matrix is given by an integral kernel with arguments in $\Gamma_{2\kappa}\times \Gamma_{2\kappa}$.
By the Perron-Frobenius theorem, the transfer matrix has a unique maximal eigenvalue, which is separated by a 
gap from the rest of the spectrum. With this input,
one concludes that there is a unique limit measure. In one dimension, phase transitions would  occur only if the interaction 
potential has a decay slower than (range)$^{-2}$, much slower than the case under consideration here. The spectral gap also implies exponential decay of correlations. 
If $\kappa = \infty$ such methods fail completely. Other techniques will have to be developed. \hfill$\blackdiamond\hspace{-1pt}\blackdiamond$
\subsection{Generalized free energy}
\label{sec3.2}
The Toda partition function is defined by
\begin{equation}\label{3.7} 
Z_{\mathrm{toda},N}(P,V)  = \int_{\Gamma_N} \prod_{j=1}^{N}  \mathrm{d}p_j  \mathrm{d}a_j \frac{2}{a_j}
(a_j)^{2P} \exp\!\big(-\mathrm{tr}[V(L_N)]\big).
\end{equation}
Accordingly, normalizing  the expression in \eqref{3.6}, one arrives at the probability measure
\begin{equation}\label{3.8} 
\mu^\mathrm{GGE}_{P,V,N} (\mathrm{d}a\mathrm{d}p) = \frac{1}{Z_{\mathrm{toda},N}(P,V)} \prod_{j=1}^{N}  \mathrm{d}p_j  \mathrm{d}a_j \frac{2}{a_j}
(a_j)^{2P} \exp\!\big(-\mathrm{tr}[V(L_N)]\big).
\end{equation}
This measure is time-stationary under the dynamics \eqref{2.5}, since the a priori measure and the eigenvalues of $L_N$ do not change in time.
The  expectations of $\mu^\mathrm{GGE}_{P,V,N}$ will be denoted by $\langle \cdot \rangle_{P,V,N}$. As  central thermodynamic object,  the free energy per lattice site is defined by 
\begin{equation}\label{3.9} 
F_\mathrm{toda}(P,V) = - \lim_{N \to \infty} \frac{1}{N}\log Z_{\mathrm{toda},N}(P,V). 
\end{equation}
For hydrodynamics, a basic input is the GGE average of the conserved fields. They are computed as  derivative with respect to $P$ and  as variational  derivative with respect to $V$ of the
free energy. In terms of the eigenvalues of the Lax matrix, the averages can be written as
 \begin{equation}\label{3.10} 
\frac{1}{N} \langle Q^{[n],N} \rangle_{P,V,N} = \frac{1}{N} \big\langle \sum_{j=1}^N (\lambda_j)^n \big\rangle_{P,V,N}
= \int_\mathbb{R} \mathrm{d}w w^n  \langle\rho_{\mathrm{Q},N} (w)\rangle_{P,V,N}, 
\end{equation}
where
 \begin{equation}\label{3.11} 
\rho_{\mathrm{Q},N} (w) = \frac{1}{N}  \sum_{j=1}^N \delta(w - \lambda_j)
\end{equation}
is the  \textit{empirical density of states} (DOS) of the Lax matrix $L_N$. Here empirical refers to the fact that the DOS is defined for every collection of eigenvalues $\{\lambda_1,...,\lambda_N\}$. Thus $\rho_{\mathrm{Q},N} (w)$ is a random function under 
$\mu^\mathrm{GGE}$.

This observation suggests a novel perspective. The Lax matrix becomes a random matrix under $\mu^\mathrm{GGE}$.
Thermal equilibrium is particularly simple. Since $\mathrm{tr}[(L_N)^2]= \sum_{j=1}^N(p_j^2 + 2a_j^2)$, the diagonal and off-diagonal matrix elements of $L_N$ are families of independent 
identically distributed  (i.i.d.) random variables. For all other GGEs the matrix elements are correlated. The DOS encodes the complete statistical information on the conserved fields. In fact $\rho_{\mathrm{Q},N} (w)$ is self-averaging with fluctuations of order
$1/ \sqrt{N}$ and the limit
 \begin{equation}\label{3.12} 
 \lim_{N\to\infty} \rho_{\mathrm{Q},N} (w) = \rho_\mathrm{Q}(w)
\end{equation}
exists with some nonrandom limiting density $\rho_\mathrm{Q}$.\bigskip\\
$\blackdiamond\hspace{-1pt}\blackdiamond$~\textit{The Dumitriu-Edelman identity}.\hspace{1pt} In 2002 I. Dumitriu and A. Edelman studied the $\beta$-ensembles of random matrix theory.
To ease a comparison, I describe their result in the original notation,  in which some symbols will reappear with a different meaning. The proper translation
will be obvious, however. Their starting point is a $n \times n$, symmetric, tridiagonal matrix $T$ with real matrix elements $T_{j,j} = a_j$, $T_{j,j+1} = 
b_j >0$, and zero otherwise, compare with $L^\diamond$ from Section \ref{sec2.2}, in particularly, $(a,b) \in \Gamma^\diamond_N
=  \mathbb{R}^{N} \times  \mathbb{R}_+^{N-1}$. The eigenvalues are ordered as $\lambda_1 <....< \lambda_n$
and the first component of an eigenvector is denoted by $\psi_j(1) = q_j > 0$ with $|q| = 1$ imposed. We set $\mathrm{d}a = \prod_{j=1}^n  \mathrm{d}a_j$,  $\mathrm{d}b = \prod_{j=1}^{n-1}  \mathrm{d}b_j$, $\mathrm{d}\lambda = \prod_{j=1}^n  \mathrm{d}\lambda_j$, and 
$\mathrm{d}q$ is the surface element of the unit $n$-sphere. As mentioned in Section \ref{sec2.2}, there is the one-to-one and onto map $\Phi: (a,b) \mapsto (\lambda, q)$. Dumitriu and Edelman managed to obtain some information on the Jacobian of $\Phi$.
Their identity has a free parameter $\beta > 0$ and reads
\begin{equation}\label{3.13} 
 \prod_{j=1}^{n-1}\big(2 (b_j)^{\beta j-1}\big) \mathrm{d}a \mathrm{d}b = \Big(c_q^\beta \prod_{j=1}^n(q_j)^{\beta-1}\mathrm{d}q\Big)\big(n!\zeta_n(\beta) 
 \Delta (\lambda)^\beta\mathrm{d}\lambda\big).
\end{equation}
 Here $c_q^\beta =  2^{n -1} \Gamma(\tfrac{1}{2}\beta n) \Gamma( \tfrac{1}{2}\beta)^{- n}$ normalizes the first factor to 1. In the second factor,
 $\Delta(\lambda)$ is the Vandermonde determinant
 \begin{equation}\label{3.14} 
\Delta(\lambda) = \prod_{1 \leq i<j\leq n}(\lambda_j - \lambda_i)
\end{equation}
and $\zeta_n(\beta)$ a nomalization constant given by
 \begin{equation}\label{3.15} 
\zeta_n(\beta) = \Gamma(\tfrac{1}{2}\beta n)^{-1} \Gamma(1+ \tfrac{1}{2} \beta)^n \prod_{j=1}^n \frac{\Gamma( \tfrac{1}{2} \beta j)}
{\Gamma(1 + \tfrac{1}{2} \beta j)}. 
\end{equation}

Below Eq. \eqref{2.4} we mentioned the free multiplicative factor in the definition of the Flaschka variables. To use the identity \eqref{3.13}  constrained by the condition that $V$ remains unchanged 
fixes our choice  in \eqref{2.4}.  \bigskip\hfill $\blackdiamond\hspace{-1pt}\blackdiamond$


We insert the Dumitriu-Edelman identity into our partition function \eqref{3.7} and use the spectral representation of $\mathrm[V(L_N)]$. The integration over the $n$-sphere is normalized
to $1$. The integration over eigenvalues is unordered, which takes care of the factor $n!$.  The free parameter $\beta$
is now chosen specifically as
 \begin{equation}\label{3.16} 
\beta = \frac{2P}{N}.
\end{equation}
Then  
\begin{eqnarray}\label{3.17} 
&&\hspace{-39pt}Z_{\mathrm{de},N}(P,V) = \int_{\Gamma^\diamond_N}\exp\!\big(-\mathrm{tr}[V(L^\diamond_N)]\big) \prod_{j=1}^{N}  \mathrm{d}p_j  
\prod_{j=1}^{N-1}\mathrm{d}a_j \frac{2}{a_j}
(a_j)^{2 (j/N)P} \nonumber\\
&&\hspace{18pt} = \zeta_N(P)\int_{\mathbb{R}^N} \mathrm{d} \lambda 
\exp\!\Big(- \sum_{j=1}^N V(\lambda_j) + P \frac{1}{N}\sum_{i,j=1,i \neq j}^N \log|\lambda_i - \lambda_j|\Big). 
\end{eqnarray}
with normalization
\begin{equation}\label{3.18} 
\zeta_N(P) = \Gamma(P)^{-1} \Gamma(1+ \tfrac{P}{N})^N \prod_{j=1}^N \frac{\Gamma(\tfrac{j}{N})}
{\Gamma(1 + \tfrac{j}{N})}.
\end{equation}
In the first identity, except for a boundary term, we note the Toda partition function with a pressure changing linearly with slope $1/N$.
This is not exactly what is required, since for the Toda chain the pressure is constant. But in a large segment, with size $\ll N$, the pressure is constant to a very good approximation.
Since GGEs have good spatial mixing properties, local free energies merely add up and one concludes that 
\begin{equation}\label{3.19} 
-\lim_{N\to\infty}\frac{1}{N}\log Z_{\mathrm{de},N}(P,V) = \int_0^1\mathrm{d}u F_\mathrm{toda}(uP,V).
\end{equation}

The term on the right side of \eqref{3.17} requires more explanations. For the normalization one obtains
\begin{equation}\label{3.20} 
- \lim_{N\to\infty}\frac{1}{N}\log \zeta_N(P) = \log P -1 .
\end{equation}
Otherwise the partition function is the one of the repulsive one-dimensional log gas. $V$ turns out to be  the confining potential of the log gas,
which is the real reason for our original choice of name. The standard log gas has an interaction term of strength $1$,
which implies that the free energy is dominated by the interaction energy. But in our case the interaction strength is $1/N$,
which is the standard mean-field scaling. In terms of the DOS,  compare with \eqref{3.11}, except for the diagonal contribution, the energy in \eqref{3.17} can be expressed as
\begin{equation}\label{3.21} 
N\Big(\int_\mathbb{R} \mathrm{d}w V(w) \rho_{\mathrm{Q},N} (w) - P\int_\mathbb{R} \mathrm{d}w\int_\mathbb{R} \mathbb{d}w'
\log|w - w'| \rho_{\mathrm{Q},N} (w)\rho_{\mathrm{Q},N} (w')\Big).
\end{equation}
One has to add the usual entropy term resulting from the $\mathrm{d}\lambda$ measure. This then leads to the mean-field free energy functional  
\begin{equation}\label{3.22}
\mathcal{F}^\mathrm{MF}(\varrho) =  \int _\mathbb{R}\mathrm{d}w \varrho(w) V(w)      - P\int _\mathbb{R}\mathrm{d}w\int _\mathbb{R}\mathrm{d}w'   \log|w - w'|\varrho(w) \varrho(w') + 
\int _\mathbb{R}\mathrm{d}w \varrho(w) \log \varrho(w).
\end{equation} 
The free energy is obtained by  minimizing over all $\varrho$ with $\varrho \geq 0$ and $\int_\mathbb{R} \mathrm{d}w \varrho(w) = 1$. As will be discussed below,
there is a unique minimizer, $\varrho^\star$, and thus
\begin{equation}\label{3.23} 
-\lim_{N\to\infty}\frac{1}{N}\log Z_{\mathrm{de},N}(P,V) = \mathcal{F}^\mathrm{MF}(\varrho^\star) +\log P -1.
\end{equation}
To obtain the Toda free energy we use
\begin{equation}\label{3.24} 
\mathcal{F}^\mathrm{MF}(\varrho^\star) +\log P -1 = \int_0^1\mathrm{d}u F_\mathrm{toda}(uP,V)
\end{equation}
and hence
\begin{equation}\label{3.25} 
F_\mathrm{toda}(P,V) = \partial_P(P\mathcal{F}^\mathrm{MF}(\varrho^\star)) +\log P.
\end{equation}

It turns out to be more convenient to absorb $P$ into $\varrho$ by setting $\rho = P\varrho$. Then 
$P \mathcal{F}^\mathrm{MF}(P^{-1}\rho)=  \mathcal{F}(\rho) -P\log P$ with 
the transformed free energy functional 
\begin{equation}\label{3.26}
\mathcal{F}(\rho) =  \int _\mathbb{R}\mathrm{d}w \rho(w) V(w)      - \int _\mathbb{R}\mathrm{d}w\int _\mathbb{R}\mathrm{d}w'   \log|w - w'|\rho(w) \rho(w') + 
\int _\mathbb{R}\mathrm{d}w \rho(w) \log \rho(w).
\end{equation} 
$\mathcal{F}$ has to be minimized under the constraint
\begin{equation}\label{3.27}
\rho(w) \geq 0,\quad  \int _\mathbb{R}\mathrm{d}w\rho(w) =P 
\end{equation}
with minimizer denoted by $\rho^\star$. Then
 \begin{equation}\label{3.28} 
 F_\mathrm{toda}(P,V) =  \partial_P \mathcal{F}(\rho^\star) - 1.
 \end{equation}
 
The constraint \eqref{3.27} is removed by introducing the Lagrange multiplier $\mu$ as
  \begin{equation}\label{3.29} 
 \mathcal{F}_\mu(\rho) =  \mathcal{F}(\rho) - \mu \int _\mathbb{R}\mathrm{d}w \rho(w).
 \end{equation}
A minimizer of  $\mathcal{F}_\mu(\rho)$ is denoted by  $\rho_\mu$ and determined as solution of the Euler-Lagrange equation
\begin{equation}\label{3.30} 
  V(w)   - \mu -  2 \int_\mathbb{R} \mathrm{d}w'  \log|w-w'| \rho_\mu(w') +\log \rho_\mu(w) = 0.
 \end{equation}
The Lagrange parameter $\mu$ has to be adjusted such that
\begin{equation}\label{3.31} 
 P =  \int _\mathbb{R}\mathrm{d}w  \rho_\mu(w).
 \end{equation}
 To obtain the Toda free energy, we differentiate as
 \begin{eqnarray}\label{3.32} 
 && \hspace{-40pt} \partial_P \mathcal{F}(\rho^\star) =
  \int_\mathbb{R} \mathrm{d}w \partial_P\rho^\star(w) V(w)      - 2 \int_\mathbb{R} \mathrm{d}w\int_\mathbb{R} \mathrm{d}w' \log|w-w'| \rho^\star(w)\partial_P\rho^\star(w')\nonumber\\
 && \hspace{40pt} 
 + \int_\mathbb{R} \mathrm{d}w( \partial_P\rho^\star(w) ) \log \rho^\star(w) +1 .
  \end{eqnarray}
 Integrating \eqref{3.30} against  $\partial_P\rho^\star$ one arrives at
 \begin{equation}\label{3.33} 
 \partial_P \mathcal{F}(\rho_\mu) = \mu +1
\end{equation}
 and thus 
 \begin{equation}\label{3.34} 
  F_\mathrm{toda}(P,V) =  \mu(P,V).
 \end{equation}
Sharing with other integrable models, the Toda lattice has the property that its free energy 
is determined by a variational problem for densities over $\mathbb{R}$, in our case  normalized to $P$.
\subsection{Lax density of states, TBA equation}
\label{sec3.3}
To obtain the hydrodynamic equations required is the GGE average of the conserved fields, $\langle Q^{[n]}_0\rangle_{P,V}$, for which purpose there are two equivalent methods.
One can start from the microscopic definition and use that  $Q^{[n],N}$ depends only on the eigenvalues of the Lax matrix. The other method, employed here, is to simply differentiate the free energy. We start with $n=0$ and note that the average stretch
\begin{equation}\label{3.35} 
\nu= \langle Q^{[\mathrm{0}]}_{0}\rangle_{P,V} = \partial_P F_\mathrm{toda}(P,V) = \partial_P\mu(P,V) = \Big(\int _\mathbb{R}\mathrm{d}w  \partial_\mu\rho_\mu(w) \Big)^{-1},
\end{equation}
where the last equality results from differentiating Eq. \eqref{3.31} as $1 = (\int \partial_\mu\rho_\mu) \mu'(P)$. For $n\geq 1$ we perturb $V$ as $V_\kappa(w) = V(w) + \kappa w^n$ and differentiate the free energy at $\kappa = 0$.  Then 
\begin{equation}\label{3.36} 
 \langle Q^{[n]}_{0}\rangle_{P,V} = \partial_\kappa F_\mathrm{toda}(P,V_\kappa)\big|_{\kappa = 0} = \partial_P \partial_\kappa \mathcal{F}(\rho^\star(P,V_\kappa))\big|_{\kappa = 0} 
 \end{equation}
 and, first introducing the linearization of $\rho^\star$ as
\begin{equation}\label{3.37} 
\partial_\kappa \rho^\star(P,V_\kappa )\big|_{\kappa = 0} =  \rho{^\star}',
\end{equation}
one obtains
\begin{eqnarray}\label{3.38} 
 && \hspace{-40pt}  \partial_\kappa \mathcal{F}(\rho^\star(P,V_\kappa))\big|_{\kappa = 0} =
\int_\mathbb{R} \mathrm{d}w \rho^\star(w,P,V)w^n + \int_\mathbb{R} \mathrm{d}w V(w) \rho{^\star}'(w) \\
 &&\hspace{-10pt} 
  - 2 \int_\mathbb{R} \mathrm{d}w\int_\mathbb{R} \mathrm{d}w'  \log|w-w'| \rho{^\star}'(w)\rho^*(w',P,V) + \int_\mathbb{R} \mathrm{d}w 
  \rho{^*}'(w) \log \rho^\star(w,P,V),\nonumber
  \end{eqnarray}
using that $\int \mathrm{d}w \rho{^\star}'(w)= 0$.
 Integrating the Euler-Lagrange equation  \eqref{3.30} at $\mu = \mu(P)$ against $\rho{^\star}'$, the terms on the right side of \eqref{3.38} vanish and
 \begin{equation}\label{3.39} 
 \langle Q^{[n]}_{0}\rangle_{P,V} 
 =  \int_\mathbb{R}\mathrm{d}w \partial_P\rho^\star(w,P,V) w^n. 
\end{equation}
Thus the Lax DOS is given by
\begin{equation}\label{3.40} 
 \rho_\mathrm{Q}(w) = \partial_P\rho^\star(w).
\end{equation}
Naively one might have guessed that the Lax DOS equals $\varrho^\star$. But the linear variation of the pressure in the Dumitriu-Edelman identity amounts to a slightly deviating  result. 

In the literature the Euler-Lagrange equation \eqref{3.30} is written differently by formally introducing a Boltzmann weight through
 \begin{equation}\label{3.41} 
 \rho_\mu(w) = \mathrm{e}^{-\varepsilon(w)}
\end{equation}
with quasi-energy $\varepsilon(w)$. Then
\begin{equation}\label{3.42} 
 \varepsilon(w) = V(w) -\mu -  2 \int_\mathbb{R} \mathrm{d}w'  \log|w-w'| \mathrm{e}^{-\varepsilon(w')}.
 \end{equation}
  
 The structure uncovered is very familiar from the Yang-Yang thermodynamics of the Lieb-Liniger $\delta$-Bose gas,
 which is a quantum many-body integrable system and solved by Bethe ansatz. Yang and Yang considered the thermal state, 
 hence only a quadratic $V$ appears in their equations. But the extension to a general confining potential is fairly straightforward. In Section \ref{sec10},
 we will compare Toda lattice and  $\delta$-Bose gas  in considerable detail. 
 For quantum integrable systems the analogue of \eqref{3.42} is called TBA 
 (thermodynamic Bethe ansatz) equation. Because of the similarities still to be discussed, we call \eqref{3.42} \textit{classical TBA equation} or simply TBA, despite the fact that no Bethe ansatz had been used in its derivation.
 
 Later on we will use identities based on TBA. We collect them here, together with introducing standard notations.
 The Hilbert space of square integrable functions on the real line is denoted by $L^2(\mathbb{R}, \mathrm{d}w)$ with scalar product
 \begin{equation}\label{3.43} 
\langle f,g\rangle = \int_\mathbb{R} \mathrm{d}w f(w)^* g(w).
 \end{equation}
 Since we work only with real functions, the complex conjugation in \eqref{3.43} can be omitted. There will be many integrals 
 over $\mathbb{R}$ and a convenient  shorthand is simply
  \begin{equation}\label{3.44} 
\langle f \rangle = 
 \langle f,1\rangle  = \int_\mathbb{R} \mathrm{d}w f(w).
 \end{equation}
 To distinguish, an average over some probability measure is denoted by  $ \langle \cdot\rangle_{P,V,N}$, carrying suitable subscripts.
 Starting  from the $Q^{[n]}$'s, so far a discrete basis has been used. Obviously any linear combination of conserved fields is still conserved
 and, as in other linear problems, the choice of basis is an important consideration. From the viewpoint of Lax DOS, the label 
 $n$ corresponds to the monomial $w^n$, which will be still used and is denoted by 
  \begin{equation}\label{3.45} 
 \varsigma_{n}(w) = w^n. 
  \end{equation}
  
 Let us define the integral operator 
\begin{equation}\label{3.46}
T\psi(w) = 2 \int_\mathbb{R} \mathrm{d}w' \log |w-w'| \psi(w'),\quad w \in \mathbb{R}.
\end{equation}
Then the  TBA equation can be rewritten as 
\begin{equation}\label{3.47} 
\varepsilon (w)  = V(w)  -  \mu  - (T \mathrm{e}^{-\varepsilon})(w).
 \end{equation}
One introduces the  dressing of a function $\psi$  through
\begin{equation}\label{3.48} 
\psi^\mathrm{dr} = \psi + T \rho_\mu \psi^\mathrm{dr},\quad \psi^\mathrm{dr} = \big(1 - T\rho_\mu\big)^{-1} \psi.
\end{equation}
where $\rho_\mu$ is regarded as multiplication operator, i.e. $(\rho_\mu\psi)(w) = \rho_\mu(w)\psi(w) $.
With our improved notation the Lax DOS in \eqref{3.40} can be written as 
\begin{equation}\label{3.49} 
\rho_{\mathrm{Q}} =  \partial_P\rho_\mu = ( \partial_P\mu)\partial_\mu \rho_\mu = \nu \rho_\mathsf{p}, \quad \partial_\mu \rho_\mu = \rho_\mathsf{p},  \quad \nu\langle\rho_\mathsf{p}\rangle = 1,
 \end{equation}
compare with \eqref{3.35}. Differentiating TBA with respect to $\mu$ we conclude
\begin{equation}\label{3.50} 
\rho_\mathsf{p}= (1 - \rho_\mu T)^{-1} \rho_\mu = \rho_\mu(1 - T\rho_\mu)^{-1}\varsigma_0 = \rho_\mu \varsigma_0^\mathrm{dr}
 \end{equation}
 with the convention $\varsigma_0(w) = 1$. The relation $ \partial_P\rho_\mu = \nu \rho_\mathsf{p}$ seems to be special for 
 the Toda lattice. But the identity $\rho_\mathsf{p} = \rho_\mu \varsigma_0^\mathrm{dr}$ generalizes to other models, compare with Sections 
 \ref{sec9.2} and \ref{sec10}. For later purposes, in we state
 \begin{equation}\label{3.51} 
q_n =   \langle Q^{[n]}_{0}\rangle_{P,V}  = \nu\langle\rho_\mathsf{p}\varsigma_n\rangle.
\end{equation}
Of physical relevance are $\nu$ and $\nu\rho_\mathsf{p}$, since they encode the GGE average of the conserved fields.
  \bigskip\\
$\blackdiamond\hspace{-1pt}\blackdiamond$~\textit{Uniqueness of solutions of the TBA equation}.\hspace{1pt}  For the Toda lattice at given $\mu$ the TBA equation has two solutions.
 At first glance this looks surprising. In fact, in a standard numerical solution scheme one follows a particular branch and encounters  
 an end-point at which instabilities arise. So, some explanations are in demand.
 
Firstly $\mu(P)$ is convex down and its derivative, $\nu(P)$, is strictly decreasing. For example, in the case of thermal equilibrium 
 $\mu(P) = \log \sqrt{\beta/2\pi} +P\log \beta -\log\Gamma(P)$,  
which has a single maximum at $P=P_\mathrm{c}$.  The physics is rather obvious. At very small $P$ the average stretch is huge and diverges
as $P \to 0$. By increasing the pressure the stretch is  decreased. Since there is no hard core, increasing $P$ even further the stretch becomes negative. In physical space, for small $P$, up to small random errors, the labelling of particles is increasing. But at large $P$ the labelling is reversed.
At $P_\mathrm{c}$ the stretch vanishes
and the typical distance between particles with adjacent index is of order $1/\sqrt{N}$.
The function  inverse to $\mu(P)$ has two branches, meaning that for given $\mu$ there are two values of $P$.

Now considering the densities, by construction $\rho_{\mu} \geq 0$,  $\nu \rho_\mathsf{p} \geq 0$, $\nu \langle\rho_\mathsf{p}\rangle =1$.
$\rho_{\mu} \geq 0$ is pointwise increasing in $P$ and  varies smoothly through $P_\mathrm{c}$, so does $\nu \rho_\mathsf{p}$. On the other hand, $\rho_\mathsf{p}(w)$ diverges to $+\infty$
as $P$ approaches $P_\mathrm{c}$ from the left, globally flips to  $-\infty$, and then flattens out as $P \to \infty$.
\hfill$\blackdiamond\hspace{-1pt}\blackdiamond$\bigskip\\
\textbf{Thermal equilibrium}.\hspace{1pt}
Thermal equilibrium corresponds to the quadratic confining potential $V(w) = \tfrac{1}{2} \beta w^2$
with $\beta$ the inverse temperature. 
Only for this particular case the diagonal entries of the Lax matrix, 
 $\{p_j, j \in \mathbb{Z}\}$,  are independent  with $p_j$ a Gaussian random variable of mean zero and variance $\beta^{-1}$.
Hence $\langle (p_j)^n \rangle_{P,\beta}  = 0$ for odd $n$ and  $\langle (p_j)^n \rangle_{P,\beta}  = (n-1)!!(\beta)^{-n/2}$ for even $n$. 
The off-diagonal entries, $\{a_j, j \in \mathbb{Z}\}$,  are also independent with $a_j$ a 
$\chi$ distributed random variable with parameter $2P$. In particular for the even moments $\langle (a_j)^{2n} \rangle_{P,\beta}  = 
P(P+1)...(P+n -1)$, $n = 1,2,...\,$. Because of independence, the free energy of the chain is easily computed with the result
\begin{equation}\label{3.52}
F_\mathrm{eq}(P,\beta) =  \log \sqrt{\beta/2\pi} +P\log \beta -\log\Gamma(P).
\end{equation}
However to figure out the entire DOS requires the TBA machinery. 

We start from the Euler-Lagrange equation for \eqref{3.22}, set $V(w) = \tfrac{1}{2} \beta w^2$, differentiate
with respect to $w$, and multiply the resulting expression by $\varrho^\star$. Then
\begin{equation}\label{3.53} 
( \beta w + \partial_w) \varrho^\star(w) -  2 P \int_\mathbb{R} \mathrm{d}w' \frac{1}{w - w'} \varrho^\star(w)\varrho^\star(w') = 0.
 \end{equation}
 Note that $\beta$  scales by setting
 \begin{equation}\label{3.54}
 \varrho^\star_{P,\beta}(w) = \sqrt{\beta} \varrho^\star_{P,1}(\sqrt{\beta}w).
  \end{equation}
 For simplicity we drop $P$ and denote  by $\varrho^\star$ the solution to \eqref{3.53} upon setting $\beta = 1$.
 
 Taking the Stieltjes transform, 
 \begin{equation}\label{3.55}
 g(z) =   \int_\mathbb{R} \mathrm{d}w\varrho{^\star}(w) \frac{1}{w -z},
  \end{equation}
yields the equation
\begin{equation}\label{3.56}
 zg(z) + \frac{d}{dz}g(z) + P g(z)^2 = -1 .
  \end{equation}
Setting $g(z) = u'(z)/u(z)$, Eq. \eqref{3.56} is transformed to the linear second order differential equation
\begin{equation}\label{3.57}
 u''(z) + zu'(z)+ Pu(z) = 0.
  \end{equation}
Since $\varrho{^*}$ is a probability density, the $|z| \to \infty$ asymptotics,
 \begin{equation}\label{3.58}
 u(z) \simeq \frac{c_1}{z^P},
  \end{equation}
follows.
  Finally changing to the function $u(z) = \mathrm{e}^{z^2/4} y(z) $ one arrives at the Schr\"{o}dinger type equation
\begin{equation}\label{3.59}
 y''(z) + \big( P - \tfrac{1}{2} - \tfrac{1}{4} z^2\big)y(z) = 0,
  \end{equation}
  which can be solved in terms of parabolic cylinder functions. The appropriate linear combination is determined by the asymptotic condition \eqref{3.58}. Somewhat unusually, the Stieltjes transform can be still inverted and yields the fairly explicit expression
  \begin{equation}\label{3.60}
\varrho{^\star}(w) = \int_0^\infty \mathrm{d}t f_P(t) \mathrm{e}^{\mathrm{i}wt}, \quad f_P(t) = (P/\Gamma(P))^\frac{1}{2}t^{P - 1}  \mathrm{e}^{-\frac{1}{2}t^2}. 
\end{equation}

For small $P$, the Lax off-diagonal matrix elements $a_j \to 0$ and hence in leading order $\varrho^\star(w) = (2\pi)^{-1/2} \exp[-\tfrac{1}{2}w^2]$. 
On the other hand, one can integrate \eqref{3.53} against $\varsigma_n(w)$ to obtain a recursion relation for the even moments of $\varrho^\star$, $c_n = \langle \varrho^\star(w) w^{2n}\rangle$, 
\begin{equation}\label{3.61}
c_n = (2n-1)c_{n-1}  +  P \sum_{j=0}^{n-1}c_{n-1-j} c_j,\quad c_0 = 1,\quad n =1,2,...\,.
\end{equation} 
For large $P$, the second term dominates implying the asymptotic result
\begin{equation}\label{3.62}
c_n =  P^n   \frac{1}{n+1}
 \begin{pmatrix}
2n\\
n
\end{pmatrix}. 
\end{equation}  
On the right side one notes the Catalan numbers. Hence $\varrho^\star$ is the normalized Wigner semi-circle probability distribution function
\begin{equation}\label{3.63}
\varrho^\star(w) = \frac{1}{2\pi P} \sqrt{4 P - w^2}, \quad w^2 \leq 4 P.
\end{equation}
To obtain the Lax DOS one still has to act with $\partial_P P$. The low pressure Gaussian does not change, while the high pressure density becomes 
 \begin{equation}\label{3.64}
\rho_\mathrm{Q}(w) \simeq \frac{1}{\pi \sqrt{(2P +w)(2P -w)}}, \quad w^2 \leq 4P.
  \end{equation}
The Lax eigenvalues concentrate close to $\pm 2P$. \bigskip\\
\textbf{\large{Notes and references}}
  \bigskip\\
  \textbf{ad 3.0}: At age twenty-four Ludwig Boltzmann wrote his fundamental contribution \cite{B68a} on  the microcanonical
  ensemble. He argued that this ensemble provides the natural description of the long time behavior of a mechanical system and, in addition,
has close links to thermodynamics. Recommended is his very readable letter-style account \cite{B68b}. The way how we teach equilibrium statistical mechanics today goes back to the must-read  book by J. Williard Gibbs \cite{G02}.\medskip\\
   \textbf{ad 3.1}: The generalized Gibbs ensemble came naturally into focus when studying the quench of integrable systems 
   starting from a translation invariant state, see the reviews \cite{PSS11,VR16}, the special volume \cite{CEM16}, and the more recent short review with many references
   \cite {AC17}. In this context the conserved charges averaged over the initial state determine the parameters of the GGE obtained in the long time limit.
    In some circles the notion ``generalized" was initially not so welcome, since there is a long tradition in the study of Gibbs measures 
    for a general class of potentials \cite{R69,L73,FV17}. In our context ``generalized" means that for the given mechanical system one has a high dimensional  set of time-invariant measures all of them having the form anticipated by Statistical Mechanics. Equivalence of ensembles in its original meaning refers to the property that the resulting thermodynamic potentials are related to each other through the respective Legendre transform. This is a widely studied subject, with close relations to the theory of large deviations. In our context we mostly use a much stronger version:  In the limit  of infinite volume, two distinct ensembles constructed from the same bulk hamiltonian yield  identical distributions of strictly local observables. In other words,  provided the respective parameters are transformed according to the rules of thermodynamics, the local correlations are identical. Such a stricter version fails  
at phase transitions and is thus a more subtle notion. 

 A very general theory for Gibbs measures in one dimension has been developed by R.L. Dobrushin in \cite{D74}.\medskip  \\
  \textbf{ad 3.2}: The generalized free energy of the Toda lattice has been obtained in \cite{S19}, see also \cite{D19a,D19b} for a complimentary discussion. 
  Originally, Dumitriu and Edelman \cite{DE02} investigated how the fully filled GUE random matrix transforms isospectrally to a 
  tridiagonal matrix. For this purpose they iteratively applied the Householder transform, which is a standard numerical scheme. Once Dumitriu  and Edelman had understood 
 how this scheme works for the GUE random matrix, they realized that their algebra holds for arbitrary values of the parameter $\beta$, not only for $\beta = 2$. Their identity has been used to obtain a fairly detailed information on the edge behavior of the general $\beta$-ensemble \cite{RRV11}, see \cite{BEY14} for further developments. A standard reference on log gases is the monumental volume by \medskip Forrester \cite{F10}. 
  \textbf{ad 3.3}:  The particular form of the TBA formalism can be better grasped upon reading Section \ref{sec10}. The double-valuedness of solutions to the TBA equation was pointed out to me by Cao  and Bulchandani \cite{BCM19,CBS19}. Numerical plots of the densities as they vary through $P_\mathrm{c}$ can be found in \cite{MS20}.
The TBA equation \eqref{3.42} with quadratic confining potential was first obtained by Opper \cite{O85}, who investigated 
the classical limit of the quantum Toda chain. He already obtained the solution \eqref{3.60}. Our discussion is based on 
\cite{ABG12}, a study of Dyson Brownian motion with weak interaction, compare with Section \ref{sec4}.   A related study has been carried out for $\beta$-Wishart ensembles \cite{ABM12}. The recursion relation \eqref{3.61} appears already in \cite{DS15,D18}.

\section{Mean-field Dyson Brownian motion}
\label{sec4}
\setcounter{equation}{0}
In the mid-sixties Freeman Dyson pioneered the study of random matrices, as initiated by Eugene Wigner and others 
as a phenomenological model for the complex energy spectra of highly excited nuclei. More specifically, Dyson studied the statistics of eigenvalues
for the Gaussian orthogonal, unitary, and symplectic ensembles. For this purpose he introduced a stochastic dynamics of eigenvalues with the property that its stationary distribution coincides with the distribution of  eigenvalues of the respective random matrix ensemble. Such a diffusion process is now called
Dyson Brownian motion. In this way, rather than focusing on eigenvalues, one studies a physically more intuitive stochastic particle system. The particles move in one dimension, their positions being denoted by
$\{x_j(t), j = 1,...,N\}$, and are governed by the coupled stochastic differential equations      
\begin{equation}\label{4.1}
dx_j(t) = -V'(x_j(t))dt + 2\beta\sum_{i = 1,i\neq j}^N \frac{1}{x_j(t) - x_i(t)} dt + \sqrt{2} db_j(t), \quad j = 1,...,N, \quad\beta \geq 0. 
\end{equation}
The particles repel each other with a $1/x$ repulsive force and are confined by the  external potential $V$, which in fact will coincide with the generalized chemical potential  \eqref{3.5} of the Toda lattice, which in the first place was the reason for using the same name and
symbol. $\{ b_j(t), j = 1,...,N\}$ is a collection of independent standard Brownian motions.
For later purposes, the generator, $\mathcal{L}_N$, of the coupled diffusion processes is defined as the linear operator
\begin{equation}\label{4.2}
\mathcal{L}_N \mathsfit{f}(x) = \sum_{j=1}^N\Big(\partial_{x_j}- V'(x_j) + 2\beta\sum_{i = 1,i\neq j}^N \frac{1}{x_j - x_i}\Big) \partial_{x_j}\mathsfit{f}(x), 
\end{equation}
acting on functions on configuration space, $\mathsfit{f}: \mathbb{R}^N \to \mathbb{R}$,
$x = (x_1,...,x_N)$. The kernel of the semigroup
\begin{equation}\label{4.3}
\mathrm{e}^{\mathcal{L}_Nt}(x,x') \mathrm{d} x', \quad t \geq 0,
\end{equation}
is the transition probability, in other words, the probability density at time $t$ given the initial configuration $x$. 

The reader might suspect a luxury detour. But this is not the case. Dyson Brownian motion is a powerful method to numerically 
solve the TBA  equations of the Toda lattice. Even more importantly, as will be discussed, the GGE averaged currents are linked to
eigenvalue fluctuations. To study such properties,  Dyson Brownian motion is a convenient tool.

Dyson investigated the particular values $\beta = 1,2,4$, which possess a large symmetry group. But recently there has 
been much progress also for general $\beta$. In our context, we will have to study the case of very small $\beta$,
specifically $\beta = \alpha/N$. In fact, $\alpha =P$ in the application to the Toda lattice. In \eqref{4.1} the drift is the gradient 
of a potential. Hence the diffusion process is reversible and its unique stationary measure is given by
\begin{equation}\label{4.4}
(Z_{\alpha,V,N})^{-1} \exp\!\Big(- \sum_{j=1}^N V(x_j) 
+\frac{\alpha}{N}\sum_{i,j =1, i \neq j}^N  \log| x_i - x_j| \Big),
\end{equation} 
 which agrees with the normalized version of the probability density function in \eqref{3.17}. Our strategy will be to first study the large $N$ limit of the dynamics and then deduce information on the stationary measure, which is our real interest. \bigskip\\
 \textbf{Macroscopic equation, law of large numbers}.\hspace{1pt} We choose some smooth test function $f$ and introduce the empirical
 density, $\varUpsilon_N(x,t)$, through
 \begin{equation}\label{4.5}
\varUpsilon_N(f,t) = \frac{1}{N} \sum_{j = 1}^N f(x_j(t))  = \int_\mathbb{R}\mathrm{d}x  \varUpsilon_N(x,t) f(x). 
\end{equation}
Then
\begin{eqnarray}\label{4.6} 
&&\hspace{-45pt}d\varUpsilon_N(f,t)= \varUpsilon_N( -V'\partial_x f + \partial_x^2f,t) dt\\
&&\hspace{10pt}+ \alpha \int_\mathbb{R}\mathrm{d}x \int_\mathbb{R} \mathrm{d}y
\frac{f'(x) - f'(y)}{x-y} \varUpsilon_N(x,t)\varUpsilon_N(y,t) dt + \frac{1}{N}  \sum_{j = 1}^N f'(x_j(t))\sqrt{2} db_j(t).\nonumber
 \end{eqnarray}
 When integrated in time, the process averaged square of the noise term becomes
\begin{equation}\label{4.7}
 \frac{2}{N^2} \sum_{j = 1}^N \int _0^t \mathrm{d} s \mathbb{E}\big(f(x_j(s))^2\big).
\end{equation}
Thus the noise is of order $1/\sqrt{N}$ and vanishes in the limit $N \to \infty$.

We assume a starting measure such that with probability one the initial empirical density has the limit
$\rho_0$,
 \begin{equation}\label{4.8}
\lim_{N \to \infty} \varUpsilon_N(f,0) = \int_\mathbb{R} \mathrm{d}x\rho_0(x) f(x). 
\end{equation} 
In the statistical physics literature this property is often referred to as ``self-averaging", meaning that the limit is non-random and no averaging is required.
Since in \eqref{4.6} only the drift term survives, one concludes  the limit
\begin{equation}\label{4.9}
\lim_{N \to \infty}  \varUpsilon_N(f,t) = \varUpsilon(f,t) = \int_\mathbb{R} \mathrm{d}x \rho(x,t) f(x)
\end{equation} 
for all $t>0$, again with probability one. The limit density $ \rho(x,t) $ then satisfies
\begin{eqnarray}\label{4.10} 
&&\hspace{-30pt}\frac{d}{dt} \int_\mathbb{R}\mathrm{d}x \rho(x,t) f(x) \\
&& =  \int_\mathbb{R}\mathrm{d}x\rho(x,t)(-V'\partial_x f + \partial_x^2f)(x)
+ \alpha \int_\mathbb{R}\mathrm{d}x \int_\mathbb{R} \mathrm{d}y
\frac{f'(x) - f'(y)}{x-y} \rho(x,t) \rho(y,t) \nonumber
\end{eqnarray}
together with the initial data $\rho_0(x)$. Written pointwise, $\rho(x,t)$ is governed by the nonlinear Fokker-Planck equation
\begin{equation}\label{4.11}
\partial_t \rho(x,t) = \partial_x\big( V_\mathrm{eff}'(x,t)\rho(x,t) + \partial_x  \rho(x,t)\big).
\end{equation} 
The bare confining potential $V$ is modified to  an effective potential given by 
\begin{equation}\label{4.12}
V_\mathrm{eff}(x,t) = V(x) - \alpha (T \rho)(x,t),
\end{equation} 
$T$ the  integral operator being defined in \eqref{3.46}.

Our particular interest is the stationary Fokker-Planck equation 
\begin{equation}\label{4.13}
\partial_x\Big( V'(x)\rho_\mathrm{s}(x) - 2\alpha\int _\mathbb{R}\mathrm{d}y\frac{1}{x-y}\rho_\mathrm{s}(y) \rho_\mathrm{s}(x)  + \partial_x  \rho_\mathrm{s}(x)\Big) = 0,\quad
  \int _\mathbb{R}\mathrm{d}x\rho_\mathrm{s}(x) = 1,
  \end{equation} 
which has a unique strictly positive solution $\rho_\mathrm{s}$.
Since Dyson Brownian motion is time-reversible, the large round bracket itself has to vanish. We compare to the mean-field free energy
functional 
$\mathcal{F}^\mathrm{MF}(\rho)$ in Eq. \eqref{3.22}, setting $P = \alpha$. Its minimizer $\rho^\star$ satisfies the Euler-Lagrange equation
\begin{equation}\label{4.14} 
  V(x)  - \mu -  2\alpha \int_\mathbb{R} \mathrm{d}x'  \log|x-x'| \rho^\star(x') +\log \rho^\star(x) + 1 = 0,
  \end{equation}
where $w$ has been substituted by $x$. Differentiating with respect to $x$ and then multiplying by $\rho^\star$ yields
\begin{equation}\label{4.15} 
  V'(x)\rho^\star(x) -  2 \alpha \int_\mathbb{R} \mathrm{d}y \frac{1}{x-y} \rho^\star(x)\rho^\star(y) + \partial_x\rho^\star(x) = 0.
 \end{equation}
Thus we conclude $\rho_\mathrm{s} = \rho^\star$.

The nonlinear Fokker-Planck equation depends smoothly on $\alpha$ and so does $\rho_\mathrm{s}$.
The issue of two-valuedness appears only in the TBA equation. \bigskip\\
\textbf{Fluctuation theory}.\hspace{1pt} The nonlinear Fokker-Planck equation is the result of a law of large numbers. Thus one would expect to have
a central limit type theorem which captures the order $1/\sqrt{N}$ fluctuations. Quite generically the fluctuations are governed
by a linear Langevin equation for the conserved field, in our case the density field. The drift term is determined by the linearized macroscopic equation. But in addition 
the dynamics generates an effective noise term, whose precise structure depends on the model.

Our interest is the stationary dynamics, hence the initial state of the $N$-particle system is the one in Eq. \eqref{4.4}. To study the fluctuations of the density it is convenient to  introduce the fluctuation field as 
\begin{equation}\label{4.16}
\phi_N(f,t) = \frac{1}{\sqrt{N}} \sum_{j=1}^N \big(f(x_j(t)) - \langle\rho_\mathrm{s} f \rangle\big) = \int_\mathbb{R} \mathrm{d}x f(x) \phi_N(x,t).
\end{equation}
As explained in more detail below,  there is a Gaussian random field $\phi(f,t)$, jointly in $f$ and $t$, such that in distribution
 \begin{equation}\label{4.17}
\lim_{N\to \infty} \phi_N(f,t) = \phi(f,t). 
\end{equation} 
Note that now the limit is still random. The limit field   $\phi$ is governed by the linear Langevin equation
\begin{equation}\label{4.18}
\partial_t \phi(x,t) = \partial_x \big(D\phi(x,t) + \sqrt{2\rho_\mathrm{s}(x)}\xi(x,t)\big).
\end{equation}  
Here  $\xi(x,t)$ is normalized spacetime Gaussian white noise, $\mathbb{E}\big(\xi(x,t)\xi(x',t')\big) =
\delta(x - x')\delta(t-t')$ and $D$ the linear operator 
\begin{equation}\label{4.19}
D =  V'_\mathrm{eff} + \partial_x - \alpha \rho_\mathrm{s} T' .
\end{equation} 
Thus $\partial_x D$ is the Fokker-Planck evolution operator linearized  at $\rho_\mathrm{s}$. The effective potential 
$V_\mathrm{eff}$ is still defined as in \eqref{4.12} upon  substituting $\rho(x,t)$  by $\rho_\mathrm{s}(x)$. 

The Gaussian process $\phi(x,t)$ is stationary in time, has  mean zero and is uniquely characterized by its covariance
$\mathbb{E}\big(\phi(x,t)\phi(x',t')\big)$. For the Toda lattice, of interest is the spatial covariance $\mathbb{E}\big(\phi(x,0)\phi(x',0)\big)
= C^\sharp(x,x')$. As a general property of linear Langevin equations with time-independent coefficients,
$C^\sharp$ is determined by 
\begin{equation}\label{4.20}
\langle D^*\partial_x f,C^\sharp g\rangle + \langle f,C^\sharp D^*\partial_x g\rangle = 2\langle \partial_x f, \rho_\mathrm{s} \partial_x g\rangle
\end{equation} 
with $D^*$ the adjoint operator to $D$. We claim that, as an operator, the solution is 
\begin{equation}\label{4.21}
C^\sharp = (1 - \alpha \rho_\mathrm{s} T)^{-1}\rho_\mathrm{s} - \big\langle(1 - \alpha\rho_\mathrm{s}T)^{-1}\rho_\mathrm{s}
\big\rangle^{-1}  \big|(1 - \alpha \rho_\mathrm{s} T)^{-1}\rho_\mathrm{s}\big\rangle \big\langle (1 - \alpha\rho_\mathrm{s}T)^{-1}\rho_\mathrm{s}\big|,
\end{equation} 
where for simplicity we use the Dirac notation $|\cdot\rangle\langle \cdot|$ for the one-dimensional projector. The subtracted  term ensures that the number of particles does not fluctuate, i.e. $C^\sharp \varsigma_0 = 0$.

We consider only the left most term of \eqref{4.20}, the other one following by symmetry, and have to show that  
\begin{equation}\label{4.22}
\langle \partial_x f,D \rho_\mathrm{s} (1 -\alpha T \rho_\mathrm{s})^{-1}g\rangle = \langle \partial_x f, \rho_\mathrm{s} \partial_x g\rangle.
\end{equation} 
Upon  replacing $g$ by $(1 -\alpha T \rho_\mathrm{s})g$, one arrives at
\begin{equation}\label{4.23}
\langle \partial_x f,D \rho_\mathrm{s} g\rangle = \langle \partial_x f, \rho_\mathrm{s} \partial_x (1 -\alpha T \rho_\mathrm{s})g\rangle.
\end{equation} 
$\rho_\mathrm{s}$ satisfies the stationary nonlinear Fokker-Planck equation \eqref{4.15},  
\begin{equation}\label{4.24}
\big(V_\mathrm{eff}'  +  \partial_x\big)\rho_\mathrm{s} = 0.
\end{equation} 
When inserted in \eqref{4.23} this leads to the condition 
\begin{equation}\label{4.25}
\rho_\mathrm{s}  \partial_x g -\alpha \rho_\mathrm{s} T' ( \rho_\mathrm{s}g) = \rho_\mathrm{s} \partial_x (1 -\alpha T \rho_\mathrm{s})g,
\end{equation} 
which is clearly valid. Thereby  our claim \eqref{4.22} is verified. \bigskip\\
$\blackdiamond\hspace{-1pt}\blackdiamond$~\textit{Martingales and central limit theorem}.\hspace{-1pt}  In probability theory tremendous efforts
have been invested to
develop tools for proving the central limit theorem in case of dependent random variables. Such techniques
can be applied to our infinite-dimensional setting, which is required, since the test functions form an infinite-dimensional linear space. 
Only the basic computational steps are explained here. A full proof would be too technical. 
For example, we would have to discuss the existence of solutions to \eqref{4.1}. In fact, if $\beta > 1$, trajectories never touch and the solution theory is standard. But for $\beta < 1$ the crossing probability is 
strictly positive and the existence of solutions becomes more intricate. The limit theorem \eqref{4.17} is based on compactness. In an appropriate function space one ensures that along some subsequence the stochastic process of fluctuations has a limit. 
The limit process satisfies an equation for which uniqueness is established, which then implies convergence.
We consider only the stationary process, but the time-dependent case can be handled in a similar fashion. 

Using \eqref{4.2}, one arrives at
\begin{equation}\label{4.26}
\mathcal{L}_N\xi_N(f) = \xi_N( -V'\partial_x f + \partial_x^2f) 
+ \alpha \int_\mathbb{R}\mathrm{d}x \int_\mathbb{R} \mathrm{d}y
\frac{f'(x) - f'(y)}{x-y} \xi_N(x)\varUpsilon_N(y) = \varkappa_{[1],N}(f).
\end{equation}
For the quadratic variation we obtain
\begin{equation}\label{4.27}
\mathcal{L}_N\xi_N(f)^2  - 2\xi_N(f)\mathcal{L}_N\xi_N(f)= 2N^{-1}\sum_{j=1}^N f'(x_j)^2 = \varkappa_{[2],N}(f).
\end{equation}
As before,  $\varkappa_{[1],N}(f,t)$ is the function $\varkappa_{[1],N}(f)$ evaluated at the random configuration\\
$(x_1(t),...,x_N(t))$, and the same for $\varkappa_{[2],N}(f,t)$. By the standard theory of Markov processes,
\begin{equation}\label{4.28}
M_{[1],N}(f,t) = \xi_N(f,t) - \xi_N(f,0) - \int_0^t \mathrm{d}s \varkappa_{[1],N}(f,s)
\end{equation}
and
\begin{equation}\label{4.29}
M_{[2],N}(f,t) =  M_{[1],N}(f,t)^2 -  \int_0^t \mathrm{d}s \,\varkappa_{[2],N}(f,s)
\end{equation}
are martingales. 

In the limit $N \to \infty$ the martingale property is preserved. For $\varkappa_{[1],N}(f)$ and $\varkappa_{[2],N}(f)$  one can use
the law of large numbers. Then the limit expressions  
\begin{eqnarray}\label{4.30}
&&\hspace{-30pt} M_{[1]}(f,t) = \xi(f,t) - \xi(f,0) -  \int_0^t \mathrm{d}s \,\xi( -V'\partial_x f + \partial_x^2f,s) \nonumber\\
&&\hspace{20pt}-  \int_0^t \mathrm{d}s \, \alpha \int_\mathbb{R}\mathrm{d}x \int_\mathbb{R} \mathrm{d}y
\frac{f'(x) - f'(y)}{x-y} \xi(x,t)\rho_\mathrm{s}(y)
\end{eqnarray}
 and   
\begin{equation}\label{4.31}
M_{[2]}(f,t) =  M_{[1]}(f,t)^2 - 2 t \int_\mathbb{R}\mathrm{d}x\rho_\mathrm{s}(x) f'(x)^2
\end{equation}
are still martingales under the limit process. The unique solution of this martingale problem is the Langevin equation of  \eqref{4.18},
as already promised.
\hfill$\blackdiamond\hspace{-1pt}\blackdiamond$\bigskip\\
\textbf{\large{Notes and references}}
  \bigskip\\
  \textbf{ad 4}:  Freeman Dyson introduced his model in \cite{D62}. An interesting account of the early history is his preface to The Oxford Handbook on Random Matrix Theory \cite{ABF11}, which serves as a useful source of information and allows one to capture the vastness of the subject. For  general $\beta > 0$, Dyson Brownian motion is properly constructed in \cite{CL97}, where also the law of large numbers is proved. We also refer to the introductory monograph on random matrices  \cite{AGZ10}. The fluctuation theory at strong coupling is established in  \cite{I01}. The methods can be extended to the case discussed
  here.
 Fluctuation theory for stochastic particle systems in greater generality is discussed in \cite{S91} with more detailed arguments and references on the  martingale method. For thermal equilibrium, eigenvalue fluctuations are proved in \cite{NT18} by a different technique.
 
 Dyson Brownian motion also serves as a technical tool in the study of universality for Wigner random matrices \cite{EYY12}
 and the edge behavior of the density of states of the beta random matrix ensembles \cite{BEY14}.
\section{Hydrodynamics for hard rods}
\label{sec5}
\setcounter{equation}{0} 
So far our focus has been the generalized free energy. But hydrodynamics relies on further building blocks.
The hard rod system will serve as an illustration of the general method with our discussion guided by  
the analogy  to the Toda lattice.\bigskip\\
 \textbf{Hard rod fluid}.\hspace{1pt} The fluid consists of hard rods of length $\mathsfit{a} > 0$ which interact through elastic collisions.
 The lattice version will be discussed below. The positions are ordered as $... < q_j < q_{j+1} < ...$\,. Besides the number of particles,
 any sum function of the velocities is conserved. This field is labelled by some general function, $\phi$, 
 and has a density given by
 \begin{equation}\label{5.1}
 Q^{[\phi]}(x) = \sum_{j \in \mathbb{Z}} \delta(q_j - x) \phi(p_j).
 \end{equation}
 In fact, also the particle number is included and corresponds to the constant function, $\phi(x) = 1$. 
 
 For the current one has to differentiate the density with respect to $t$. Then there is the flow term resulting from the $\delta$-function and a collision term resulting  from  $\phi$. 
 For the latter
 we consider a small time $t>0$ and the change
$\phi(p_j(t)) - \phi(p_j)$. If $r_j -\mathsfit{a} < \delta$ with sufficiently small $\delta$ and if $p_{j+1}< p_j$, then $p_j(t) = p_{j+1}$.
Similarly, if $r_{j-1} -\mathsfit{a} < \delta$ and if $p_{j-1} >  p_j$, then $p_j(t) = p_{j-1}$, which implies
 \begin{eqnarray}\label{5.2}
&&\hspace{-34pt}\phi(p_j(\mathrm{d}t)) - \phi(p_j) 
= \delta(r_j - \mathsfit{a})(\phi(p_{j+1}) - \phi(p_j))\chi(\{p_{j+1}< p_j\})(p_j - p_{j+1}) \mathrm{d}t  \nonumber\\
&&\hspace{66pt}
+\delta(r_{j-1}- \mathsfit{a}) (\phi(p_{j-1}) - \phi(p_j))\chi(\{p_{j} < p_{j-1}\})(p_{j-1} - p_{j}) \mathrm{d}t,
\end{eqnarray}
where $\chi(\{\cdot\})$ is the indicator function of the set $\{\cdot\}$. Thus  
 \begin{eqnarray}\label{5.3}
 &&\hspace{-25pt}J^{[\phi]}(x) = \sum_{j \in \mathbb{Z}} \big(\delta(q_j - x) p_j \phi(p_j) \\
 &&\hspace{0pt} +  (\theta(q_{j} -x) - \theta(q_{j-1} -x)) 
\delta(r_{j-1}- \mathsfit{a}) (\phi(p_{j-1}) - \phi(p_{j}))\chi(\{p_{_j} < p_{j-1}\})(p_{j-1} - p_{j})\big)\nonumber
 \end{eqnarray}
 with $\theta$ the step function, $\theta(x) =0$ for $x\leq 0$ and $\theta(x) =1$ for $x > 0$.
 
 A GGE is characterized by the distribution function $f(v) = \rho h(v) \geq 0$. The velocities $\{p_j, j \in \mathbb{Z}\}$
 are a family of i.i.d. random variables with common probability density $h$ and $\rho$ is the macroscopic density of particles. 
 The positions $\{q_j, j \in \mathbb{Z}\}$ are statistically invariant under spatial shifts and 
 $\{r_j = q_{j+1} - q_j \}$ are i.i.d., such that $r_j$ is exponentially distributed with density $P\exp\big(-P(x -\mathsfit{a})\big)\chi(\{x\geq \mathsfit{a}\})$. Then $\rho = P/(1 + \mathsfit{a}P)$. 
In the sense of thermodynamics, $\rho, h$ are extensive parameters, while, setting $h = \mathrm{e}^{-V}/\langle  \mathrm{e}^{-V}\rangle$,
$P,V$ are intensive parameters. 
Slightly deviating from our standard  convention the GGE average is denoted by 
$\langle \cdot \rangle_{\rho,h}$. The required GGE averages are easily computed. For the fields one obtains 
\begin{equation}\label{5.4}
 \langle Q^{[\phi]}(0) \rangle_{\rho,h} = \rho \langle  h \phi\rangle.
 \end{equation}
For the currents one uses that
\begin{equation}\label{5.5}
\int_\mathbb{R}\mathrm{d}v \int_\mathbb{R}\mathrm{d}w h(v) h(w)
\big(\phi(w) - \phi(v)\big) \chi(\{v< w\})(w-v) = \int_\mathbb{R}\mathrm{d}v \phi(v)(v-u)h(v)
\end{equation}
with the mean velocity $u = \langle vh(v)\rangle$. Then
\begin{equation}\label{5.6}
 \langle J^{[\phi]}(0) \rangle_{\rho,h} = \rho\langle h  v^\mathrm{eff} \phi\rangle 
 \end{equation}
and the effective velocity
\begin{equation}\label{5.7}
  v^\mathrm{eff}(v)   = \frac{1}{1 - \mathsfit{a} \rho} ( v -  \mathsfit{a} \rho u).
\end{equation}

The hydrodynamic evolution equations can be build from the input \eqref{5.4}, \eqref{5.6}, and \eqref{5.7}. For the initial state one assumes to have  locally the statistics of  some GGE. In our case, one example would be to maintain independence but let $f = \rho h $ change very slowly  on the scale set  by the maximal correlation length. On a more formal level one introduces the dimensionless  parameter $\epsilon$,  $\epsilon \ll 1$, of a typical scale for the ratio ``microscopic length\,/\,macroscopic length''. Microscopically the physically relevant length is the  maximal correlation length of the initial GGE. Translated to the initial $f$, the scale separation is ensured through the functional form $f(\epsilon x; v)$. 
With such random initial conditions, the hard rod fluid is evolving through free motion and collisions. Independence is lost immediately. Yet on the macroscopic scale nothing is moving,
 at least not for short times. Because of the conservation laws, changes can be observed only on microscopic times of order $\epsilon^{-1}t$.
The hydrodynamic scaling has to be ballistic, since  changes in space and in time are both $\mathcal{O}(\epsilon ^{-1})$ in microscopic units. Hydrodynamics then postulates that    
even for such long times local GGE is approximately valid. If propagation of local GGE holds, one can write down the appropriate hydrodynamic equations. The starting point is the exact microscopic conservation law
\begin{equation}\label{5.8}
\partial_t Q^{[\phi]}(x,t) +\partial_x J^{[\phi]}(x,t)=0.
\end{equation}
Since this law holds for any dynamical trajectory, one can average over an arbitrary ensemble, in particular the initial slowly varying state. 
Instead of averaging the time $t$ fields over initial conditions, we average the physical fields over the statistical state at time $t$, average denoted by $\langle \cdot \rangle_{t,\epsilon}$ with the subscript $\epsilon$ indicating the parameter of initial slow variation,
\begin{equation}\label{5.9}
\partial_t\langle Q^{[\phi]}(x)\rangle_{t,\epsilon} +\partial_x  \langle J^{[\phi]}(x)\rangle_{t,\epsilon} =0,
\end{equation}
which is still exact. Now, switching to macroscopic spacetime $(\epsilon^{-1}x, \epsilon^{-1}t)$ propagation of local GGE amounts to 
the approximation
\begin{equation}\label{5.10}
 \langle Q^{[\phi]}(\epsilon^{-1}x) \rangle_{\epsilon^{-1}t,\epsilon} \simeq \langle Q^{[\phi]}(0)\rangle_{f(x,t)},\qquad
 \langle J^{[\phi]}(\epsilon^{-1}x) \rangle_{\epsilon^{-1}t,\epsilon} \simeq \langle J^{[\phi]}(0) \rangle_{f(x,t)},
\end{equation}
where on the right hand side of each equation  $Q^{[\phi]}(x)$ is replaced by $Q^{[\phi]}(0)$, resp.  $J^{[\phi]}(x)$ by $J^{[\phi]}(0)$, because of the spatial translation invariance of the GGE average
$\langle \cdot \rangle_{f(x,t)}$. Using \eqref{5.4}, \eqref{5.6}, 
 and \eqref{5.7}, we arrive at a closed equation for $f(t)$,
\begin{equation}\label{5.11}
\partial_t f(x,t;v) +\partial_x \big((1 - \mathsfit{a} \rho(x,t))^{-1} ( v -  \mathsfit{a} \rho(x,t) u(x,t))f(x,t;v)\big) = 0.
\end{equation}
Here the argument of $f$ indicates that $(x,t)$ is the spacetime patch under consideration and $v$ is the label of the conserved
field in that patch. Eq. \eqref{5.11} is the hydrodynamic equation of the hard rod fluid.  Through the nonlinearity the hydrodynamic fields are coupled and decouple only in the ideal gas limit
$\mathsfit{a} \to 0$. 

As an apparently general feature of generalized hydrodynamics, the hydrodynamic equation \eqref{5.11} can be written in quasilinear form through the transformation 
\begin{equation}\label{5.12}
\rho_\mu(v) = \frac{1}{\nu - \mathsfit{a}}f(v), \quad \nu \rho = 1.
\end{equation}
Then
\begin{equation}\label{5.13}
 	\partial_t \rho_\mu(x,t;v) +v^\mathrm{eff}(x,t;v)\partial_x \rho_\mu(x,t;v) = 0.
\end{equation}
The fields $\rho_\mu(v)$ are the normal modes of the hard rod fluid equations and the convective equation \eqref{5.13} identifies their local propagation velocities as $v^\mathrm{eff}(v)$, which  depend nonlinearly on  $\rho_\mu$ through \eqref{5.7}.

Hydrodynamic equations are expected to become exact in the limit $\epsilon \to 0$. But in practice, hydrodynamics becomes applicable much before the actual limit. The precise value of $\epsilon$ depends on the particular physical system and its initial conditions. The range of validity for hydrodynamics is mostly a qualitative rule of thumb, rather than a sharp error  \bigskip estimate.\\
$\blackdiamond\hspace{-1pt}\blackdiamond$~\textit{Hydrodynamics of simple fluids, the issue of mathematical rigor}.\hspace{1pt} In 1757 Leonhard Euler published his monumental
treatise on fluids, which in particular included the compressible Euler equations.  In this case the locally conserved fields are 
number, momentum, and total energy. Euler argued for the appropriate form of the currents on a phenomenological basis. It was a triumph of early Statistical Mechanics to understand that these currents can also be obtained from averaging the microscopic currents
over a Gibbs ensemble. The required technical tool is an identity known as virial theorem. Since the resulting equations 
agreed with what had been known already for a long time, the discussion was mostly confined to the textbook level.   
Still, the derivation was a further compelling evidence for the universal validity of Statistical Mechanics. For  integrable many-body
systems one cannot rely on phenomenological evidence. The only available method is to figure out the general form of the GGE averaged  currents. 

A much harder question is the propagation of local GGEs. For the hard rod fluid Roland Dobrushin and collaborators 
proved such a result in the mid 80ies. But for more complicated integrable systems, in particular the Toda lattice, no
results are available yet. Because of their physical importance, for simple fluids there have been many attempts to establish the compressible Euler equations in a mathematically convincing way. From that perspective the best result is still the work of S. Olla, 
S.R.S. Varadhan, and H.-T. Yau. They investigated the kind of dynamical mixing which would be needed for propagation of local equilibrium. In fact OVY modified the mechanical model by assuming that the deterministic collisions between particles are substituted by random ones, still respecting the mechanical conservation laws. Thus the macroscopic equations are the compressible version of Euler. 
For the stochastic dynamics OVY prove the required mixing. But for the mechanical system such mixing would have to come from deterministic chaos. The resulting gap is so huge that progress is unlikely, even allowing for a long term perspective. 
\hfill$\blackdiamond\hspace{-1pt}\blackdiamond$\bigskip\\
 \textbf{Hard rod lattice}.\hspace{1pt}
Physically the fluid picture can be more easily visualized. But as in the case of the Toda chain, one can switch to the lattice version.
Then the dynamics is formulated in terms of the stretches $r_j$, where  $r_j(t) \geq \mathsfit{a}$ because of the hard core.
If $p_j(t) -p_{j+1}(t) < 0$, then $r_j(t)$ is decreasing and reaches collision at
$r_j(t_\mathrm{c}) = \mathsfit{a}$, $t_\mathrm{c}$ the collision time.   Instantaneously the momenta $(p_j,p_{j+1})$ are exchanged and $r_j(t)$ is increasing for
$t> t_\mathrm{c}$. Such local collisions happen for every stretch throughout time. Obviously the densities of the conserved fields are
 \begin{equation}\label{5.14}
 Q^{[0]}_j = r_j, \qquad Q^{[\phi]}_j = \phi(p_j).
 \end{equation}
 Note that $\phi =1$ is just the constant function, hence not admissible. Thus, in contrast to the fluid, the index $0$ will have to be treated  differently 
 from $n\geq 1$. The currents are
\begin{equation}\label{5.15}
 J^{[0]}_j = -p_j
 \end{equation}
 and
 \begin{equation}\label{5.16}
 J^{[\phi]}_j = \delta(r_{j-1}- \mathsfit{a}) (\phi(p_{j-1}) - \phi(p_{j}))\chi(\{p_{j} < p_{j-1}\})(p_{j-1} - p_{j}).
 \end{equation} 
 Compared to \eqref{5.3}, there is no flow term and the factor $(\theta(q_{j} -x) - \theta(q_{j-1} -x))$ is missing.
 
 In a hard rod lattice GGE, the velocities are i.i.d. with probability density $h(v)$, $h \geq 0$, $\langle h \rangle = 1$ and 
 the $r_j$'s  are also i.i.d.  with  probability density $P\exp\big(-P(x -\mathsfit{a})\big)\chi(\{x\geq \mathsfit{a}\})$. The average stretch is then
$\nu = P^{-1} +\mathsfit{a}$. Now $\nu, h$, are extensive parameters, while $P,V$ are still intensive parameters. The GGE average is denoted by  $\langle \cdot \rangle_{\nu,h}$. The averaged conserved fields are given by 
\begin{equation}\label{5.17}
 \langle Q^{[0]}_0  \rangle_{\nu,h} = \nu, \qquad \langle Q^{[\phi]}_0  \rangle_{\nu,h} = \langle h \phi \rangle
 \end{equation}
and the averaged currents by
\begin{eqnarray}\label{5.18}
&& \langle J^{[0]}_0 \rangle_{\nu,h} = - \langle h v\rangle = -u, \nonumber\\[1ex] 
&&\langle J^{[\phi]}_0 \rangle_{\nu,h} = (\nu - \mathsfit{a})^{-1}\big( \langle hv\phi\rangle - u\langle h \phi\rangle\big)
= \nu^{-1}\langle h(v^\mathrm{eff} - u) \phi\rangle,
 \end{eqnarray}
 where we used the same computation as in \eqref{5.5} and in Eq. \eqref{5.7} rewritten as  
$v^\mathrm{eff}(v)   = (\nu - \mathsfit{a})^{-1} ( \nu v -  \mathsfit{a}  u)$.
As a result the hydrodynamic equations of the hard rod lattice take the form
\begin{equation}\label{5.19}
\partial_t \nu(x,t) - \partial_x u(x,t) = 0, \quad
 \partial_t h(x,t;v) +\partial_x \big((\nu(x,t) - \mathsfit{a})^{-1}(v -u(x,t))h(x,t;v)\big) = 0.
\end{equation}
Compared to \eqref{5.11}, there are now two equations rather than one.
But switching to normal modes through
\begin{equation}\label{5.20}
\rho_\mu(v) = \frac{1}{\nu- \mathsfit{a}}h,
\end{equation}
one arrives again at the quasilinear equation \eqref {5.13}.\bigskip\\
\textbf{TBA, collision rate assumption}.\hspace{1pt}  As commonly agreed, for integrable many-body systems the structure of the hydrodynamic equations is in essence independent of the microscopic system. Thus the hard rod fluid serves as an interesting test case.
So for the moment, we ignore the particle structure and only use the property that the hard rod two-particle scattering shift equals $-\mathsfit{a}$. 
Hence the integral operator $T$ of  \eqref{3.46} becomes the projection
 \begin{equation}\label{5.21}
 T = -\mathsfit{a} |1\rangle\langle 1|,
 \end{equation}
in other words $T\psi(w) = - \mathsfit{a} \langle \psi \rangle$, and the TBA equation for the hard rod lattice reads
\begin{equation}\label{5.22}
 V- \mu + \mathsfit{a} \langle \varrho_\mu \rangle +\log \varrho_\mu = 0, \quad \langle \varrho_\mu \rangle =P.
 \end{equation}
 Since $T$ is such a simple operator, the solution is still explicit with
 \begin{equation}\label{5.23}
\mu = \log P + \mathsfit{a} P - \log   \langle \mathrm{e}^{-V} \rangle
 \end{equation}
 and 
 \begin{equation}\label{5.24}
h =  \mathrm{e}^{-V} /\langle\mathrm{e}^{-V} \rangle, \quad \varrho_\mu  = (\nu - \mathsfit{a})^{-1}h,\quad  \varrho_\mathsf{p}
 =  \nu^{-1}h.
 \end{equation}
The average of the conserved fields is determined by 
\begin{equation}\label{5.25}
\langle Q_0^{[\phi]}\rangle_{\nu,h}  =  \langle h\phi\rangle = \nu \langle \varrho_\mathsf{p}\phi\rangle.
 \end{equation} 
 in agreement with \eqref{5.17}.
 
We proceed with a conventional argument for the GGE averaged currents, which is based on the notion of quasiparticles.  
They are most easily visualized in the fluid picture, which is adopted for a little while. A quasiparticle retains
 its velocity when undergoing a collision. While a physical particle rattles back and forth between its two neighbors, the quasiparticle moves with constant velocity interrupted by jumps of size $\pm\mathsfit{a}$ at collisions. We now prepare a GGE characterized
 by $\rho,h$ and initially place  at the origin a tracer quasiparticle with velocity $v$. Viewed on a somewhat coarser scale, when 
 time-averaging over  
 many collisions, the quasiparticle travels with an effective velocity $v^\mathrm{eff}(v)$, assumed to be increasing in $v$.
 The tracer quasiparticle collides with a fluid quasiparticle of velocity $w$. Such fluid particles have density $\rho_\mathsf{p}(w)$.  If the collision is from the left, $v< w$, then the tracer jumps by
 $-\mathsfit{a}$.  Hence the collision rate is  $\rho_\mathsf{p}(w)(v^\mathrm{eff}(w) - v^\mathrm{eff}(v))$. On the other hand,   if the collision is from the right, $v>w$, then the tracer jumps by
 $\mathsfit{a}$ and the collision rate is  $\rho_\mathsf{p}(w)(v^\mathrm{eff}(v) - v^\mathrm{eff}(w))$. When integrating over all fluid quasiparticles, the bare tracer quasiparticle velocity is modified through the collisions according to
 \begin{eqnarray}\label{5.26}
&&\hspace{-20pt} v^\mathrm{eff}(v) 
 = v - \mathsfit{a}\int_v^\infty \mathrm{d}w  \rho_\mathsf{p}(w)\big(v^\mathrm{eff}(w) - v^\mathrm{eff}(v)\big)
+ \mathsfit{a}\int_{-\infty} ^v \mathrm{d}w  \rho_\mathsf{p}(w)\big(v^\mathrm{eff}(v) - v^\mathrm{eff}(w)\big)\nonumber\\
&&\hspace{17pt} = v - \mathsfit{a}\int_\mathbb{R} \mathrm{d}w  \rho_\mathsf{p}(w)\big(v^\mathrm{eff}(w) - v^\mathrm{eff}(v)\big) \nonumber\\
&&\hspace{17pt}= v + T(\rho_\mathsf{p}v^\mathrm{eff})(v) - (T\rho_\mathsf{p})(v) v^\mathrm{eff}(v).
\end{eqnarray}
 For hard rods
the rate equation is easily solved with the result
\begin{equation}\label{5.27}
v^\mathrm{eff}(v)  =  v + \frac{\mathsfit{a}\rho}{1 -  \mathsfit{a}\rho} \int_\mathbb{R} \mathrm{d}w h(w) (v-w)   =      (1 - \mathsfit{a}\rho)^{-1}
( v -  \mathsfit{a} \rho u).
\end{equation}
The average current equals $\rho h v^\mathrm{eff} $, in agreement with the microscopic computation \eqref{5.6}.

With hindsight, the expression \eqref{5.26} for the effective velocity  carries already a recipe of how to extend to other integrable models. 
The two-particle scattering shift for the Toda fluid equals $2 \log|v - v'|$. Thus one might hope to obtain the correct average currents for the Toda fluid 
by substituting the operator $T$ in \eqref{5.27} by the operator $T$ from \eqref{3.46}. \bigskip\\
\textbf{\large{Notes and references}}
  \bigskip\\
  \textbf{ad 5}:  The hydrodynamic limit for hard rods, i.e. the asymptotic validity of \eqref{5.11}, is proved in  \cite{BDS83} under the assumption of a
  sufficiently random initial state. The limit holds with probability one. A short account is provided in \cite{S91}.  From the perspective 
  of generalized hydrodynamics, hard rods are discussed in \cite{DS17a,D19a}. There is much work on the hydrodynamic limit for stochastic lattice  gases and interacting diffusions. While not hamiltonian,  they still provide mathematical models on what to expect for
  a mechanical particle system. 
  
  For a many-particle hamiltonian systems with short range interactions \cite{M55} is an interesting early attempt. The modification of adding some extra randomness in collisions is studied in \cite{OVY93}.  The hydrodynamic limit on the Euler scale can be viewed as a stability result, 
  in the sense that on the ballistic scale there is no deviation yet from local equilibrium, resp. local GGE. 

\section{Generalized hydrodynamic equations }
\label{sec6}
\setcounter{equation}{0} 
\subsection{Average currents}
\label{sec6.1}
The microscopic version of the currents has been obtained already, see \eqref{2.21} and \eqref{2.23}. Now the goal is to compute their 
GGE average in the limit $N \to \infty$. As before the lattice size is $N$ and we use the pressure ensemble \eqref{3.8}. Then in terms of the spectral resolution of $L_N$ the $n$-th microscopic current reads
\begin{equation}\label{6.1}
J^{[n],N}  = \mathrm{tr}\big[((L_N)^nL_N^{\scriptscriptstyle \downarrow})\big] = N \langle  \rho_{\mathrm{J},N} \varsigma_n\rangle
\end{equation}
 with 
 \begin{equation}\label{6.2} 
\rho_{\mathrm{J},N} (w) = \frac{1}{N}  \sum_{j=1}^N \delta(w - \lambda_j)\Big(\sum_{i=1}^N a_i\psi_j(i) \psi_j(i+1)\Big).
\end{equation}
The $\delta$-peaks of the Lax DOS are  weighted by coefficients with arbitrary sign. Not to duplicate names we call
$\rho_{\mathrm{J},N}$ the current DOS. For given GGE the current DOS is self-averaging and has the deterministic limit 
$\rho_\mathrm{J}$. In particular 
\begin{equation}\label{6.3}
\lim_{N\to \infty}\frac{1}{N} \langle J^{[n],N} \rangle_{P,V,N} = \langle J^{[n]}_0 \rangle_{P,V} = \langle\rho_{\mathrm{J}} \varsigma_n\rangle.
\end{equation}
Setting $n=0$ one concludes $\langle \rho_\mathrm{J}\rangle = 0$. Hence, as expected, $\rho_\mathrm{J}$ cannot have a definite sign.
From the few numerical simulations available, very qualitatively   $\rho_\mathrm{J}(w)$ has the shape of $-\partial_w\rho_\mathrm{Q}(w)$.

Since the Dumitriu-Edelman identity worked so well for the conserved fields, one is tempted to use the same strategy 
for the currents. But now not only the distribution of the eigenvalues is in demand. While we have an explicit form for the joint distribution of $\{\lambda_j,\psi_j(1), j = 1,...,N\}$, to apply the scattering map $\Phi$ to the weights $\sum_{i=1}^N a_i\psi_j(i) \psi_j(i+1)$ seems to be a complicated enterprise. A new 
approach is needed.
Unexpectedly, the key idea will come from the susceptibility matrices.

We start with the fields and define the infinite volume correlator 
\begin{equation}\label{6.4} 
C_{m,n}(i-j) = \langle Q_i^{[m]}Q_j^{[n]} \rangle_{P,V}^\mathrm{c},
\end{equation}
where the superscript $^\mathrm{c}$ denotes truncation, respectively connected correlation, $\langle gf \rangle^\mathrm{c} = \langle gf \rangle -\langle g\rangle\langle f \rangle$. Truncated correlations decay rapidly to zero and the field-field susceptibility matrix is given by
\begin{equation}\label{6.5} 
C_{m,n} = \sum_{j\in \mathbb{Z}}C_{m,n}(j) = \langle Q^{[m]};Q^{[n]}\rangle_{P,V}, 
\end{equation}
where the second equality is merely a convenient notation. $C_{m,n}$ is the matrix of second derivatives of the generalized free energy and has been discussed already in Section \ref{sec4}.
Correspondingly we introduce the field-current correlator
\begin{equation}\label{6.6} 
B_{m,n}(j-i) = \langle J_j^{[m]} Q_i^{[n]} \rangle_{P,V}^\mathrm{c}, \quad B_{m,n} =  \sum_{j\in \mathbb{Z}} B_{m,n}(j).
\end{equation}
Despite its apparently asymmetric definition, $B$ satisfies
\begin{equation}\label{6.7} 
B_{m,n}(j) = B_{n,m}(-j).
\end{equation}
To prove, we employ the conservation law and spacetime stationarity to arrive at
\begin{eqnarray}\label{6.8} 
&&\hspace{-70pt}\partial_j \langle J_j^{[m]}(t) Q_0^{[n]}(0) \rangle_{P,V}^\mathrm{c} 
= -\partial_t  \langle Q_j^{[m]}(t) Q_0^{[n]}(0) \rangle_{P,V}^\mathrm{c}\nonumber\\[0.5ex]
&&\hspace{37pt} = -\partial_t \langle Q_0^{[m]}(0) Q_{-j}^{[n]}(-t) \rangle_{P,V}^\mathrm{c} 
= \partial_j \langle Q_0^{[m]}(0) J_{-j}^{[n]}(-t) \rangle_{P,V}^\mathrm{c},
\end{eqnarray}
denoting the difference operator by $\partial_jf(j) = f(j+1) - f(j)$.
Setting $t=0$, the difference $\langle J_j^{[m]} Q_0^{[n]} \rangle_{P,V}^\mathrm{c} -  \langle J_{-j}^{[n]} Q_0^{[m]} \rangle_{P,V}^\mathrm{c}$ is constant in $j$. Since truncated correlations decay to zero,
this constant has to vanish, which yields \eqref{6.7}. In particular, the field-current susceptibility matrix is symmetric,
\begin{equation}\label{6.9} 
B_{m,n} = B_{n,m}.
\end{equation}

Using  this symmetry we consider the $P$-derivative of the average current 
 \begin{equation}\label{6.10}
   \partial_P \langle( L^nL^{\scriptscriptstyle \downarrow})_{0,0} \rangle_{P,V} = - B_{n,0} = - B_{0,n}
   =  \langle Q^{[1]};Q^{[n]}\rangle_{P,V},
\end{equation}
since $J^{[0]} = - Q^{[1]}$ by \eqref{2.23}. We easily arrived at a very surprising identity. The $P$-derivative of the average current equals a particular
matrix element of the field-field susceptibility. Such susceptibility depends only on the eigenvalues of the Lax matrix.
In fact,  this quantity has been studied already in Section \ref{sec4} and one only has to adjust the results from there, setting $\alpha = P$.

As in the case of the free energy, since the pressure is varying as $1/N$, the fluctuation covariance is adding up, resulting in
\begin{equation}\label{6.11}
\langle \varsigma_1, C^\sharp \varsigma_n\rangle = \int_0^1 \mathrm{d}u \langle Q^{[1]};Q^{[n]} \rangle_{uP,V}.
\end{equation}
Integrating \eqref{6.10} over $P$, noting that the average current vanishes at $P=0$, one concludes that
\begin{equation}\label{6.12}
    \langle J^{[n]}_0 \rangle_{P,V} = P\langle \varsigma_1, C^\sharp \varsigma_n\rangle
\end{equation}
with  the operator $C^\sharp$ defined in \eqref{4.21}. Setting $\alpha \rho_\mathrm{s} = \rho_\mu$ and inserting there,
\begin{equation}\label{6.13}
    \langle J^{[n]}_0 \rangle_{P,V} = \langle \varsigma_1, (1 -  \rho_\mu T)^{-1}\rho_\mu \varsigma_n\rangle
    -    \nu \langle \varsigma_1 (1 -  \rho_\mu T)^{-1} \rho_\mu\rangle  \langle \varsigma_n (1 -  \rho_\mu T)^{-1} \rho_\mu\rangle.
\end{equation}
Since $(1- \rho_\mu T)^{-1} \rho_\mu$ is a symmetric operator, one finally arrives at 
 \begin{equation}\label{6.14}
 \langle J^{[n]}_0 \rangle_{P,V} =   \langle \rho_\mu \varsigma_1^\mathrm{dr}   \varsigma_n\rangle -  q_1\langle \rho_\mathsf{p}\varsigma_n\rangle,\qquad q_1 = \nu \langle \rho_\mathsf{p}\varsigma_1\rangle.
\end{equation}

While \eqref{6.14} provides already the formula for the averaged currents, there is a rewriting which yields a physically more intuitive identity. For hard rods we introduced already the concept of an \textit{effective velocity}, which refers to the motion  of a tracer quasiparticle. To adjust to the Toda  lattice one merely substitutes for  $- \mathsfit{a}$ 
 the  two-particle Toda scattering shift, resulting in  the integral equation
 \begin{equation}\label{6.15}
v^\mathrm{eff}(v) 
 = v +2 \int_\mathbb{R} \mathrm{d}w \log(|v-w| \rho_\mathsf{p}(w)\big(v^\mathrm{eff}(w) - v^\mathrm{eff}(v)\big). 
\end{equation} 
As in \eqref{5.26}, the integral term is the combined sum of collisions of the tracer particle with fluid particles from the right and  from the left.   For some other many-body integrable system, in essence one would have to substitute the appropriate two-particle scattering shift.

Eq. \eqref{6.15} is called \textit{collision rate assumption}, since it is based on postulating a rate equation for the motion of a tracer
quasiparticle, when moving in a fluid which initially is in a GGE characterized by $\rho_\mathsf{p}$.  Using the integral operator $T$ from \eqref{3.46},  
the collision rate assumption can be reexpressed as
 \begin{equation}\label{6.16} 
( 1 + T\rho_\mathsf{p}) v^\mathrm{eff} = \varsigma_1 + T(\rho_\mathsf{p} v^\mathrm{eff}).
 \end{equation}
 From the TBA equation, see \eqref{3.50}, we have $1 + T\rho_\mathsf{p} =  \rho_\mathsf{p}/\rho_\mu$. Hence
 \begin{equation}\label{6.17} 
 \rho_\mu  \varsigma_1 =  (1 -  \rho_\mu T) (\rho_\mathsf{p} v^\mathrm{eff}) 
 \end{equation}
 and
 \begin{equation}\label{6.18} 
 \rho_\mathsf{p} v^\mathrm{eff} = (1- \rho_\mu T)^{-1}  (\rho_\mu  \varsigma_1) = \rho_\mu \varsigma_1^\mathrm{dr}.
 \end{equation}
Using again \eqref{3.50}, one concludes
 \begin{equation}\label{6.19} 
v^\mathrm{eff} = \frac{\varsigma_1^\mathrm{dr}}{\varsigma_0^\mathrm{dr}}\,. 
 \end{equation}
Comparing with \eqref{6.14}, the average currents can thus be written as
\begin{equation}\label{6.20} 
\langle J^{[0]}_0\rangle_{P,V} = - q_1 \quad \langle J^{[n]}_0 \rangle_{P,V} =  \langle \rho_\mathsf{p}(v^\mathrm{eff} - q_1)  \varsigma_n\rangle, \quad n = 1,2,...\,,
\end{equation}
with $v^\mathrm{eff}$ the solution of  \eqref{6.15} or even more conveniently the ratio \eqref{6.19}.

To establish directly the ansatz \eqref{6.15} seems to be difficult. But since we have computed the GGE averaged microscopic currents, indirectly we confirm the validity of \eqref{6.15}. In this sense the collision rate assumption is proved for the Toda lattice.

 To mimic the reasoning in case of a hard rod fluid, one first needs to introduce the notion of a quasiparticle, which is meaningful in the  small pressure limit. Then particles are far apart and interact mostly through isolated two-particle collisions. Quasiparticles are defined as  maintaining their velocity upon excluding the time spans for collisions. The tracer quasiparticle jumps by a distance regulated by the two-particle scattering shift. One can then argue for the validity of the rate equation, as we did already for a hard rod fluid. Moving away 
from the dilute limit, already our notion of tracer quasiparticle becomes blurry because of many-particle interactions. The collisions between the tracer quasiparticle and fluid particles are highly  
correlated and the assumed approximate statistical independence tends to be dubious. However, the form \eqref{6.20} 
holds over the entire parameter range, no exceptions. In fact, while the usually adopted pictorial arguments are helpful, more importantly the collision rate equation tells us  that there is an underlying effective
particle model, which on a large scale is governed exactly by the Toda generalized hydrodynamics. We are allowed to think
of  particles which instantaneously switch from incoming to outgoing velocities and jump by the distance $\phi_{12}(v_1-v_2)$, compare with
 Section \ref{sec2.3}.  In fact, the true dynamics is more complicated, because a third particle or even more particles might interfere with the collision between particles 1 and 2. To properly  set up the particle model requires some thoughts. But, for the Toda fluid, and also quantum many-body integrable systems,
 one can use such a scheme as an alternative to direct numerical simulations of the hydrodynamic equations.  
\subsection{Hydrodynamic equations}
\label{sec6.2}
On the hydrodynamic scale the local GGE is characterized by the stretch $\nu$ and the Lax DOS $\nu\rho_\mathsf{p}$, both of which now become spacetime dependent. Merely inserting the average currents, one arrives at the Euler type hydrodynamic evolution equations,
\begin{eqnarray}\label{6.21} 
&&\partial_t \nu(x,t) -\partial_x q_1(x,t) = 0,\nonumber\\[0.5ex]
&& \partial_t\big(\nu(x,t) \rho_\mathsf{p}(x,t;v)\big) + \partial_x\big((v^\mathrm{eff}(x,t;v) - q_1(x,t))\rho_\mathsf{p}(x,t;v)\big) = 0.
\end{eqnarray}
As a most remarkable feature of generalized hydrodynamics,  the equations can be  transformed explicitly to a quasilinear 
form. 
For this purpose we rewrite the identity  \eqref{3.50} and thereby introduce the nonlinear mapping $\rho_\mathsf{p} \mapsto \rho_\mu$,  defined by
\begin{equation}\label{6.22} 
\rho_\mu =  \rho_\mathsf{p}(1+ (T\rho_\mathsf{p}))^{-1}.
\end{equation}
Then Eq. \eqref{6.21} takes the normal form
\begin{equation}\label{6.23} 
 \partial_t \rho_\mu + \nu^{-1}(v^\mathrm{eff} - q_1)\partial_x \rho_\mu = 0.
\end{equation}
Thus the linearization operator is in a fact merely a multiplication by $\nu^{-1}(v^\mathrm{eff} - q_1)$, in other words the operator is diagonal. 
For hydrodynamic equations with only a few, say $\bar{n}$, conservation laws, the linearization operator is 
a $\bar{n} \times \bar{n}$ matrix, which in general is not diagonal. In our case $\nu^{-1}(v^\mathrm{eff} - q_1)$
is expected to be smooth function and thus the spectrum of the linearization operator has no eigenvalues. Hence for smooth initial data
the solution of \eqref{6.23} is expected to stay smooth. No mathematical analysis has been attempted to verify such conjecture.  
In fact, the situation might be more subtle, since  $\nu$ can take either sign and $\nu^{-1}$ might be singular.

To verify \eqref{6.23} we  start from 
\begin{equation}\label{6.24} 
 \rho_\mathsf{p}\partial_t\nu  + \nu\partial_t\rho_\mathsf{p} + \partial_x\big((v^\mathrm{eff} - q_1)\rho_\mathsf{p}\big) = 0,
\end{equation}
which together with the continuity equation yields
\begin{equation}\label{6.25}
\nu\partial_t \rho_\mathsf{p} + \partial_x(v^\mathrm{eff} \rho_\mathsf{p})  - q_1 \partial_x \rho_\mathsf{p}= 0.
\end{equation}
Using this identity together with \eqref{6.22}, one obtains
\begin{eqnarray}\label{6.26}
&&\hspace{-20pt} \nu\partial_t \rho_\mu = \frac{\rho_\mu}{\rho_\mathsf{p}}\big(-\partial_x(v^\mathrm{eff}\rho_\mathsf{p}) +q_1\partial_x \rho_\mathsf{p}\big)
-\frac{\rho_\mu^2}{\rho_\mathsf{p}}\big(T(-\partial_x(v^\mathrm{eff}\rho_\mathsf{p}) + q_1\partial_x \rho_\mathsf{p})\big)\nonumber\\
&&\hspace{12pt} =  \frac{\rho_\mu}{\rho_\mathsf{p}}\big(- \partial_x(v^\mathrm{eff}\rho_\mathsf{p}) + \rho_\mu \partial_x T(v^\mathrm{eff}\rho_\mathsf{p})\big) + q_1 \partial_x\rho_\mu.
\end{eqnarray}
The effective velocity solves the integral equation
\begin{equation}\label{6.27}
v^\mathrm{eff} = v + T(v^\mathrm{eff}\rho_\mathsf{p}) - (T\rho_\mathsf{p})v^\mathrm{eff}
\end{equation}
and
\begin{equation}\label{6.28}
\partial_x\big( (1+ T\rho_\mathsf{p}) v^\mathrm{eff} \big) = \partial_x  T(v^\mathrm{eff}\rho_\mathsf{p}).
\end{equation}
Inserting this identity in the first bracket of \eqref{6.26},
\begin{equation}\label{6.29}
- \partial_x(v^\mathrm{eff}\rho_\mathsf{p}) + \rho_\mu \partial_x T(v^\mathrm{eff}\rho_\mathsf{p}) = 
-\frac{\rho_\mathsf{p}}{\rho_\mu}v^\mathrm{eff}\partial_x \rho_\mu,
\end{equation}
thereby yielding \eqref{6.23}. 

The normal form \eqref{6.23} of the hydrodynamic equations is a crucial insight.  A priori, the coupled conservation laws might be so complicated that it becomes an impossible task to extract information of interest. Here the normal form is of considerable help. It allows to obtain some partially analytic solutions, as the domain wall problem and the linearized version to be discussed in the next section. The normal form suggests also 
how to devise a numerical scheme for solving the hydrodynamic equations. 
In its most basic version, at the current time, $t$,
one keeps 
$\nu^{-1}(v^\mathrm{eff} - q_1)$ fixed and solves the resulting linear equation for one further time step $\mathrm{d}t$. With the so 
obtained $\rho_\mu(t +\mathrm{d}t)$ one updates $\nu^{-1}(v^\mathrm{eff} - q_1)$ according to \eqref{6.19} for $v^\mathrm{eff}$, $\nu^{-1} = \langle \rho_\mathsf{p}\rangle$, $\nu^{-1}q_1 = \langle \rho_\mathsf{p}\varsigma_1\rangle$, and $\rho_\mathsf{p} = \rho_\mu  \varsigma_0^\mathrm{dr}$.  
No explicit use of TBA is involved in this step. The iteration is then repeated many times until the desired final time is realized.
\bigskip\\
\textbf{\large{Notes and references}}
  \bigskip\\
  \textbf{ad 6.1}: The computation \eqref{6.8} is taken from \cite{DBD19}, see also \cite{KS19}. The symmetry of the 
  $B$-matrix is discussed in
 \cite{GS11,S14}. Rather likely there is earlier work. The connection between the symmetry of the $B$-matrix 
 and the average current was first noted in \cite{S20} and extended to quantum systems in \cite{YS20}. For integrable quantum systems, the appropriately 
 adjusted rate equation \eqref{6.15} has been discovered in  \cite{CDY16,BCDF16}. Very quickly it was understood that 
 the scheme  developed for the specific models in   \cite{CDY16,BCDF16} has a much wider applicability \cite{DYC18}.
 \medskip \\  
 \textbf{ad 6.2}:  The transformation to normal form is a standard result from that time. The flea gas algorithm is discussed in 
 \cite{DYC18,MA20} as an alternative to more standard PDE discretization schemes. A more numerical oriented contribution is \cite{MS20a}.

\section{Linearized hydrodynamics and GGE dynamical correlations}
\label{sec7} 
\setcounter{equation}{0}
In equilibrium statistical mechanics a central theme is to understand the structure of static correlations, with particular emphasis
on thermodynamic parameters for which the correlation length is large on microscopic scales. A natural extension are time-dependent correlations, which can be viewed as the propagation of initially small perturbations in the equilibrium state.  Now conservation laws
and broken symmetries will play a crucial role. 
The most elementary approach is the Landau-Lifshitz theory which uses the link to  macroscopic equations
linearized at thermal equilibrium, as will be discussed below. Near critical points the more sophisticated techniques of critical dynamics  
will come into play.
\subsection{Equilibrium time correlations for nonintegrable chains}
\label{sec7.1}
The general structure 
behind the Landau-Lifshitz theory can be explained already in the context of nonintegrable systems with a few conservation laws.
For concreteness we consider nonintegrable anharmonic chains.  In essence, the extension to integrable chains consists in
substituting the respective $3\times 3$ matrices by operators on Hilbert spaces with infinite dimension. To distinguish from the infinite dimensional case, we use $\vec{u}$ for $3$\,-vectors and $\mathsfit{A}$ for $3\times3$ matrices.

The infinitely extended chain is governed by the hamiltonian
\begin{equation}\label{7.1}
  H=\sum_{j \in \mathbb{Z}}\big( \tfrac{1}{2}p^2_j +V_\mathrm{ch}(r_j)\big)
\end{equation}
with equations of motion
\begin{equation}\label{7.2}
\frac{d}{dt}r_j=p_{j+1}-p_j\,,\qquad
\frac{d}{dt}p_j=V_\mathrm{ch}'(r_j)-V_\mathrm{ch}'(r_{j-1}).
\end{equation}
The chain potential, $V_\mathrm{ch}$, is bounded from below and, so to have a finite partition function, must increase linearly 
at infinity at least one-sidedly. Stretch, momentum, and energy,
\begin{equation}\label{7.3}
\vec{Q}_j= \big(r_j, p_j,e_j\big), \qquad e_j=\tfrac{1}{2}p^2_j + V_\mathrm{ch}(r_j),
\end{equation}
are always locally conserved and the respective currents are
\begin{equation}\label{7.4}
\vec{J}_j = \big( -p_j,-V_\mathrm{ch}'(r_{j-1}), - p_jV_\mathrm{ch}'(r_{j-1})\big).
\end{equation}
It is assumed that there are no further local conservation laws, which is the essence of nonintegrability.

The canonical  equilibrium state factorizes with one factor given by
\begin{equation}\label{7.5}
Z(P,u,\beta)^{-1} \exp\!\big(-\beta \big(\tfrac{1}{2}(p_0- u)^2 +V_\mathrm{ch}(r_0)\big)  - Pr_0\big),
\end{equation}
where we introduced the intensive dual parameters $P, u ,\beta$. $P/\beta$ is the physical pressure, $u$ the mean velocity,  and
$\beta$ the inverse temperature. The chain free energy per site  equals
\begin{equation}\label{7.6}
F_\mathrm{ch} = - \log Z(P,u,\beta). 
\end{equation}
Of course, more explicit expressions could be provided, but this is not needed at the moment. In accordance with the Toda chain, 
the indices run over $m,n = 0,1,2$, throughout. The static field-field correlator is defined by 
\begin{equation}\label{7.7}
\mathsfit{C}_{m,n}(j) = \delta_{0,j}\langle Q_0^{[m]}Q_0^{[n]}\rangle_{P,u,\beta}^\mathrm{c}
\end{equation}
and the static field-field susceptibility matrix by
\begin{equation}\label{7.8}
\mathsfit{C}_{m,n} = \sum_{j \in \mathbb{Z}}\mathsfit{C}_{m,n}(j) = \langle Q_0^{[m]}Q_0^{[n]}\rangle_{P,u,\beta}^\mathrm{c}.
\end{equation}
$\mathsfit{C}$ is the matrix of second derivatives of the chain free energy $F_\mathrm{ch}$. In the same fashion the field-current correlator is given by
\begin{equation}\label{7.9}
\mathsfit{B}_{m,n}(j) = \langle J_j^{[m]} Q_0^{[n]}\rangle_{P,u,\beta}^\mathrm{c}
\end{equation}
and the static field-current susceptibility matrix by
\begin{equation}\label{7.10}
\mathsfit{B}_{m,n} = \sum_{j \in \mathbb{Z}}\mathsfit{B}_{m,n}(j).
\end{equation}

The average fields and currents are denoted by $\vec{\mathsfit{q}}= \langle \vec{Q}_0\rangle_{P,u,\beta}$,
$\vec{\mathsfit{j}}= \langle \vec{J}_0\rangle_{P,u,\beta}$. Then the hydrodynamic equations for the chain read
\begin{equation}\label{7.11}
\partial_t \vec{\mathsfit{u}}(x,t) +  \partial_x \vec{\mathsfit{j}}(x,t) = 0
\end{equation}
with quasilinear form
\begin{equation}\label{7.12}
\partial_t \vec{\mathsfit{u}}(x,t) + \mathsfit{A}(x,t)\partial_x \vec{\mathsfit{u}}(x,t) = 0.
\end{equation}
By the chain rule, $ \mathsfit{A} = \mathsfit{B}\mathsfit{C}^{-1}$. These matrices depend parametrically on
$P,u,\beta$. Thus, for example, 
$\mathsfit{A}(x,t) = \mathsfit{A}\big(P(x,t),u(x,t),\beta(x,t)\big)$. 

The field-field time-correlator is the matrix
\begin{equation}\label{7.13}
\mathsfit{S}_{m,n}(j,t)  = \langle Q_j^{[m]}(t)Q_0^{[n]}(0)\rangle_{P,u,\beta}^\mathrm{c},
\end{equation}
which for $t \neq 0$ is no longer $\delta$-correlated in $j$. It is  convenient to also introduce its Fourier transform
\begin{equation}\label{7.14}
\hat{\mathsfit{S}}_{m,n}(k,t) = \sum_{j \in \mathbb{Z}}\mathrm{e}^{\mathrm{i}kj}\mathsfit{S}_{m,n}(j,t), \quad k \in [-\pi,\pi]. 
\end{equation}
The hydrodynamic equation covers the ballistic scale. So more precisely one expects, for small $\epsilon$,
\begin{equation}\label{7.15}
\epsilon^{-1} \mathsfit{S}_{m,n}(\lfloor\epsilon^{-1}x\rfloor, \epsilon^{-1}t) \simeq \mathsfit{S}^\diamond_{m,n}(x,t), \quad x \in \mathbb{R},
\end{equation}
wich defines $\mathsfit{S}^\diamond_{m,n}$. Here $\lfloor\cdot\rfloor$ denotes the integer part and the prefactor is chosen such that the sum rule 
\begin{equation}\label{7.16}
\sum_{j\in \mathbb{Z}} \mathsfit{S}(j,t) = \sum_{j\in \mathbb{Z}} \mathsfit{S}(j,0)
\end{equation}
holds. In Fourier space, more compactly, 
\begin{equation}\label{7.17}
\lim_{\epsilon \to 0}\hat{\mathsfit{S}}_{m,n}(\epsilon k,\epsilon^{-1}t) =  \hat{\mathsfit{S}}^\diamond_{m,n}(k,t).
\end{equation}

We now view the average \eqref{7.13} as a small initial perturbation of equilibrium. Then to leading order the correlator  may be 
approximated by Eq. \eqref{7.12} linearized around a constant background with parameters  $P,u,\beta$. 
Denoting this perturbation again by $\vec{\mathsfit{u}}$ the nonlinear hydrodynamic equation turns to its linearized version
\begin{equation}\label{7.18}
\partial_t \vec{\mathsfit{u}}(x,t) +  \mathsfit{A}\partial_x \vec{\mathsfit{u}}(x,t) = 0,
\end{equation}
where the $\mathsfit{A}$\,-matrix is now evaluated at the background, hence constant in spacetime. However the initial conditions, 
$\vec{\mathsfit{u}}(x)$, are random according to thermal equilibrium and, as a consequence of \eqref{7.7}, on the macroscopic scale,
\begin{equation}\label{7.19}
\langle \mathsfit{u}_m(x)\mathsfit{u}_n(0)\rangle = \mathsfit{C}_{m,n} \delta(x) = \mathsfit{S}^\diamond_{m,n}(x,0).
\end{equation}
In conclusion, on the Euler scale one obtains 
\begin{equation}\label{7.20}
 \hat{\mathsfit{S}}^\diamond_{m,n}(k,t) = (\mathrm{e}^{-\mathrm{i}k\mathsfit{A}t}\mathsfit{C})_{m,n}.
\end{equation}

In general the $\mathsfit{A}$\,-matrix is not symmetric. But $\mathsfit{B}$ is symmetric and also  $\mathsfit{C}$ with $\mathsfit{C}>0$. Since $\mathsfit{A} = \mathsfit{B}\mathsfit{C}^{-1}$, this implies that $\mathsfit{A}$ has real eigenvalues, $c_\alpha$, $\alpha = 0,1,2$,  and a complete system of right, $|\psi_\alpha\rangle$, and left 
eigenvectors, $\langle \tilde{\psi}_\alpha  | $. Hence in position space \eqref{7.15} has the explicit form
\begin{equation}\label{7.21}
\mathsfit{S}^\diamond_{m,n}(x,t) = \sum_{\alpha = 0}^2 \delta(x - c_\alpha t) (|\psi_\alpha\rangle \langle \tilde{\psi}_\alpha | \mathsfit{C})_{m,n}.
\end{equation}
The eigenvectors of $\mathsfit{A}$ are the normal modes. The $\alpha$-th mode travels with velocity $c_\alpha$ and the 
initial condition determines the particular linear combination of normal modes as encoded by the susceptibility matrix $\mathsfit{C}$.  The spacetime correlator has three 
delta peaks moving ballistically. If the background has zero average momentum, then the eigenvalues are $\vec{c} =
(-c,0,c)$ with $c$ the isentropic speed of sound. There are two sound peaks with equal speed moving in opposite directions and a heat peak standing
still. For an integrable system, there are so to speak infinitely many peaks, each moving with its own velocity. We thus expect the corresponding time correlator to have a broad spectrum which expands ballistically. 
\subsection{Equilibrium time correlations for the Toda lattice}
\label{sec7.2}
For the Toda chain we follow step by step the road map provided by the finite mode case, where we recall that  $\varsigma_n(w) = w^n$, including $n = 0$. The field-field and field-current correlator have been introduced before in \eqref{6.5}, resp. \eqref {6.6}.
Since we have computed already the GGE average of fields in \eqref{3.51} and of currents in \eqref{6.14}, the matrix elements of $C,B$
can be determined 
by differentiating these averages with respect to $P$ and $\mu_n$. The computation is somewhat lengthy and stated is merely the final result.
For the field-field correlator one obtains
\begin{eqnarray}\label{7.22}
&&\hspace{-20pt} C_{0,0} =\nu^3 \langle \rho_\mathsf{p} \varsigma_0^\mathrm{dr} \varsigma_0^\mathrm{dr} \rangle,\nonumber\\[0.5ex]
&&\hspace{-20pt}C_{0,n}= C_{n,0}= -\nu^2  \langle \rho_\mathsf{p} \varsigma_0^\mathrm{dr} (\varsigma_n- q_n\varsigma_0)^\mathrm{dr} \rangle,\nonumber\\[0.5ex]
&&\hspace{-20pt} C_{m,n}= \nu  \langle \rho_\mathsf{p} (\varsigma_m- q_m\varsigma_0)^\mathrm{dr} (\varsigma_n- q_n\varsigma_0)^\mathrm{dr} \rangle
\end{eqnarray}
 and for the field-current correlator
\begin{eqnarray}\label{7.23}
&&\hspace{-20pt} B_{0,0} = \nu^2 \langle \rho_\mathsf{p}(v^\mathrm{eff} -q_1) \varsigma_0^\mathrm{dr}   \varsigma_0^\mathrm{dr}\rangle,\nonumber\\[0.5ex]
&&\hspace{-20pt}B_{0,n}= B_{n,0}=  -\nu \langle \rho_\mathsf{p}(v^\mathrm{eff} -q_1)  \varsigma_0^\mathrm{dr} ( \varsigma_n- q_n \varsigma_0)^\mathrm{dr}\rangle,\nonumber\\[0.5ex]
&&\hspace{-20pt} B_{m,n} =  \langle \rho_\mathsf{p}(v^\mathrm{eff} -q_1)( \varsigma_m- q_m \varsigma_0)^\mathrm{dr} ( \varsigma_n- q_n \varsigma_0)^\mathrm{dr}\rangle.
\end{eqnarray}

Both matrices have a two-block structure. Rather than using the discrete index, $n = 1,2,...$, we introduce some real valued function,
$\phi$.
Our linear space consists now of two-vectors $(r,\phi)$ with $r$ the coefficient for the index $0$ and $\phi$ for all other modes.
More formally, the appropriate linear space is $\mathbb{C} \oplus \big(L^2(\mathbb{R}, \mathrm{d}w)\ominus\{\varsigma_0\}\big)$,
the second summand consisting of all square-integrable functions orthogonal to $\varsigma_0$.
It will be convenient to use the abbreviation  
\begin{equation}\label{7.24}
h = \nu\rho_\mathsf{p}, \quad \langle h \rangle = 1, \quad h \geq 0.
\end{equation}
We also introduce  the linear operator, $F$, acting on a general function $\phi$ by
\begin{equation}\label{7.25}
F\phi = (\phi - \langle h\phi\rangle \varsigma_0)^\mathrm{dr}  = (1 - T\rho_\mu)^{-1}(\phi - \langle h\phi\rangle).
\end{equation}
Note that $F\varsigma_0 = 0$. Then, in Dirac notation, 
\begin{equation}\label{7.26}
C = 
\begin{pmatrix}
\nu^2 \langle h  \varsigma_0^\mathrm{dr} \varsigma_0^\mathrm{dr}\rangle &-\nu \big\langle h \varsigma_0^\mathrm{dr}\big|F\\[0.5ex]
-\nu F^*\big| h\varsigma_0^\mathrm{dr}\big\rangle& F^*h F
\end{pmatrix}.
\end{equation}
For some general operator $A$, as introduced before, $A^*$ stands for the adjoint operator of $A$
with respect to the standard $L^2$ inner product $\langle \cdot,\cdot\rangle$. In particular,
\begin{equation}\label{7.27}
F^* \psi = (1 - \rho_\mu T)^{-1}\psi - h \langle \varsigma_0^\mathrm{dr} \psi\rangle.
\end{equation} 
 With the same notation, the matrix $B$ is given by
\begin{equation}\label{7.28}
B = 
\nu^{-1}\begin{pmatrix}
\nu^2 \langle h (v^\mathrm{eff} -q_1)\varsigma_0^\mathrm{dr}\varsigma_0^\mathrm{dr}\rangle &-\nu \big\langle h (v^\mathrm{eff}-q_1)\varsigma_0^\mathrm{dr}\big|F\\[0.5ex]
-\nu F^*\big| h(v^\mathrm{eff} - q_1)\varsigma_0^\mathrm{dr}\big\rangle& F^*h(v^\mathrm{eff} -q_1) F
\end{pmatrix}.
\end{equation}

Following the road map we are supposed to compute $\mathrm{e}^{At}C$ with $A = B C^{-1}$, which does not seem to be completely
straightforward. But instead one might guess the entire solution $S(x,t)$ by noting that $S(x,0) = \delta(x)C$ and $\partial_t S(x,t)\big|_{t=0}
= -\delta'(x)B$. Indeed, the two conditions can be satisfied by setting
\begin{equation}\label{7.29}
S(x,t) = 
   \begin{pmatrix}
\nu^2 \langle h \,\delta(x - t\nu^{-1}(v^\mathrm{eff} -q_1)) \varsigma_0^\mathrm{dr}\varsigma_0^\mathrm{dr}\rangle &-\nu \big\langle h\,\delta(x - t\nu^{-1}(v^\mathrm{eff} -q_1))\varsigma_0^\mathrm{dr}\big|F\\[0.5ex]
-\nu F^*\big| h\,\delta(x - t\nu^{-1}(v^\mathrm{eff}  -q_1))\varsigma_0^\mathrm{dr}\big\rangle& F^*h\delta(x - t\nu^{-1}
(v^\mathrm{eff} -q_1)) F
\end{pmatrix}.
\end{equation}
Our computation misses is that matching simply the value and first derivative at $t=0$ does not determine the full solution. 
Leaving  for the moment a more convincing argument aside, let us reflect on the resulting predictions. For example, within the stated approximations, the time-dependent stretch-stretch correlation function is predicted as
\begin{equation}\label{7.30}
S_{0,0}(x,t) = \nu^2 \int_\mathbb{R}\mathrm{d}w   \delta(x- t\nu^{-1}(v^\mathrm{eff}(w) - q_1)) \nu\rho_\mathsf{p}(w)\varsigma_0^\mathrm{dr}(w)^2,
\end{equation}
and similarly for the momentum-momentum correlation, 
\begin{equation}\label{7.31}
S_{1,1}(x,t) =  \int_\mathbb{R}\mathrm{d}w  \delta(x- t\nu^{-1}(v^\mathrm{eff}(w)-q_1)) \nu\rho_\mathsf{p}(w)\varsigma_1^\mathrm{dr}(w)^2.
\end{equation}
Our result is in perfect analogy to the finite mode case. The modes are labelled by the rapidity $w$ and propagate with velocity
$\nu^{-1}(v^\mathrm{eff}(w)-q_1)$. The last factor provides the weights and depends on the particular choice for the 
matrix elements of $S(x,t)$.

To confirm the guess, one has to figure out the spectral representation of the operator $A = BC^{-1}$. 
As our key observation,  in \eqref{6.22} we already introduced a nonlinear map which transforms the system of conservation laws into its quasilinear version in such a way that the  operator corresponding to $A$ is manifestly diagonal. This property is expected to persist 
when linearizing the map \eqref{6.22}. In the $\nu,h$ variables the map is given by
\begin{equation}\label{7.32}
\rho_\mu = h(\nu + (Th))^{-1}.
\end{equation} 
Both sides are linearized as $\rho_\mu +\epsilon g$, $\nu +\epsilon r$, $h +\epsilon \phi$, $\langle \phi \rangle = 0$. To first order in $\epsilon$
this yields the linear map   
$R: g \mapsto (r,\phi)$  given by  
\begin{equation}\label{7.33}
Rg = \nu
\begin{pmatrix} 
- \nu \langle g (\varsigma_0^\mathrm{dr})^2\rangle\\[0.5ex]
F^*(g \varsigma_0^\mathrm{dr})
\end{pmatrix},
\end{equation} 
with $F^*$ as in \eqref{7.27}. 
 Note that indeed $\langle F^* \psi\rangle = 0$.  \eqref{7.33} can be inverted as  
\begin{equation}\label{7.34}
\nu = \langle \rho_\mathsf{p} \rangle^{-1},\quad h = \langle \rho_\mathsf{p} \rangle^{-1}  \rho_\mathsf{p}, \quad \rho_\mathsf{p} = (1- \rho_\mu T)^{-1}\rho_\mu,
\end{equation} 
thereby deriving, by a similar argument as before,
\begin{equation}\label{7.35}
R^{-1} 
\begin{pmatrix} 
r\\
\phi
\end{pmatrix}
= (\nu \varsigma_0^\mathrm{dr})^{-1}(- \rho_\mu r + (1-\rho_\mu T )\phi). 
\end{equation} 
Indeed one checks that
\begin{equation}\label{7.36}
RR^{-1} = 1,\quad R^{-1}R = 1,
\end{equation}
the first $``1"$ standing for the identity operator as a $2\times 2$ block matrix and the second $``1"$ for the identity operator in the space of scalar functions.

The operators $C,B$ can be written in the new basis with the result
\begin{equation}\label{7.37}
R^{-1}CR =   \nu^2 \big| h \big\rangle \big\langle ( \varsigma_0^\mathrm{dr})^2\big| + ( \varsigma_0^\mathrm{dr})^{-1}h FF^*\varsigma_0^\mathrm{dr}
\end{equation}
and 
\begin{eqnarray}\label{7.38}
&&\hspace{0pt}R^{-1}BR
 = \nu^{-1}\big(\big| (v^\mathrm{eff}-q_1)\nu^2h\big\rangle\big\langle  (\varsigma_0^\mathrm{dr})^2\big|
   +   (v^\mathrm{eff} -q_1)(\varsigma_0^\mathrm{dr})^{-1}hFF^* \varsigma_0^\mathrm{dr} \big)\nonumber\\[0.5ex]
&&\hspace{44pt} = \nu^{-1} (v^\mathrm{eff} - q_1) R^{-1}CR.  
\end{eqnarray}
As anticipated, in the new basis  $R^{-1}AR$ is simply multiplication by $\nu^{-1} (v^\mathrm{eff} - q_1)$.
We conclude that 
\begin{equation}\label{7.39}
\mathrm{e}^{At}C =  R\exp\!\big(\nu^{-1}(v^\mathrm{eff} - q_1)t\big)R^{-1}C.
\end{equation}
Working out the algebra, one arrives at 
\begin{equation}\label{7.40}
\mathrm{e}^{At}C =  \begin{pmatrix}
\nu^2 \langle h \exp\!\big(\nu^{-1}(v^\mathrm{eff} - q_1)t\big)   (\varsigma_0^\mathrm{dr})^2\rangle &-\nu \big\langle h
\exp\!\big(\nu^{-1}(v^\mathrm{eff} - q_1)t\big)  \varsigma_0^\mathrm{dr}\big|F\\[0.5ex]
-\nu F^*\big| h \exp\!\big(\nu^{-1}(v^\mathrm{eff} - q_1)t\big) \varsigma_0^\mathrm{dr}\big\rangle& F^*h \exp\!\big(\nu^{-1}(v^\mathrm{eff} - q_1)t\big) F
\end{pmatrix}.
\end{equation}
Adding the spatial dependence, first in Fourier space, the propagator $\mathrm{e}^{At}C$ is modified to 
$\mathrm{e}^{\mathrm{i}kAt}C$, which in position space yields \eqref{7.29}. 
For completeness the matrix $A$ is recorded as 
\begin{equation}\label{7.41}
A =  \nu^{-1}R(v_\mathrm{eff} - q_1)R^{-1} = 
\begin{pmatrix}
\langle \rho_\mu (v^\mathrm{eff} -q_1) \varsigma_0^{\mathrm{dr}}\rangle &- \big\langle  (v^\mathrm{eff}- q_1) \varsigma_0^\mathrm{dr}
(1 -\rho_\mu T)\big|(1 -\rho_\mu T)\\[0.5ex]
-\nu^{-1} F^*\big| \rho_\mu (v^\mathrm{eff} - q_1)\big\rangle& \nu^{-1}F^*(v^\mathrm{eff} -q_1)(1 -\rho_\mu T)
\end{pmatrix}.
\end{equation}

Actually, the peak structure on the Euler scale is only the starting step of the Landau-Lifshitz theory. Its main focus is the broadening of the peaks due to dissipation and molecular noise. For a simple fluid in three dimensions the broadening is of order $\sqrt{t}$ with a Gaussian shape function. In one dimension the standard Landau-Lifshitz theory would make the same prediction. But the linear 
fluctuation theory has to be corrected by expanding the nonlinear Euler term to second order, which then results in a well-confirmed superdiffusive spreading as $t^{2/3}$. More details can be found in Section \ref{sec12.2}.
But even before there is a generic obstacle. For a well-isolated peak the broadening is easily observed. But for the Toda chain the Euler term yields already  a large background. So one has to specifically tune to physically situations for which the anticipated dissipative corrections can be detected.\bigskip\\
\textbf{\large{Notes and references}}\bigskip\\
 \textbf{ad 7.1}:  The existence of the infinite volume dynamics is proved in \cite{LLL77}. The book by Forster \cite{F75} is still a very readable account of the Landau-Lifshitz fluctuation theory \cite{LL75}.
 Critical dynamics is covered in the classic review \cite{HH77}. Fluctuation theory for simple fluids is discussed in  \cite{S91}. 
 As a crucial distinction, the hydrodynamic equations for a few conservation laws may develop shocks. A most readable account is
  \cite{B13}.
 \medskip\\
  \textbf{ad 7.2}: A more complete account of GGE time correlations can be found  in \cite{S19b}. Molecular dynamics of the Toda chain are available for the thermal state \cite{KD16}.  At various temperatures and pressures, measured are the $00$, $11$, $22$, and some cross correlations. The numerical data are in good agreement with theoretical predictions \cite{MS21}.

\section{Domain wall initial states}
\label{sec8} 
\setcounter{equation}{0}
From a theory perspective, a very natural nonequilibrium initial state is enforced by joining  two semi-infinite systems, the left half line in one GGE and the right half line in another GGE. The fields are spatially constant except for a  single jump at the origin. On the level of GHD, the solution scales exactly as 
$x/t$, a simplification which promises to yield exact solutions. In the mathematical theory of finitely many hyperbolic conservation laws, this set-up is known as Riemann problem. Its solution consists of  flat pieces, smooth rarefaction waves, and jump discontinuities, called shocks.
Their precise spatial sequence can be complicated and is, in principle, encoded  by the eigenvectors and eigenvalues of the matrix $\mathsfit{A}$ in their dependence on the thermodynamic parameters, compare with \eqref{7.12}. 
As discussed in Section \ref{sec6.2} in normal form the matrix $A$ is multiplication by $\nu^{-1} (v^\mathrm{eff} - q_1)$. Thus $A$ has continuous spectrum, which will make the domain wall solution different from the case of finite number of modes.

Besides its intrinsic interest, we also pick the example to illustrate the predictive power of GHD. Domain wall is concerned with  macroscopic behavior. One is interested how stretch, momentum, and energy vary spatially given their particular initial jump discontinuity. 
There does not seem to be any other theoretical scheme available through which one could obtain such information.

To conform with conventional notation,
we  set $\rho_\mu = n$ and slightly rewrite Eq. \eqref{6.23} as
\begin{equation}\label{8.1} 
 \partial_t n(x,t;v) + \tilde{v}^\mathrm{eff}(x,t;v) \partial_x n(x,t;v) = 0,\quad \tilde{v}^\mathrm{eff}(v) = \nu^{-1}(v^\mathrm{eff}(v) - q_1). 
\end{equation}
Since $\nu$ could vanish, the hydrodynamic equation looks singular at first sight.
The domain wall initial conditions are 
\begin{equation}\label{8.2}
 n(x,0;v) = \chi(\{x <0\}) n_-(v)+ \chi(\{x \geq 0\})n_+(v).
\end{equation}
Instead of $n_\pm$, physically it might be more natural to prescribe the DOS of the Lax matrix and the average stretch. But mathematically the normal form \eqref{8.1} is more accessible. 

Since the solution to  \eqref{8.1}, \eqref{8.2} scales ballistically, we set $n(x,t;v) = \mathsf{n}(t^{-1}x;v)$ and $\tilde{v}^\mathrm{eff}(x,t;v) = \tilde{\mathsf{v}}^\mathrm{eff}(t^{-1}x;v)$. Without loss of generality one adopts $t=1$ and arrives at  
\begin{equation}\label{8.3}
( x - \tilde{\mathsf{v}}^\mathrm{eff}(x;v))\partial_x \mathsf{n}(x;v) =0, \qquad \lim_{x \to \pm\infty} \mathsf{n}(x;v) = n_\pm(v).
 \end{equation}
Therefore  $x \mapsto \mathsf{n}(x;v)$  for fixed $v$ has to be constant except for jumps at the zeros of $ x - \tilde{\mathsf{v}}^\mathrm{eff}(x;v)$ as function of $x$. 
At this stage, it is not clear which level of generality to adopt. Physically, one would expect to have a unique solution. Thus 
in the $(x,v)$-plane there should be a \textit{contact line} which divides the plane into two domains, one containing the set $\{-\infty\}\times \mathbb{R}$ and the other the set
$\{\infty\}\times \mathbb{R}$. The solution  $\mathsf{n}(x;v)$ is constant in either domain and jumps from $n_-(v)$ to $n_+(v)$ across the contact line. Uniqueness means an unique contact point for every $v$.
Then the contact line is the graph of the function $v \mapsto \tilde{\phi}(v)$. As discussed below, for hard rods, $\tilde{\phi}$ is invertible,  inverse denoted by $\phi$. Hence  the contact line is the graph of $x \mapsto\phi(x)$.  We write our argument for the latter case, since hard rods can be considered as a limiting case of the Toda lattice. More general contact lines could be handled in a similar fashion. 

With our assumptions, for every $x$ there is a contact point $\phi(x)$. Then the solution ansatz reads
\begin{equation}\label{8.4}
n^\phi(v) = \chi(\{v > \phi\}) n_-(v)+ \chi(\{v \leq \phi\})n_+(v) \,\,\,\mathrm{and}\,\,\, \mathsf{n}(x;v) =  n^{\phi(x)}(v).
\end{equation}
The superscript $\phi$ will be used to generically  indicate that in the TBA formalism $n^\phi$ is substituted for $n= \rho_\mu$ and similarly the subscript
$\pm$ signals the substitution of  $n_\pm$. For example,  
\begin{equation}\label{8.5} 
\psi^{\mathrm{dr},\phi} = \big(1 - Tn^\phi\big)^{-1} \psi,\quad (\nu_-)^{-1} = \langle \rho_{\mathsf{p}-} \rangle
= \langle n_- \varsigma_{0}^{\mathrm{dr},-}\rangle, 
\end{equation}
compare with \eqref{3.48}. We set
\begin{equation}\label{8.6}
\tilde{\mathsf{v}}^\mathrm{eff}(x;v) = \tilde{\mathsf{v}}^{\mathrm{eff}, \phi(x)}(v).
\end{equation}
Then the condition on the left hand side of \eqref{8.3} translates to 
\begin{equation}\label{8.7}
x =  \tilde{\mathsf{v}}^\mathrm{eff} (x,\phi(x)).
\end{equation}
Put differently, one defines
\begin{equation}\label{8.8}
G(\phi) = \tilde{\mathsf{v}}^{\mathrm{eff},\phi} (\phi),
\end{equation}
then
\begin{equation}\label{8.9}
x = G(\phi(x)),\qquad \phi(x) = G^{-1}(x),
\end{equation}
which is means that $G$ is the inverse of the contact line $\phi$.

While $G(\phi)$ itself has to be obtained numerically, the large $|\phi|$ asymptotics
can still be argued analytically. We start from 
\begin{equation}\label{8.10} 
v^{\mathrm{eff},\phi}(\phi)\big(1 + T\rho_\mathsf{p}^\phi(\phi)\big) = \phi + \big(T\rho_\mathsf{p}^\phi v^{\mathrm{eff},\phi}\big)(\phi),
 \end{equation}
Setting
\begin{equation}\label{8.11} 
n^\phi(v) = n_+(v)+ \chi(\{v > \phi\})(n_-(v) - n_+(v)),
 \end{equation}
we note that for large $\phi$ the second term is exponentially small and can be neglected. The second summand on the left
of \eqref{8.10} then reads 
\begin{equation}\label{8.12} 
2 \int_\mathbb{R} \mathrm{d}w \log|\phi - w|\rho_{\mathsf{p},+}(w) \simeq2 \log \phi \int_\mathbb{R} \mathrm{d}w\rho_{\mathsf{p},+}(w)
= 2(\nu_+)^{-1}\log \phi .
 \end{equation}
Similarly for the right side of \eqref{8.10} one finds a logarithmic increase with a prefactor $c_+$, which could be determined by a 
further iteration.  The term $q_1^\phi/\nu^\phi$ converges to $q_{1,+}/ \nu_+$.
Thus for $\phi \to \infty$ one arrives at
 \begin{equation}\label{8.13} 
 G(\phi) \simeq  \frac{\phi + c_+\log |\phi|}{\nu_+ + 2 \log|\phi|} - \frac{q_{1,+}}{ \nu_+}.
 \end{equation}
For $\phi \to -\infty$, the same asymptotics holds upon substituting $\nu_-, c_-$ for   $\nu_+,c_+$. If the boundary terms $n_\pm$
have exponential decay, the error in \eqref{8.13} would be of the same order.

From numerical solutions with two thermal GGEs in contact, one finds indeed that the contact line is smooth and monotone increasing. In fact, qualitatively the contact line seems to be pretty much independent of the particular choice of $n_\pm$. If one imposes left and right pressures as satisfying
$P_- < P_\mathrm{c} < P_+$, then $\nu_- > 0$ and $\nu_+ <0$. Numerically one observes that $\nu(x)$ is a smooth function and that there is a unique point, $x_\mathrm{c}$,
such that $\nu(x_\mathrm{c}) = 0$. Somehow the  apparent $\nu^{-1}$ singularity in $ \tilde{v}^\mathrm{eff}(v)$ 
is cancelled. IF AVAILABLE: behavior of particles.\bigskip\\
$\blackdiamond\hspace{-1pt}\blackdiamond$~\textit{Hard rod domain wall}.\hspace{1pt} For the hard rod lattice the function $G$ can still be computed
analytically. We recall  that $\nu$ is the average stretch, $\nu>\mathsfit{a}$, and $h(v)$ denotes the normalized velocity distribution,
 \begin{equation}\label{8.14} 
\rho_\mathsf{p}(v) = \nu^{-1} h(v),\quad n(v) = (\nu- \mathsfit{a})^{-1} h(v), \quad q_1 = \int_\mathbb{R} \mathrm{d}w h(w)w. 
 \end{equation} 
In normal form the hard rod GHD reads
\begin{equation}\label{8.15} 
\partial_t n(x,t;v) + (\nu(x,t) -\mathsfit{a})^{-1}(v - q_1(x,t))\partial_x n(x,t;v) = 0, 
 \end{equation}
compare with the structurally identical Eq. \eqref{8.1}. The initial condition is the same as in \eqref{8.2} and the $G$-function is computed by the method explained above with the result
\begin{equation}\label{8.16} 
G_\mathrm{hr}(\phi) 
= \Big( \int_{\phi}^\infty \hspace{-4pt}\mathrm{d}v \, n_-(v) + \int^{\phi}_{-\infty} \hspace{-4pt}\mathrm{d}v \, n_+(v)\Big) \phi
- \int_{\phi}^\infty \hspace{-4pt}\mathrm{d}v v \,n_-(v) - \int^{\phi}_{-\infty} \hspace{-4pt}\mathrm{d}v \,v n_+(v).
 \end{equation}
Clearly, $dG_\mathrm{hr}(\phi)/d\phi >0$. Thus  $G_\mathrm{hr}$ is invertible. For $\phi \to \infty$ one obtains 
\begin{equation}\label{8.17} 
G_\mathrm{hr}(\phi) \simeq \frac{1}{\nu_+ -a}(\phi - q_{1,+})
 \end{equation}
and correspondingly for $\phi \to - \infty$ with $\nu_+, q_{1,+}$ replaced by $\nu_-, q_{1,-}$. If the boundary velocity distributions have an
exponential decay, then the error in \eqref{8.17} is also exponentially small. The contact line is asymptotically linear with a localized monotone interpolation. \hfill$\blackdiamond\hspace{-1pt}\blackdiamond$\bigskip\\
\textbf{\large{Notes and references}}\bigskip\\
 \textbf{ad 8}: The section is based on \cite{MS20}, where in particular numerical simulations of GHD with domain wall initial conditions are reported. Domain wall for hard rods are studied in \cite{DS17}.  On the Euler scale the step at the contact line is sharp. One expects that because of dissipation the step is actually broadened to an error-like function. Earlier work \cite{BCM19} considers the case of low pressure, $ P_- < P_+ < P_\mathrm{c}$. Molcecular dynamics is compared with predictions of GHD. 
Also the lowest order quantum mechanically corrections are studied. 
 For the Toda lattice, the structure of diffusive corrections are discussed in Section \ref{sec12}. The broadening is convincingly observed in the XXZ model \cite{DBD18}. 

Domain wall initial conditions have been studied numerically for the discrete sinh-Gordon model \cite{BDWY18}.
Most detailed studies are available for the XXZ model \cite{PNCBF17,MMK17}. In this case the spectral parameter space contains in addition the type of string states and the contact line refers to a larger space. 

Because of momentum conservation the contact line of the Toda lattice is supported by the full real line. On the other hand, for XXZ, and other discrete models, one usually observes a ``light cone'', i.e. the contact line is supported by an interval and there are left and right edges up to which  boundary values remain
constant. The behavior near the edge often shows intricate oscillatory decay, which has been elucidated in considerable detail
\cite{CLV18,BK19}.

 \section{Toda fluid}
\label{sec9}
\setcounter{equation}{0}
So far the lattice field theory picture prevailed, except for scattering theory, mostly because of the convenient handling of GGEs. For this section we view the Toda particles as moving on the real line. The name ``fluid" is somewhat misleading, since particles are distinguishable. Still the dynamical behavior is fluid-like. In a way, we  have to start from the beginning and redo the computation for average fields and their currents.   Since 
we rely on the chain results, the discussion can be more compressed. To avoid duplicating the symbols,
the fluid is distinguished from the chain through the index ``${}_\mathsf{f}$''. $x \in \mathbb{R}$ stands for the coordinate of the one-dimensional physical space.
\subsection{Euler equations}
\label{sec9.1}
The conserved fields have a density given by  
\begin{equation}\label{9.1} 
Q^{[0]}_\mathsf{f}(x) = \sum_{j \in \mathbb{Z}} \delta(q_j - x),\qquad Q^{[n]}_\mathsf{f}(x) = \sum_{j \in \mathbb{Z}} 
\delta (q_j- x) Q^{[n]}_j,\;n\geq1.
 \end{equation}
  Taking their time derivative
 \begin{eqnarray}\label{9.2} 
&&\frac{d}{dt}Q^{[0]}_\mathsf{f}(x) = \sum_{j \in \mathbb{Z}} \delta'(q_j - x)p_j,\nonumber\\
&&\frac{d}{dt}Q^{[n]}_\mathsf{f}(x) = \sum_{j \in \mathbb{Z}} \big(\delta'(q_j - x)p_jQ^{[n]}_j + \delta(q_j - x)(J_j^{[n]}
- J_{j+1}^{[n]})\big),  \quad n\geq 1,
 \end{eqnarray}
and hence the fluid current densities read
 \begin{eqnarray}\label{9.3} 
&&\hspace{0pt}J^{[0]}_\mathsf{f}(x) = \sum_{j \in \mathbb{Z}}\delta(q_j - x)p_j ,\\
&&\hspace{0pt} J^{[n]}_\mathsf{f}(x) = \sum_{j \in \mathbb{Z}}\delta(q_j - x)\big(p_jQ^{[n]}_j + (\theta(q_{j} -x) - \theta(q_{j-1} -x))J_j^{[n]}\big),\; n \geq 1,\nonumber
\end{eqnarray}
with $\theta$ the step function, $\theta(x) =0$ for $x\leq 0$ and $\theta(x) =1$ for $x > 0$. The index $0$ has been treated separately.
But in fact, setting formally $Q^{[0]}_j =1$, the case $n \geq 1$ naturally extends to  $n=0$.

The next item is GGE, where we start from the microcanonical ensemble with partition function
\begin{equation}\label{9.4} 
Z_\mathrm{mic}(N,\ell)= \int_{\tilde{\Gamma}_N}\prod_{j=1}^N\mathrm{d}r_j \mathrm{d}p_j \delta\big( Q^{[0],N} -\ell\big)\exp\!\big(-\mathrm{tr}[V(L_N)]\big). 
\end{equation} 
In our discussion the confining potential will play a passive role and only the $P$-dependence is indicated. For the chain we considered $\ell = \nu N$ and, by the equivalence of ensembles, switched to $ \exp\!\big( - P Q^{[0],N}\big)$
with the pressure $P$ dual to the stretch $\nu$. In the  limit $N\to \infty$  the free energy per site,
$F_\mathrm{toda}(P)$, has been obtained already, compare with  \eqref{3.34}. For the fluid we want to keep the volume fixed and allow for fluctuations in $N$.
The corresponding chemical potential is denoted by $\mu$. Then
\begin{equation}\label{9.5} 
Z_\mathsf{f}(\mu,\ell) = \sum_{N=1}^\infty \mathrm{e}^{\mu N} \int_{\tilde{\Gamma}_N} \prod_{j=1}^N\mathrm{d}r_j \mathrm{d}p_j \delta\big( Q^{[0],N} -\ell\big)\exp\!\big(-\mathrm{tr}[V(L_N)]\big) .
\end{equation}
One has to distinguish the cases $\ell >0$ and $\ell <0$, corresponding to  either increasing or decreasing particle labelling,
on average.
Thus the particle number $Q_\mathsf{f}^{[0],N} >0$, while the fluid density $\rho_\mathsf{f} = N/\ell$ can take either sign. Since $\nu = \ell/N$,
the relation  $\rho_\mathsf{f} \nu = 1$ holds always. Close to $P_\mathrm{c}$, defined by $\nu(P_\mathrm{c}) = 0$, the fluid density $\rho_\mathsf{f}$ jumps from $\infty$ to 
$- \infty$ when increasing $P$. This looks singular but merely reflects the particular choice of coordinates.  $\nu(P)$ is a smooth function. 
Of interest is the limit
\begin{equation}\label{9.6} 
\lim_{\ell \to \pm \infty} -\frac{1}{\ell}\log Z_\mathsf{f}(\mu,\ell) = F_{\mathsf{f},\pm}(\mu),\quad \mathrm{sgn}(\ell) = \pm 1 .
\end{equation}
The fluid free energy has two branches depending on  $\mathrm{sgn}(\ell)$. Correspondingly there are two distinct infinite volume GGE averages denoted by
$\langle \cdot \rangle_{\mu,\pm,V}$. For a convergent partition function, one has to require $\mu < \mu_\mathrm{max}$
for either branch with $\mu_\mathrm{max}= F_\mathrm{toda}(P_\mathrm{c})$, as explained below. 

The dual for $P$ is the stretch $\nu$ and for $\mu$ the fluid density $\rho_\mathsf{f}$. Let us denote by $\mathcal{L}F$ the Legendre
transform of some function $F$. By the equivalence of ensembles on the level of  free energies one concludes that 
\begin{equation}\label{9.7}
\lim_{N \to \infty} - N^{-1} \log Z_\mathsf{mic}(N,\nu N) = \mathcal{L}F_\mathrm{toda}(\nu)
\end{equation}
and 
 \begin{equation}\label{9.8}
\lim_{\ell \to \pm\infty} - \ell^{-1} \log Z_\mathrm{mic}(\rho_\mathsf{f}\ell,\ell) = \mathcal{L}F_{\mathsf{f},\pm}(\rho_\mathsf{f}).
\end{equation}
Thus the two free energies are related by
\begin{equation}\label{9.9} 
\mathcal{L}F_\mathrm{toda}(\nu) = \rho_\mathsf{f}^{-1}\mathcal{L}F_{\mathsf{f},\pm}(\rho_\mathsf{f}), \quad \nu \rho_\mathsf{f} =1,
\end{equation}
which implies
\begin{equation}\label{9.10} 
F_\mathrm{toda}(P) = \mu,\quad - F_{\mathsf{f},\pm}(\mu) = P.
\end{equation}
In other words, $F_\mathrm{toda}$ and $- F_{\mathsf{f},\pm}$ are inverse functions of each other. The first relation was already noted in
\eqref{3.34} and the argument of $F_{\mathsf{f},\pm}$ is in fact the parameter appearing in the TBA equation. $F_\mathrm{toda}(P)$ is convex down
with maximum at $P_\mathrm{c}$, while $F_{\mathsf{f},+}$  is convex down and   $F_{\mathsf{f},-}$ convex up with both graphs 
being smoothly joined at $\mu_\mathrm{max}$. 

The mean value of the conserved fields are first derivatives of $F_{\mathsf{f}}$ with respect to $\mu$ and, perturbing the 
confining potential as  $V +\kappa \varsigma_n$, with respect to $\kappa$ at $\kappa = 0$. For example, 
\begin{equation}\label{9.11} 
\rho_\mathsf{f} = \frac{1}{\ell} \langle N \rangle = - \frac{\partial F_{\mathsf{f}}}{\partial \mu} =  
 -\frac{\partial F_{\mathsf{f}}}{\partial F_\mathrm{toda}} = - \big(\frac{\partial F_\mathrm{toda}}{\partial F_{\mathsf{f}}}\big)^{-1}
= \big(\frac{\partial F_\mathrm{toda}}{\partial P}\big)^{-1} = \nu^{-1}
\end{equation}
and similarly for the $\kappa$-derivative. One concludes that
\begin{equation}\label{9.12}
\langle Q_{\mathsf{f}}^{[n]} (0)\rangle_{\mu,\mathrm{sgn}(\nu),V} = \nu^{-1}\langle Q_0^{[n]}\rangle_{P,V} = 
\langle \rho_\mathsf{p}\varsigma_n\rangle,  \quad n = 0,1,...\,.
\end{equation}

The current averages still have the form of a weighted DOS. Hence there exists a yet unknown function $\bar{v}$ such that 
\begin{equation}\label{9.13} 
\langle J_{\mathsf{f}}^{[n]} (0)\rangle_{\mu,\pm,V} = \langle \rho_\mathsf{p} \bar{v}  \varsigma_n\rangle, 
\end{equation}
compare with (6.20). As for the lattice we introduce the field-current susceptibility matrix through
\begin{equation}\label{9.14} 
B_{\mathsf{f},m,n} = \int_\mathbb{R} \mathrm{d}x \langle J_{\mathsf{f}}^{[m]} (x)Q_{\mathsf{f}}^{[n]} (0)\rangle^\mathrm{c}_{\mu,\mathrm{sgn}(\nu),V},
\end{equation}
which, by the same argument as in \eqref{6.8}, is a symmetric matrix, $B_{\mathsf{f},m,n} = B_{\mathsf{f},n,m}$. Now
\begin{equation}\label{9.15} 
B_{\mathsf{f},n,0} = \partial_\mu \langle J_{\mathsf{f}}^{[n]} (0)\rangle_{\mu,\pm,V}. 
\end{equation}
Since $J_{\mathsf{f}}^{[0]} =   Q_{\mathsf{f}}^{[1]}$, we also determine
\begin{equation}\label{9.16} 
B_{\mathsf{f},0,n} = -\partial_\kappa  \langle Q_{\mathsf{f}}^{[n]} (0)\rangle_{\mu,\pm,V+ \kappa w}\big|_{\kappa = 0}, 
\end{equation}
which uses \eqref{9.12}.  The confining potential is
$V(w)+ \kappa w$ and the corresponding DOS is denoted by $\nu\rho_\mathsf{p}(\kappa)$. Then by symmetry,
\begin{equation}\label{9.17} 
\partial_\mu \langle J_{\mathsf{f}}^{[n]} (0)\rangle_{\mu,\pm,V} 
= -\partial_\kappa\langle \rho_\mathsf{p}(\kappa)  \varsigma_n \rangle\big|_{\kappa = 0}. 
\end{equation}
 
The corresponding  $\rho_\mu(\kappa)$ is solution of the TBA equation 
\begin{equation}\label{9.18} 
V +\kappa  \varsigma_1 -\mu - T \rho_\mu(\kappa) + \log \rho_\mu(\kappa)  =0.
\end{equation}
Hence $\rho_\mu'$, the derivative at $\kappa =0$, is determined by
\begin{equation}\label{9.19} 
\rho_\mu' = - \rho_\mu (1 - T \rho_\mu)^{-1}  \varsigma_1
\end{equation}
and
\begin{equation}\label{9.20} 
\rho_\mathsf{p}' = \partial_\mu \rho_\mu' = - \partial_\mu \rho_\mu \varsigma_1^\mathrm{dr}
= -\partial_\mu \big(\rho_\mathsf{p} v^\mathrm{eff}\big) ,
\end{equation}
according to \eqref{3.49} and \eqref{6.19}. We conclude that
\begin{equation}\label{9.21}
\partial_\mu \big(\rho_\mathsf{p}(\bar{v} -   v^\mathrm{eff})\big) = 0
\end{equation}
implying that the difference $(\bar{v} -   v^\mathrm{eff})$ does not depend on $\mu$. To fix the constant, 
the current vanishes in the limit of vanishing density of particles, in other words for $\mu \to -\infty$. In conclusion the Toda fluid currents are given by
\begin{equation}\label{9.22} 
\langle J_{\mathsf{f}}^{[n]} (0)\rangle_{\mu,\pm,V} = \langle \rho_\mathsf{p}v^\mathrm{eff} \varsigma_n \rangle.
\end{equation}
This formula holds also for $n = 0$.
Compared to the lattice currents \eqref{6.20}, the changes are minimal. Only the shift $q_1$ has been removed. 
This can be understood by starting from the fluid side. Switching to the lattice dynamical equations  the lateral motion has been subtracted and such a term has to appear in the average currents. 

Since  $n =0$ is included, 
there is only a single hydrodynamic equation which reads
\begin{equation}\label{9.23} 
\partial_t\rho_\mathsf{p}(x,t;v) + \partial_x\big(v^\mathrm{eff}(x,t;v)\rho_\mathsf{p}(x,t;v)\big) = 0.
\end{equation}
Transforming to quasilinear form yields
\begin{equation}\label{9.24} 
\partial_t\rho_\mu(x,t;v) + v^\mathrm{eff}(x,t;v)\partial_x\rho_\mu(x,t;v) = 0,
\end{equation}
which is identical to the lattice version, except for the trivial change in the effective velocity.\bigskip\\
$\blackdiamond\hspace{-1pt}\blackdiamond$~\textit{Negative volume}.\hspace{1pt}
The reader might have noticed that in \eqref{9.12} the $\nu$-dependence of $\rho_\mathsf{p}$ is hidden. 
By construction $\rho_\mathsf{p} \geq 0$ for $\nu >0$ and $\rho_\mathsf{p} \leq 0$ for $\nu < 0$. Eq. \eqref{9.20}
preserves positivity. If initially $\rho_\mathsf{p} (x) \geq 0$, this will remain so at later times, and  correspondingly 
so for the negative sign. Microscopically this reflects that approximate order, resp. anti-order, of particle's positions is preserved on 
hydrodynamic time scales. But one could imagine to prepare an infinitely extended thermal state with $\nu_+ > 0$ 
and tune in a large, finite segment with parameter $\nu_- <0$. Microscopically, ignoring fluctuations, this would amount to  
the order  $ ...<q_{-1} < q_0 <q_N <...<q_1 < q_{N+1} < q_{N+2}< ...$\,.
But then $r_0$ and $r_{N+1}$ are of order $N$. Such a state is far away from a local equilibrium state 
and the hydrodynamic equation \eqref{9.20} is no longer applicable. Of course it could be that for such initial conditions 
the Toda fluid still reaches a hydrodynamic regime, which can be clarified only through further studies. 
\hfill$\blackdiamond\hspace{-1pt}\blackdiamond$
\subsection{Preparing for integrable quantum systems}
\label{sec9.2}
When turning to integrable quantum systems, the TBA equations are organized somewhat differently. To establish the connection
we rewrite the Toda fluid with some additional notions. Of course generalized hydrodynamics is not affected.

We introduce various densities. However they do \textit{not} refer to physical space, rather to momentum space or more abstractly to the space of spectral parameters, also called rapidities or Bethe roots.
But most importantly we rename $\rho_\mu$ as
\begin{equation}\label{9.25} 
\rho_\mu = \rho_\mathsf{n},
\end{equation}
which is referred to as \textit{number density}. Secondly there is a \textit{space density},  $\rho_\mathsf{s}$, satisfying
\begin{equation}\label{9.26} 
\rho_\mathsf{s} = 1 + T  \rho_\mathsf{p}
\end{equation}
which then leads to the \textit{particle density},  $\rho_\mathsf{p}$, through
\begin{equation}\label{9.27} 
 \rho_\mathsf{p} = \rho_\mathsf{n} \rho_\mathsf{s}
\end{equation}
and the \textit{hole density} as 
\begin{equation}\label{9.28} 
 \rho_\mathsf{p} +  \rho_\mathsf{h} =  \rho_\mathsf{s}
\end{equation}
Using \eqref{3.50} one deduces 
\begin{equation}\label{9.29} 
 \rho_\mathsf{s} = \varsigma_1^\mathrm{dr}.
\end{equation}

To capture this standard terminology, it is useful to revisit
the one-dimensional ideal Fermi gas in the periodic volume of size $L$. Setting $\hbar =1$, momentum space is $2\pi L^{-1} \mathbb{Z}$. Since this is a uniform lattice the space density $\rho_\mathsf{s} = \tfrac{1}{2\pi}$. The creation and annihilation operators are
$\{ a(k)^*, a(k), k \in 2\pi L^{-1} \mathbb{Z}\}$. Then the number density is  $a(k)^*a(k)$, the particle density $(2\pi)^{-1}a(k)^*a(k)$,
and the hole density $(2\pi)^{-1}(1 - a(k)^*a(k))$, either as operators or as average over some state of the Fermi gas.
In an interacting integrable system, the regular lattice becomes distorted, leading to a nonuniform $\rho_\mathsf{s}$, 
and there are more complicated linear relations between the various densities. 
 
The TBA equation remains intact, only we absorb the parameter $\mu$ into the confining potential, with the result
\begin{equation}\label{9.30} 
\varepsilon = V - T\mathrm{e}^{-\varepsilon}, \qquad  \mathrm{e}^{-\varepsilon} = \rho_\mathsf{n}.
\end{equation}
In Section \ref{3.2} we discussed a variational problem determining $\rho_\mathsf{n}$, We now switch to a variational problem for $\rho_\mathsf{p}$, for which we again write a free energy functional. This time energy is just due to the confining potential and entropy is the entropy of  $\rho$ relative to $\rho_\mathsf{s}$. Regarding $\rho$
as the free parameter this leads to 
\begin{equation}\label{9.31} 
\mathcal{F}(\rho) = \int_\mathbb{R} \mathrm{d}w \rho(w) (V(w) - 1) + \int_\mathbb{R} \mathrm{d}w\rho_\mathsf{s}(w)\Big(\frac{\rho}
{\rho_\mathsf{s}} \log\frac{\rho}{\rho_\mathsf{s}}\Big)(w), 
\end{equation}
with the constraint  
\begin{equation}\label{9.32} 
\rho_\mathsf{s} = 1 + T  \rho,
\end{equation}
which states the way how the particle density  changes the underlying geometry.  The Euler-Lagrange equation determining 
$\rho_\mathsf{p}$ is then
\begin{equation}\label{9.33} 
V + \log\rho_\mathsf{n} - T  \rho_\mathsf{n} = 0,
\end{equation}
in agreement with the TBA equation \eqref{9.30}.

For quantum models the relation between $\rho_\mathsf{n}$ and the quasi-energies will have to be modified.
As most striking aspect,  only the two-particle scattering shift appears in the variational formula for the generalized free energy.
For the Lieb-Liniger model, this feature originates
 from the Bethe ansatz, in which  the eigenfunctions carry already the information on the two-particle phase shift. In our Toda lattice computation,
 this feature comes as a surprise and looks more accidental at first sight.
\bigskip\\
\textbf{\large{Notes and references}}\bigskip\\
 \textbf{ad 9.1}:  A more detailed discussion of the Toda fluid is supplied in \cite{D19,D19a}. Reported is on employing the  scattering map
  to establish a direct link between two-particle scattering shift and the generalized free energy. \medskip\\
 \textbf{ad 9.2}: Our terminology originates from the thermodynamics of Lieb-Liniger $\delta$-Bose gas
 \cite{YY69}. However, the mathematical structure seems to be generic for all integrable many-body systems.
 
 \section{Hydrodynamics for  the Lieb-Liniger $\delta$-Bose gas}
\label{sec10}
\setcounter{equation}{0} 
Viewed from a distance the Toda lattice is only a tiny side branch of generalized hydrodynamics,
the major efforts being concentrated on integrable many-body quantum systems as the XXZ chain and the spin-$\tfrac{1}{2}$ Fermi-Hubbard model. Quantum models are already well-covered in other publications. Still the structural similarities between diverse integrable models is so striking that this topic has to be addressed.
We first discuss in detail the Lieb-Liniger model of bosons in one dimension interacting through a repulsive point-like
pair potential, which will serve also as a timely preparation for the quantum Toda lattice in Section \ref{sec11}.
\subsection{Bethe ansatz}
\label{sec10.1}
The quantum field theory under consideration is the scalar bosonic field $\Psi(x)$, $x \in \mathbb{R}$, with canonical commutation
relations
\begin{equation}\label{10.1} 
[\Psi(x),\Psi(x')^*] = \delta(x-x'), \quad [\Psi(x),\Psi(x')] = 0, \quad [\Psi(x)^*,\Psi(x')^*] = 0.
\end{equation}
Normally ordered the Lieb-Liniger hamiltonian reads  
\begin{equation}\label{10.2} 
H_\mathrm{li} = \tfrac{1}{2}\int_\mathbb{R} \mathrm{d}x \big(\partial_x \Psi(x)^*\partial_x \Psi(x) + c \Psi(x)^*\Psi(x)^*\Psi(x)\Psi(x)\big).
\end{equation}
$c$ is the coupling constant with $c \geq 0$ imposed.  The attractive case, $c <0$, is considerably more complicated
because of bound states. In the widely accepted standard convention  the factor $\tfrac{1}{2}$ in front of the integral is omitted. We inserted here so have an  energy-momentum relation of the free theory identical to the one of the noninteracting Toda fluid. 
The Lieb-Liniger hamiltonian is formal because of point-interactions. Fortunately, there is  the first quantized version,
which in the $N$-particle sector reads
\begin{equation}\label{10.3}
H_{\mathrm{li},N} = - \sum_{j = 1}^N  \tfrac{1}{2} (\partial_{x_j})^2 + c\sum_{1 \leq i <  j\leq N}\delta(x_i -  x_j).
 \end{equation}
Bosonic means that the operator acts on wave functions symmetric under particle exchange.
The case of interest is a spatial interval $[0,L]$, $L> 0$, with periodic boundary conditions, which will be distinguished by either 
a sub\,- or superscript $L$. Of course, one could also consider a finite number of particles moving on the real line, which would be the natural set-up  for multi-particle scattering.  

Let us first recall, how to handle the delta potential.  The simplest example is $N =2$, for which the relative motion is governed by  the hamiltonian 
\begin{equation}\label{10.4} 
H_\mathrm{rel} = - \partial_x^2 + c\,\delta(x).
 \end{equation}
Away from $x=0$, $H_\mathrm{rel} \psi(x) = - \partial_x^2\psi(x)$
with $\psi \in \mathcal{C}^2(\mathbb{R}\setminus\{0\})$, the twice continuously differentiable functions on 
$\mathbb{R}\setminus\{0\}$ with bounded derivatives. The $\delta$-potential translates to the boundary condition
\begin{equation}\label{10.5}
 \partial_x\psi(0_+) -  \partial_x\psi(0_-) = 2c \psi(0). 
\end{equation}
The same mechanism works for the $N$-particle hamiltonian $H_{\mathrm{li},N,L}$
acting on symmetric wave functions.
For them one can restrict the construction to the Weyl chamber $\mathbb{W}_{N,L} = \{0 \leq x_1 \leq ... \leq x_N \leq L\}$. By permutation symmetry the hamiltonian can then be extended to all other sectors. Away from the boundary the hamiltonian is the $N$-dimensional Laplacian,
\begin{equation}\label{10.6} 
H_{\mathrm{li},N,L} = - \sum_{j = 1}^N  \tfrac{1}{2}(\partial_{x_j})^2,
\end{equation}
which acts on $\psi \in \mathcal{C}^2(\mathbb{W}_{N,L}\setminus\partial \mathbb{W}_{N,L})$. Periodic boundary conditions mean
\begin{equation}\label{10.7}
\psi(0,x^\bot) = \psi(x^\bot, L),\quad \partial_{x_1}\psi(0,x^\bot) = \partial_{x_N}\psi(x^\bot, L),
\end{equation}
and the interaction is encoded by 
\begin{equation}\label{10.8} 
(\partial_{x_{j+1}} - \partial_{x_{j}})\psi(x) = c \psi(x)\big|_{x_j = x_{j+1}},\quad j = 1,...,N-1,
\end{equation}
where the limit is taken from the interior of $\mathbb{W}_{N,L}$. It can be shown that the Laplacian with these boundary conditions defines a unique self-adjoint operator on $L^2(\mathbb{W}_{N,L} )$. 

On  $\mathbb{W}_{N,L}$ an eigenfunction of $H_{\mathrm{li},N,L}$ has the Bethe ansatz form
\begin{equation}\label{10.9} 
\psi(x_1,...,x_N) = \sum_{\sigma \in \mathcal{S}_N} \Big[\exp\Big(\mathrm{i} \sum_{j=1}^N \lambda_{\sigma(j)}x_j\Big) 
\prod_{\substack{j < k\\ \sigma(j) >\sigma(k)}} A(\lambda_j, \lambda_k)  \Big],
\end{equation}
where the sum is over all permutations $\sigma$ of $(1,...,N)$.
$\psi$ has to satisfy the boundary conditions \eqref{10.8}, which leads to
\begin{equation}\label{10.10} 
A(\lambda_1, \lambda_2) = \frac{\mathrm{i}(\lambda_1 - \lambda_2) - c}{\mathrm{i}(\lambda_1 - \lambda_2) + 
c}  = \mathrm{e}^{-\mathrm{i} \theta_\mathrm{li}(\lambda_1 - \lambda_2)},\quad j = 1,...,N.
\end{equation}
Here  two-particle phase shift is $\theta_\mathrm{li}(w) = 2\arctan (w/c)$,
implying the scattering  shift 
\begin{equation}\label{10.11}
\theta_\mathrm{li}'(w) = \frac{2c}{w^2 + c^2}.  
\end{equation}
Imposing the boundary conditions  \eqref{10.7},  the $\lambda$ coefficients  are constrained to satisfy the Bethe equations 
\begin{equation}\label{10.12} 
2\pi I_jL\lambda_j = L\lambda_j + \sum_{i = 1}^N \theta_\mathrm{li}(\lambda_j - \lambda_i). 
\end{equation}
The quantum states are labelled by vectors $I = (I_1,...,I_N)$, which  have entries from $\mathbb{Z}$ in case of odd and from $\mathbb{Z}+\tfrac{1}{2}$ in case of even $N$ and are constrained by $I_1 < ...< I_N$. 
For each such  $I$, the Bethe
equations have a unique solution denoted by $\lambda = (\lambda_1,...,\lambda_N)$ ordered as $\lambda_1 <...<\lambda_N$, 
also called Bethe roots or rapidities.
It is known that these eigenfunctions constitute a complete orthonormal basis in $L^2(\mathbb{W}_{N,L})$. 

The symmetric eigenfunctions on $L^2([0,L]^N)$ can be written as 
\begin{equation}\label{10.13} 
\psi_\lambda(x_1,...,x_N)  =  \sum_{\sigma \in \mathcal{S}_N} \exp\Big(\mathrm{i} \sum_{j=1}^N \lambda_{\sigma(j)}x_j\Big)
 \prod_{1 \leq j <k \leq N} \Big(1 - \frac{\mathrm{i} c \,\mathrm{sgn}(x_k- x_j)}{\lambda_{\sigma(k)} -\lambda_{\sigma(j)}}\Big). 
 \end{equation}
$\psi_\lambda$ is not normalized, the normalized state vector being denoted by  $|\lambda\rangle = \langle\psi_\lambda,\psi_\lambda\rangle^{-\frac{1}{2}}  \psi_\lambda$. The normalization constants have been computed and the vectors $|\lambda\rangle$ span the bosonic subspace of 
$L^2([0,L]^N)$. 
In particular
\begin{equation}\label{10.14} 
H_{\mathrm{li},N,L} = \sum_{\{\lambda,N\}}\Big(\tfrac{1}{2}\sum_{j=1}^N(\lambda_j)^2\Big) |\lambda\rangle \langle\lambda |,
\end{equation}
where $\{\lambda,N\}$ indicates that the sum is over all $N$-vectors of rapidities. Up to the prefactor  $\tfrac{1}{2}$, if the power $2$ is replaced by $0$, one arrives at the number operator. Similarly power $1$ results in the total momentum. This suggests 
to introduce the $n$-th conserved field (= charge) by 
\begin{equation}\label{10.15} 
Q^{[n],N,L} = \sum_{\{\lambda,N\}}\Big(\sum_{j=1}^N(\lambda_j)^n \Big)|\lambda\rangle \langle\lambda |.
\end{equation}
It will be convenient to switch to a variable particle number by introducing the Fock space of symmetric wave functions,
\begin{equation}\label{10.16} 
\mathcal{F}_{\mathrm{sym},L} = \bigoplus_{N=0}^\infty L^2([0,L]^N)_{\mathrm{sym}}.
\end{equation}
Then the $n$-th charge operator, $Q^{[n],L}$, restricted the $N$-particle subspace of 
$\mathcal{F}_{\mathrm{sym},L} $ equals $Q^{[n],N,L}$. 

Obviously, the operator $Q^{[n],L}$ is well  defined.
But what about its charge density? Put differently, in what sense is $Q^{[n],L}$ the integral over a quasi-local density, a property
which lies at the foundations of the hydrodynamic approach. One natural strategy would be to start from the classical nonlinear
Schr\"{o}dinger equation and its known sequence of charge densities involving higher order field derivatives. As naive guess, imposing normal order will yield the charges of the Lieb-Liniger model in second quantized form, which would then be manifestly local.
But this scheme  works only up to $n= 3$. Already for $Q^{[4],L}$ more complicated subtraction terms appear.  
For the total charge currents, denoted by $J^{[n],L}$, similar difficulties will appear. Of course, $J^{[0],L} = Q^{[1],L} $.
  As discussed in more detail below, the momentum current equals
\begin{equation}\label{10.17} 
J^{[1],L} = \int_0^L \mathrm{d}x\big(\partial_x\Psi(x)^*\partial_x\Psi(x) + \tfrac{1}{2}c\Psi(x)^*\Psi(x)^*\Psi(x)\Psi(x) \big). \end{equation}
But higher currents have not yet been studied in detail. We still use $Q^{[n],L}$ and $J^{[n],L}$, assuming that at some point these difficulties can be resolved. 
\subsection{Free energy, variational principle}
\label{sec10.2}
The Boltzmann weight involves a linear combination of charges. Therefore, again following the Toda lattice blue-print, we introduce  the confining potential $V$ and set
\begin{equation}\label{10.18} 
Q^{V,L} = \sum_{\{\lambda\}}\Big(\sum_{j=1}^N V(\lambda_j)\Big) |\lambda\rangle \langle\lambda |.
\end{equation}
Now the sum is over $\lambda$ vectors with an arbitrary number of components and the particular $N$ depends on $\lambda$. 
Note that  the conventional chemical potential, $\mu$,  corresponds to the constant term $V(w) = -\mu$. The unnormalized  GGE density matrix is defined by
\begin{equation}\label{10.19}
\exp\big[- Q^{V,L}\big]
\end{equation}
as operator on Fock space. Hence the normalizing  partition function is given by
\begin{equation}\label{10.20}
Z_L(V) = \mathrm{tr}\big[\mathrm{e}^{- Q^{V,L}} \big] = \sum_{\{\lambda\}} \exp\Big[ - \sum_{j=1}^N V(\lambda_j)\Big], 
\end{equation}
trace over Fock space, and the GGE average of some operator $\mathcal{O}$ is defined by
\begin{equation}\label{10.21}
\langle \mathcal{O}\rangle_{V,L} =  Z_L(V)^{-1}\mathrm{tr}\big[\mathrm{e}^{- Q^{V,L}} \mathcal{O}\big].
\end{equation}

For the free energy per unit length we have
\begin{equation}\label{10.22} 
F_\mathrm{yy}(V) = - \lim_{L\to \infty}  \frac{1}{L} \log Z_L(V).
\end{equation}
This free energy turns out to be  determined by a variational problem, namely through minimizing the  Yang-Yang free energy functional
\begin{equation}\label{10.23}
\mathcal{F}_\mathrm{yy}(\rho) = \int_\mathbb{R}\mathrm{d}w \big( \rho V +\rho \log \rho +  \rho_\mathsf{h} \log \rho_\mathsf{h} - 
(\rho +  \rho_\mathsf{h})\log (\rho +  \rho_\mathsf{h})\big).
\end{equation}
The similarity to Eq. \eqref{9.31} is striking. Here the variation is over all densities $\rho$, $\rho \geq 0$, and $\rho_\mathsf{h}$ depends linearly on $\rho$ through
\begin{equation}\label{10.24} 
\rho + \rho_\mathsf{h} = \frac{1}{2\pi} + T\rho,
\end{equation}
where $T$ is the integral operator 
 \begin{equation}\label{10.25}
Tf(w)  = \frac{1}{2\pi} \int_\mathbb{R} \mathrm{d}w'  \frac{2c}{(w - w')^2 +c^2} f(w'). 
\end{equation}
Note that the kernel of $2\pi T$ equals the scattering shift $ \theta' (w - w')$. Thus $2\pi T$ will play the same role as the $T$ operator
already familiar from the Toda lattice.  Note also that $T\varsigma_0 = \varsigma_0$ and  
$T \to 1$ for $c \to 0$, while $T \to 0$ for $c \to \infty$. 

Actually Yang and Yang worked with the fixed $N$ canonical ensemble, as in the case of the Toda lattice.
This leads to the additional constraint 
\begin{equation}\label{10.26} 
\int_\mathbb{R}\mathrm{d}w \rho(w) = \frac{N}{L},
\end{equation}
which afterwards is  removed at the expense of introducing $\mu$ as Lagrange multiplier. In the grand-canonical ensemble this step can be circumvented.

 The functional $\mathcal{F}_\mathrm{yy}$ is strictly convex. Following the argument by Yang and Yang, we start from  two  arbitrary 
spectral densities $\rho_0,\rho_1$ of well-defined Yang-Yang free energy and consider the linear interpolation
$\rho_u = \rho_0(1-u) + u \rho_1$, $0 \leq u \leq 1$. We plan to show that 
$\mathcal{F}_\mathrm{yy}(\rho_u)'' >0$, the prime denoting derivative with respect to $u$. Note that
  $\rho_u'  = \rho_1- \rho_0$  and, in general, $\rho' + \rho_\mathsf{h}' = T\rho'$. With this input, differentiating $\mathcal{F}_\mathrm{yy}(\rho(u))$
  is straightforward with the result
\begin{equation}\label{10.27} 
\mathcal{F}_\mathrm{yy}(\rho_u)''  = \int_\mathbb{R}\mathrm{d}w \Big(\frac{1}{\rho_u} + \frac{1}{\rho_{\mathsf{h},u}}\Big)\Big((\rho_1 -\rho_0)
- \frac{\rho_u}{\rho_u+ \rho_{\mathsf{h},u}}T (\rho_1- \rho_0)\Big)^2 \geq 0.
\end{equation}
Vanishing of the curvature would imply $(\rho_u+ \rho_{\mathsf{h},u})(\rho_1 - \rho_0) = \rho_u T(\rho_1- \rho_0)$, which cannot be satisfied because of \eqref{10.24}.
As a consequence, under the constraints \eqref{10.24} and $\rho \geq 0$,  the free energy functional $\mathcal{F}_\mathrm{yy}$ has a unique minimizer, denoted by $\rho_\mathsf{p}$, and the free energy of the $\delta$-Bose gas turns out to be given by
\begin{equation}\label{10.28} 
F_\mathrm{yy}(V) = \mathcal{F}_\mathrm{yy}(\rho_\mathsf{p}).
\end{equation}
 The minimizer is determined by
 the saddle-point equation
\begin{equation}\label{10.29} 
V - \log \frac{\rho_\mathsf{h}}{\rho_\mathsf{p}} - T \log \Big(1 +   \frac{\rho_\mathsf{p}}{\rho_\mathsf{h}}\Big) =0.
\end{equation}

For the $\delta$-Bose gas the Bethe root density $\rho_{\mathsf{p}}$ is often called quasiparticle momentum density, in analogy to the momentum density of a noninteracting
Fermi gas.  As introduced already in Section \ref{9.2}, $\rho_\mathsf{h}$ is the hole density, $\rho_\mathsf{s} = \rho_\mathsf{p} + \rho_\mathsf{h}$
the space density, and $\rho_\mathsf{n} = \rho_\mathsf{p} / \rho_\mathsf{s}$ the number density. In contrast to a classical model, for a quantum system with fermionic excitations, the quasi-energies are defined through  
\begin{equation}\label{10.30} 
\rho_\mathsf{n} = \frac{1}{1 + \mathrm{e}^\varepsilon},
\end{equation}
yielding the quasi-energy  $\varepsilon$ as
\begin{equation}\label{10.31}
\mathrm{e}^\varepsilon
= \frac{\rho_\mathsf{h}}{\rho_\mathsf{p}}. 
\end{equation}
With these conventions Eq. \eqref{10.29} turns into  the quantum TBA equation
\begin{equation}\label{10.32} 
\varepsilon = V  - T\log( 1 + \mathrm{e}^{-\varepsilon}).
\end{equation}
Very roughly, in the semiclassical limit $\log( 1 + \mathrm{e}^{-\varepsilon}) \simeq \mathrm{e}^{-\varepsilon}$, which is of the same structure as \eqref{9.30}.  
  The dressing of some function $\psi$ is defined through
\begin{equation}\label{10.33} 
\psi^\mathrm{dr} = \psi + T \rho_\mathsf{n} \psi^\mathrm{dr},\quad \psi^\mathrm{dr} = \big(1 - T\rho_\mathsf{n}\big)^{-1} \psi.
\end{equation} 
Then the constraint \eqref{10.24} can be written as
\begin{equation}\label{10.34}
 2\pi \rho_\mathsf{p} = \rho_\mathsf{n}\varsigma_0^\mathrm{dr}.
\end{equation} 

These formulas look familiar.  For convenience we tabulate the precise correspondence
with the convention of absorbing the chemical potential $\mu$ into the confining potential $V$:\\
\begin{center}
\begin{tabular}[h]{l|c|c}
&\hspace{10pt} \!\!\!Toda fluid & $\hspace{10pt}\delta$-Bose gas \\[1ex]
\hline
&&\\
2-particle scattering shift  & $2\log|w|$ & $ 2c(w^2 + c^2)^{-1}$ \\[1ex]
free energy functional& $\mathcal{F}(\rho)  $& $\mathcal{F}_\mathrm{yy}(\rho)$ \\[1ex]
saddle point, TBA& $ \varepsilon = V - T  \mathrm{e}^{-\varepsilon}$ & $\varepsilon = V  - T \log(
1 +  \mathrm{e}^{-\varepsilon})$\\[1ex]
occupation number, quasi-energy& $\rho_\mathsf{n}= \rho_\mu = \mathrm{e}^{-\varepsilon}$ &  $\rho_\mathsf{n} = (1 + \mathrm{e}^\varepsilon)^{-1}$\\[1ex]
spectral/root density & $\rho_\mathsf{p}= \rho_\mu\varsigma_0^\mathrm{dr}$& $\rho_\mathsf{p} = \tfrac{1}{2\pi} \rho_\mathsf{n}\varsigma_0^\mathrm{dr} $.   
\end{tabular}\bigskip
\end{center}

Adding the small perturbation $V+ \kappa \varsigma_n$ and computing the linear response of the free energy in $\kappa$ one obtains
\begin{equation}\label{10.35} 
\lim_{L\to \infty} L^{-1} \langle Q^{[n],L} \rangle_{V,L}   = \langle \rho_\mathsf{p}\varsigma_n\rangle, \quad n = 0,1,...\,.
\end{equation}
This identity suggests the proper interpretation of $\rho_\mathsf{p}$. Under the GGE  the rapidities
$\lambda_1,...,\lambda_N$ with variable  $N$ are random and combine into a empirical density as
\begin{equation}\label{10.36} 
\rho_{\mathsf{p},L}(w) = L^{-1}\sum_{j=1}^N\delta(w - \lambda_j).
\end{equation}
Obviously, $ \rho_{\mathsf{p},L}(w) \geq 0$ with normalization $\langle  \rho_{\mathsf{p},L}\rangle = N/L$, where $N$ is still random.
Eq. \eqref{10.35} states that on average the $n$-th moment of $\rho_{\mathsf{p},L}$ converges to the one of $\rho_{\mathsf{p}}$
as $L \to \infty$. In Section \ref{sec4}, we explained that typical fluctuations of the empirical density of Lax matrix eigenvalues are of the order $1/\sqrt{N}$. For the Lieb-Liniger model we cannot point at some published article. But since large deviations are well established and the Yang-Yang free energy functional is strictly convex, Gaussian fluctuations are  to be expected. But then
\begin{equation}\label{10.37}
\lim_{L \to \infty} \rho_{\mathsf{p},L}(w) = \rho_\mathsf{p}(w)
\end{equation}
with probability one. To be more precise, one integrates both sides of the equation over some smooth rapidly decaying test function.
Then this sequence of random variables converges with probability one to the stated non-random limit.

\subsection{Charge currents, hydrodynamic equations}
\label{sec10.3}
We still have to figure out the average currents. One approach is to observe that the number current is momentum,
which is a conserved charge. Since valid in such a great generality, we claim that for the $\delta$-Bose gas the identity \eqref{9.14}
still holds. To have a convenient shorthand,  the confining potential is perturbed as $V + \kappa_0\varsigma_0 + \kappa_1\varsigma_1$ and differentiating with respect to $\kappa_i$ at $\kappa_i=0$ is denoted by $\partial_i$. The charge current DOS is written as
$\rho_\mathsf{p}\bar{v}$ with a yet unknown function $\bar{v}$. With this notation, Eq. \eqref{9.17}, in conjunction with \eqref{9.13}, becomes
\begin{equation}\label{10.38}
    \partial_0\big(\rho_\mathsf{p}\bar{v}\big) = \partial_1\rho_\mathsf{p}.
\end{equation} 
To be shown is $\bar{v} = v^\mathrm{eff}$, where our obvious guess for the effective velocity is
\begin{equation}\label{10.39} 
 v^\mathrm{eff}(w)= w + 2\pi\big( T(\rho_\mathsf{p}v^\mathrm{eff})(w) -  T\rho_\mathsf{p}(w)v^\mathrm{eff}(w)\big).
\end{equation}
According to \eqref{10.25}, the kernel of $2\pi T$ is defined through the two-particle scattering shift. Since, compared to the Toda lattice, TBA and dressing transformation have been modified, one has to reconsider the argument in Section \ref{sec9.1}.

Differentiating the TBA equation with respect to $\kappa_0, \kappa_1$ one obtains the relations 
\begin{equation}\label{10.40}
 \partial_0\rho_\mathsf{n} = - \rho_\mathsf{n} (1-\rho_\mathsf{n} ) \varsigma_0^\mathrm{dr}, \quad \partial_1\rho_\mathsf{n}  = - \rho_\mathsf{n} (1-\rho_\mathsf{n} ) \varsigma_1^\mathrm{dr} 
\end{equation}
and, because  $\partial_0 \partial_1\rho_\mathsf{n}  = \partial_1 \partial_0\rho_\mathsf{n}$ for mixed derivatives,
\begin{equation}\label{10.41}
\partial_1 \varsigma_0^\mathrm{dr} =  \partial_0\varsigma_1^\mathrm{dr}.
\end{equation}
Next we write \eqref{10.39} as
\begin{equation}\label{10.42}
 (1+ 2\pi T\rho_\mathsf{p})v^\mathrm{eff}= \varsigma_1 + 2\pi T(\rho_\mathsf{p}v^\mathrm{eff}). 
\end{equation}
Since  $2\pi(\rho_\mathsf{p} + \rho_\mathsf{h}) = 1 + 2\pi T\rho_\mathsf{p}$ according to \eqref{10.24}, one obtains
\begin{equation}\label{10.43} 
2\pi(\rho_\mathsf{p}
+ \rho_\mathsf{h})v^\mathrm{eff} =  \varsigma_1 + T(\rho_\mathsf{p}v^\mathrm{eff})
\end{equation}
and hence
\begin{equation}\label{10.44}
2\pi\rho_\mathsf{p} v^\mathrm{eff} =\rho_\mathsf{n} \varsigma_1^\mathrm{dr},
\end{equation}
which together with \eqref{10.34} implies
\begin{equation}\label{10.45}
  v^\mathrm{eff} = \frac{\varsigma_1^\mathrm{dr}}{\varsigma_0^\mathrm{dr}}.
\end{equation}

Therefore, using \eqref{10.44} and \eqref{10.41}, 
\begin{eqnarray}\label{10.46}
 && \hspace{-50pt}  \partial_0(\rho_\mathsf{p} v^\mathrm{eff})=   \tfrac{1}{2\pi} \partial_0 (\rho_\mathsf{n}\varsigma_1^\mathrm{dr}  ) = \tfrac{1}{2\pi}\left(\varsigma_1^\mathrm{dr} \partial_0\rho_\mathsf{n}  +\rho_\mathsf{n}   \partial_0 \varsigma_1^\mathrm{dr}\right) \nonumber\\[1ex]
 &&= \tfrac{1}{2\pi}\left(\varsigma_0^\mathrm{dr} \partial_1\rho_\mathsf{n} +\rho_\mathsf{n}  \partial_1 \varsigma_0^\mathrm{dr}\right)=  \tfrac{1}{2\pi}\partial_1( \varsigma_0^\mathrm{dr}\rho_\mathsf{n})  = \partial_1\rho_\mathsf{p}.
  \end{eqnarray}
Altogether we have arrived at 
\begin{equation}\label{10.47} 
\partial_0\big(\rho_\mathsf{p}(\bar{v}-v^\mathrm{eff})\big)=0. 
\end{equation}
Hence, for all $w$, the difference $\rho_\mathsf{p}(\bar{v}(w)-v^\mathrm{eff})(w)$  does not depend on the chemical potential $\mu$. 
From the TBA equation \eqref{10.31} and \eqref{10.34} one infers that $v^\mathrm{eff} \to 0$ as $\mu \to -\infty$. For large negative $\mu$ there are no particles and the microscopic current $\rho_\mathsf{p}\bar{v}$ should also vanish. If so 
the free constant must be zero, establishing 
\begin{equation}\label{10.48}
\bar{v}=v^\mathrm{eff} 
\end{equation}
and hence
\begin{equation}\label{10.49} 
\lim_{L\to \infty} L^{-1}\langle J^{[n],L}\big \rangle_{V,L} = \langle \rho_\mathsf{p} v^\mathrm{eff} \varsigma_n \rangle.
\end{equation}

The hydrodynamic equation for the $\delta$-Bose gas has the same form as for the Toda fluid, namely
\begin{equation}\label{10.50} 
\partial_t\rho_\mathsf{p}(x,t;v) + \partial_x\big(v^\mathrm{eff}(x,t;v)\rho_\mathsf{p}(x,t;v)\big) = 0.
\end{equation}
According to our table the transformation to quasilinear form should be based on $\rho_\mathsf{n}$ and indeed
\begin{equation}\label{10.51} 
\partial_t\rho_\mathsf{n}(x,t;v) + v^\mathrm{eff}(x,t;v)\partial_x\rho_\mathsf{n}(x,t;v) = 0.
\end{equation}
\subsection{Gaudin matrix}
\label{sec10.4}
To determine the GGE averaged currents an interesting  entirely disjoint approach has been developed recently, for which the quantum mechanical structure of the Lieb-Liniger model enters
in a more fundamental way. While we cannot write down so easily explicit formulas, abstractly there 
must exist the total current operator $J^{[n],L}$ for the total charge $Q^{[n],L}$ as in Eq. \eqref{10.15}. For this operator
one conjectures the matrix elements $\langle \lambda|J^{[n],L}|\lambda\rangle$, which suffices for the computation of 
GGE expectations. The central object is the Gaudin matrix defined by 
 \begin{equation}\label{10.52} 
(G^L)_{i,j} = \delta_{i,j}\big(L + \sum_{k=1}^N2\pi T(\lambda_j, \lambda_k)\big) - 2\pi T(\lambda_i,\lambda_j), 
\end{equation} 
where $T(\lambda_j, \lambda_k)$ is the kernel of the integral operator defined in \eqref{10.25} and  $N$ is implicitly fixed through the dimension of the vector $\lambda$. Then the claim is
\begin{equation}\label{10.53} 
\langle \lambda |J^{[n],L}|\lambda \rangle  = L\sum_{i,j = 1}^N\big((G^L)^{-1}\big)_{i,j} (\lambda_i)^n\lambda_j .
\end{equation}
The prefactor $L$ arises because by stationarity the current density has to be independent of  $x$.  

Trusting in \eqref{10.53}, the GGE  averaged currents can be obtained. We fix a GGE with some confining potential $V$. 
Under this GGE the Gaudin matrix is a random matrix. Presumably its large $N,L $ limit does not exist. But we have to invert the Gaudin matrix only on very special vectors. Therefore we pick test functions $f,g$ and consider the quadratic form
\begin{equation}\label{10.54} 
L^{-2}\sum_{i,j=1}^N f(\lambda_i) (G^L)_{i,j} g(\lambda_j)  =   \int_{\mathbb{R}^2}\!\!\mathrm{d}w\mathrm{d}w' f(w)\mathsf{G}_L(w,w')
g(w') .
\end{equation}
The left hand side can be expressed as quadratic functional of the empirical density integrated against some smooth functions. Therefore, relying on \eqref{10.37}, one obtains the limit  
\begin{equation}\label{10.55} 
\lim_{L \to \infty}  \langle f,\mathsf{G}_Lg\rangle =  \langle \rho_\mathsf{p} f,\mathsf{G}g\rangle
\end{equation}
with probability one, where the limiting operator is defined through
\begin{equation}\label{10.56} 
\mathsf{G} g(w) = g(w) + 2\pi \int_{\mathbb{R}}\mathrm{d}w'  T(w,w')\rho_\mathsf{p}(w')(g(w') - g(w)).
\end{equation}
Since the limit is non-random, unless badly behaved near zero, also the inverse converges to the inverse of the limit
with probability one, i.e.
\begin{equation}\label{10.57} 
\lim_{L \to \infty}  \langle f,(\mathsf{G}_L)^{-1}g\rangle =  \langle \rho_\mathsf{p} f,\mathsf{G}^{-1}g\rangle.
\end{equation}
In our particular application $f(w) = w^n$ and $g(w) = w$.
Hence 
\begin{equation}\label{10.58} 
\lim_{L\to \infty} L^{-1}\langle J^{[n],L}\big \rangle_{V,L} = \langle \rho_\mathsf{p} \varsigma_n,\mathsf{G}^{-1}\varsigma_1\rangle.
\end{equation}
Defining $\mathsf{G}^{-1} \varsigma_1 = \bar{v}$, one obtains 
\begin{equation}\label{10.59} 
 \bar{v}(w)= w + 2\pi\big( T(\rho_\mathsf{p} \bar{v})(w) -  T\rho_\mathsf{p}(w) \bar{v}(w)\big)
\end{equation}
and 
\begin{equation}\label{10.60} 
\lim_{L\to \infty} L^{-1}\langle J^{[n],L}\big \rangle_{V,L} = \langle \rho_\mathsf{p} v^\mathrm{eff} \varsigma_n \rangle,
\end{equation}
in agreement with \eqref{10.49}.\bigskip\\
$\blackdiamond\hspace{-1pt}\blackdiamond$~\textit{Number and momentum current}.\hspace{1pt} In second quantization the particle current equals the
momentum,
\begin{equation}\label{10.61} 
Q^{[1],L}(x) = -\mathrm{i}\Psi(x)^* \partial_x\Psi(x),\quad -\mathrm{i} \int_0^L \!\!\mathrm{d}x \Psi(x)^* \partial_x\Psi(x)
= Q^{[1],L} = J^{[0],L}. 
\end{equation}
Hence in \eqref{10.53} we have to set $n =0$ with the result 
\begin{equation}\label{10.62} 
\langle \lambda |J^{[0],L}|\lambda \rangle  = L\sum_{i,j = 1}^N((G^L)^{-1})_{i,j}(\lambda_i)^0 \lambda_j  = \sum_{j=1}^N \lambda_j,
\end{equation}
which agrees with \eqref{10.61}, since $(G_L1)_i = 1$.

To determine the momentum current,  
the left hand part of \eqref{10.61} is integrated over some test function $g$. Working out the time derivative yields
\begin{eqnarray}\label{10.63} 
&&\hspace{-40pt}\mathrm{i} \big[H_{\mathrm{li},L}, -\mathrm{i}\int_0^L\!\! \mathrm{d}x g(x)\Psi(x)^* \partial_x\Psi(x)\big] \\
&&\hspace{-30pt}= \tfrac{1}{2}\int_0^L\!\!\mathrm{d}x \Big(\big(g''(x)
\Psi(x)^* + 2g'(x) \partial_x\Psi(x)^* \big)\partial_x\Psi(x) + g'(x)c\Psi(x)^*\Psi(x)^*\Psi(x)\Psi(x) \Big).\nonumber
\end{eqnarray}
By the conservation law, the total current is  obtained by setting $g' = 1$,
\begin{eqnarray}\label{10.64} 
&&\hspace{-48pt}J^{[1],L} = \int_0^L \!\!\mathrm{d}x\big(\partial_x\Psi(x)^*\partial_x\Psi(x) +\tfrac{1}{2}c\Psi(x)^*\Psi(x)^*\Psi(x)\Psi(x) \big) \nonumber\\
&&\hspace{-20pt}= 2 H_{\mathrm{li},L} -\tfrac{1}{2}c \int_0^L \mathrm{d}x\Psi(x)^*\Psi(x)^*\Psi(x)\Psi(x).
\end{eqnarray}
Taking expectations with respect to $|\lambda\rangle$ with $N=2$ on both sides yields 
\begin{eqnarray}\label{10.65} 
 &&\hspace{-43pt} \langle \lambda|J^{[1],L} |\lambda\rangle = 2\langle \lambda| H_{\mathrm{li},L} |\lambda\rangle-\tfrac{1}{2}c \int_0^L \!\!\mathrm{d}x\langle\lambda|\Psi(x)^*\Psi(x)^*\Psi(x)\Psi(x)|\lambda\rangle\nonumber\\
 &&\hspace{16pt}= \lambda_1^2 +\lambda_2^2 - \frac{1}{L + 2 T(\lambda_1,\lambda_2)}\frac{2c (\lambda_1 - \lambda_2)^2}{(\lambda_1 - \lambda_2)^2 +c^2}.
\end{eqnarray}
Since the Gaudin matrix is a $2 \times 2$ matrix,  \eqref{10.64} agrees \eqref{10.52} for $N=2,n=1$.

To proceed to $N>2$ a brute force computation no longer works, which is a generic experience in this area. Instead we recall that  
the momentum current equals the pressure $P$ and for an energy $E(L)$ one has $P = -dE/dL$.
Surprisingly  these a priori macroscopic relations extend to a single eigenstate of the Lieb-Liniger model. We start from 
\begin{equation}\label{10.66} 
E(L) = \tfrac{1}{2} \sum_{j=1}^N \lambda_j^2,\qquad - \frac{d}{dL}E(L) = \sum_{j,k=1}^N ((G_{N,L})^{-1})_{j,k}\lambda_j \lambda_k,  
\end{equation}
where the second identity is obtained by differentiating Eq. \eqref{10.12} with respect to $L$. Let us consider a volume of length $L$ and allow in the Lieb-Liniger hamiltonian a kinetic energy of strength $\kappa$,  as $H_{\mathrm{li},L,\kappa,c}
 =\kappa T_{\mathrm{kin},L }+ c T_{\mathrm{pot},L}$. By continuity, one can follow one particular vector of Bethe roots, $\lambda$, at fixed $N$ in their dependence on $L,\kappa, c$.
By spatial dilation the corresponding energy then scales as $E(L,\kappa,c) =  E(\alpha L,  \alpha^{-2} \kappa, \alpha^{-1}c)$, where 
$\alpha^{-2}$ results from the Laplacian and $\alpha^{-1}$ from the first order derivative of the boundary condition in \eqref{10.8}.
To first order in the deviation from $L$ there is an overall factor of $-L^{-1}$ and a factor $2$ for the kinetic energy and a factor 1 
for the potential energy, in agreement with \eqref{10.17}. Hence
 \begin{equation}\label{10.67} 
\frac{d}{dL} \langle \lambda| H_{\mathrm{li},L} |\lambda \rangle = - L^{-1} \langle \lambda |J^{[1],L}| \lambda \rangle.
\end{equation}
Combining \eqref{10.66} and \eqref{10.67}, we conclude that Eq. \eqref{10.53} spezialized to $n=1$ holds for arbitrary $N$.

 It is surprising, when averaging over a single energy eigenstate, 
one finds already a structural dependence which persists in the limit of large $L,N$. No such features seem to be known for the 
Toda lattice or other classical integrable many-body systems.
\hfill $\blackdiamond\hspace{-1pt}\blackdiamond$
\bigskip\\
\textbf{\large{Notes and references}}\bigskip\\
 \textbf{ad 10.0}: Transport in one-dimensional quantum lattice models, including integrable chains, is reviewed in \cite{BHK20}.
 Specifically conformal field theories are studied in \cite{BD16}. Generalized hydrodynamics for XXZ chain is studied in 
 \cite{BCDF16} and the spin-$\frac{1}{2}$ Fermi-Hubbard model in \cite{ID17}. The lecture notes \cite{D19a} provide a much more extensive list of references. \medskip\\
  \textbf{ad 10.1}: Elliott H. Lieb, together with Werner Liniger,  discovered the Bethe ansatz for the $\delta$-Bose gas and analysed ground state properties \cite{LL63a,LL63b}. For the repulsive case the completeness of Bethe wave functions has been proved by T. Dorlas \cite{D93}. The attractive 
  $\delta$-Bose gas is more complicated because bound states. An exhaustive discussion  of the structure of eigenfunctions can be found in \cite{D10}.  In this case the completeness is established in the thesis 
  \cite{O79}, see also \cite{PS11}. The local structure of higher charges is studied in \cite{DK11}. For computations more powerful is often the algebraic Bethe ansatz. An example is the norm of the eigenfunctions as explained in \cite{PC15}.
  
In Section \ref{sec2.3} we defined the scattering shift assuming that the incoming momenta are ordered as 
$p_1 <...<p_N$ and the same convention is adopted for the phase shift of $\delta$-Bose gas. One could have chosen also the reverse order, $p_N <...<p_1$, in which case the phase shift would be the negative of the one 
in \eqref{2.39}. Such ordering is used in \cite{YY69}. Both conventions can be found in the literature. Of course, the resulting TBA equations 
are identical. \medskip\\
  \textbf{ad 10.2}: The variational type solution for the thermodynamics has been obtained by C.N. Yang and C.P. Yang \cite{YY69}. 
  T. Dorlas, J. Lewis and Pul\'{e} \cite{DLP93} established a proof using methods from large deviations. The Bethe ansatz is very well covered in the literature \cite{T84,F96,S04,G14}. At the time the conventional thermodynamics was in focus. The GGE is discussed in \cite{MC12}. It would be interesting to find out whether and how the methods in \cite{DLP93} extends to a more general class of confining potentials. \medskip\\
 \textbf{ad 10.3}: The GHD of the $\delta$-Bose gas is discussed in \cite{D19a}. For the collision rate ansatz we follow \cite{YS20}. An experimental realization consists of a narrow tube to which  $^{87}$Rb atoms are magnetically confined, approximately 4000 in number
   \cite{SBDD19}. An expanding cloud is observed. For a single initial hump conventional hydrodynamics works fine. But for an initial double hump the full GHD is required in order to fit the observed density profiles. In fact, during the time span of the experiment 
  roughly 25\% of the atoms are lost and GHD has to be  adjusted correspondingly \cite{BDD20}.\medskip\\
  \textbf{ad 10.4}: For the discussion of the Gaudin matrix and its relation to the currents we follow \cite{BPP19}. Other methods rely on long range deformations \cite{P20} and properties of the boost operator \cite{YS20}.

 \section{Quantum Toda lattice}
\label{sec11}
\setcounter{equation}{0} 
The classical Toda hamiltonian can be quantized according to standard rules. Thereby one obtains a system of $N$ particles with
 Hilbert space $\mathcal{H}_N =  
L^2(\mathbb{R})^{\otimes N} = L^2(\mathbb{R}^{N})$. The particles have position $x_j$ and momenta $p_j$, $ j = 1,...,N$, satisfying the commutation relations 
$[p_i,x_j] = -\mathrm{i}\hbar\delta_{i,j}$. In position space representation, complex-valued wave functions  of the form  $\psi(x_1,...,x_N)$,
position of the $j$-th particle is multiplication by $x_j$ and momentum $p_j = -\mathrm{i}\hbar\partial_{x_j}$. The conventional notation for the interaction term 
is 
\begin{equation}\label{11.1} 
X_j  = \mathrm{e}^{x_j  - x_{j+1}}, \quad j = 1,...,N-1,\qquad X_N = \mathrm{e}^{x_N - x_1},
\end{equation}
the latter condition imposing periodic boundary conditions. Then the hamiltonian of the quantum Toda lattice reads
\begin{equation}\label{11.2} 
H_{\mathrm{qt},N} = \sum_{j=1}^N \big(\tfrac{1}{2} p_j^2 +  X_j\big). 
\end{equation}
Planck's constant $\hbar$ regulates the relative strength of kinetic and potential energy. In the same spirit, for the classical Toda chain we could have introduced a mass parameter, which however can be absorbed through an appropriate  rescaling of  spacetime. In this sense, the classical chain has no free model parameter. However, quantum mechanically $\hbar$ cannot not be scaled and must be maintained as relevant parameter. The semi-classical limit corresponds to $\hbar \to 0$. Another common choice is to explicitly introduce an interaction strength, $\eta$,
and the inverse decay length of the potential through $\mathrm{e}^{-\gamma x}$. Then the standard form \eqref{11.2} is recovered upon replacing $\hbar^2$ by $\hbar^2\gamma^2/\eta$.

For the harmonic lattice, i.e. $X_j = (x_j - x_{j+1})^2$, one usually introduces creation and annihilation operators, satisfying $[a_i,a_j^*] = \delta_{i,j}$. The hamiltonian  is then quadratic in $\{a_i,a_j^*, i,j = 1,...,N\}$ and describes bosonic excitations of the ground state. Formally such a transformation can be
implemented also for the Toda lattice, leading to a nonlinear interaction between bosons. For our purposes this representation does not seem to be so useful.

Our strategy is to follow the trail laid out by the classical model. With some confidence, we will arrive at the appropriate hydrodynamic Euler equations. 
As might  have been anticipated there is a strong similarity with the Lieb-Liniger model.
\subsection{Integrability, monodromy matrix}
\label{sec11.1}
While there are examples for which integrability is maintained under quantization, no guarantee can be issued. Thus our first task is to establish $N$ local conservation laws.  The elegant approach relies on the monodromy matrix, which is familiar from other integrable many-body systems.  Let us start with the classical chain and consider
\begin{equation}\label{11.3}
\det(\lambda - L_N) = \lambda^N\Big(\sum_{m=0}^{N-1}  (-\lambda)^{-m} I_m  + (-1)^N(\lambda)^{-N} (I_N- 2)\Big),
\end{equation}
$I_0 = 1$. The coefficients $I_m$ are the \textit{H\'{e}non invariants} which are in involution, i.e. $\{I_m,I_n\} = 0$. Since the hamiltonian 
is proportional to $I_1^2 - I_2$, the  H\'{e}non invariants are in fact conserved. Their explicit form is given by
\begin{equation}\label{11.4} 
I_m = \sum_{\{k+2\ell = m\}} p_{i_1}p_{i_2}\cdots p_{i_k}(-X_{j_1})\cdots (-X_{j_\ell}),
\end{equation}
where the sum is over all strings of indices $i_1,i_2,...,i_k, j_1,j_1+1,..., j_\ell ,j_\ell +1$. The  indices are distinct and satisfy  
the constraint $ m = k +2\ell$, $k\geq 0,l\geq0$. Two summands differing in the order of indices are counted only once.
More visually, one considers a ring of  $N$ sites and has available $k$ singletons and $\ell$ nearest neighbor pairs, the dominos. The ring is partially covered by singletons and dominos with no overlap allowed, which then defines the string  of indices in \eqref{11.4}. As an example, for $N=4$ one obtains
\begin{eqnarray}\label{11.5} 
&&\hspace{-20pt}I_1 = p_1+p_2+p_3 +p_4, \,\,
I_2 = p_1 p_2 + p_1p_3 + p_1p_4 + p_2p_3 + p_2p_4+ p_3p_4 - X_1 - X_2 -X_3 - X_4,\nonumber\\
&&\hspace{-20pt}I_3 = p_1p_2p_3 + p_1p_2p_4 + p_1p_3 p_4 + p_2 p_3 p_4\nonumber\\
&&\hspace{24pt} - p_1X_2 -  p_1 X_3 - p_2X_3  - p_2X_4 - p_3X_4 - p_3X_1  - p_4X_1  - p_4X_2,\nonumber\\
&&\hspace{-20pt}I_4 = p_1p_2p_3p_4 - p_1p_2X_3 - p_2p_3X_4 - p_3p_4X_1 -  p_4p_1X_2  + X_1X_3  +X_2X_4. 
\end{eqnarray}
Except for $m=1$, the $I_m$'s are not local and have no meaningful density in the limit $N\to \infty$. 
In the quantum setting, $[p_j, X_m] = 0$, except for $j = m,m+1$. Such terms do not appear
in the sum \eqref{11.4} and in the quantized version of the $I_m$'s no ambiguity in the operator ordering arises. 
This property strongly suggests that they will
continue to be conserved quantities quantum mechanically.
To convert the H\'{e}non invariants into a local form we use the identity
\begin{equation}\label{11.6} 
\log \det(1 - \lambda^{-1}L_N) = - \sum_{m=1}^\infty\frac{1}{m} \lambda^{-m}\mathrm{tr}[(L_N)^m] =  - \sum_{m=1}^\infty\frac{1}{m} \lambda^{-m}Q^{[m], N}.
\end{equation}
From Section \ref{sec2} we know already that  the $Q^{[m],N}$'s have a local density.

There is a different route to arrive at the same result. We first define the $2\times 2$ matrix
\begin{equation}\label{11.7}
L_j(\lambda) = 
\begin{pmatrix}
\lambda - p_j & \mathrm{e}^{q_j}\\
-  \mathrm{e}^{-q_j}& 0
 \end{pmatrix}
\end{equation}
and, as product, the monodromy matrix
\begin{equation}\label{11.8} 
T(\lambda) = L_1(\lambda) \cdots L_N(\lambda).
\end{equation}
Then
\begin{equation}\label{11.9} 
\mathrm{tr}[T(\lambda)] = \det(\lambda - L_N) + 2(-1)^N
\end{equation}
and this time expanding $\log  \mathrm{tr}[T(\lambda)]$ in a power series we obtain the locally conserved fields.

Switching to the \textit{quantum} Toda chain, there seems to be no analogue of the Lax pair equation \eqref{2.11}.
However, as first observed by E.K. Sklyanin in 1985, the definition of monodromy matrix stays intact. As common usage, we switch from $\lambda$ to $u$ as spectral parameter. The entries of the matrix $L_j(u)$ become now operators. 
As a consequence the monodromy matrix $T(u)$ is a $2\times2$ matrix with operator entries.  We introduce the $R$-matrix acting on $\mathbb{C}^2\otimes\mathbb{C}^2$ by 
\begin{equation}\label{11.10} 
R(u) = u1\hspace{-3pt}\mathrm{l}\otimes 1\hspace{-3pt}\mathrm{l} -\mathrm{i}\hbar \mathsfit{P},
\end{equation}
where $\mathsfit{P}$ permutes the indices 1 and 2, i.e. $\mathsfit{P} \varphi_1\otimes \varphi_2 = \varphi_2\otimes \varphi_1$.
Noting the commutation relations
\begin{equation}\label{11.11} 
[p_j,X_j] = -\mathrm{i} \hbar X_j, \quad [p_{j+1},X_j] = \mathrm{i} \hbar X_{j},
\end{equation}
one confirms that 
\begin{equation}\label{11.12}
R(u-v) (L_j(u)\otimes 1\hspace{-3pt}\mathrm{l})( 1\hspace{-3pt}\mathrm{l} \otimes L_j (v)) =  (1\hspace{-3pt}\mathrm{l} \otimes L_j (v))( L_j(u)\otimes 1\hspace{-3pt}\mathrm{l}) R(u-v)
\end{equation}
and therefore also for the product
\begin{equation}\label{11.13}
R(u-v) (T(u)\otimes 1\hspace{-3pt}\mathrm{l})( 1\hspace{-3pt}\mathrm{l} \otimes T(v)) =  (1\hspace{-3pt}\mathrm{l} \otimes T(v))( T(u)\otimes 1\hspace{-3pt}\mathrm{l}) R(u-v).
\end{equation}
Setting $\hat{t}(u) = \mathrm{tr}[T(u)]$, the trace of a  $2\times 2$-matrix, one obtains the commutation relation 
 \begin{equation}\label{11.14} 
 \hat{t}(u) \hat{t}(v) = \hat{t}(v)\hat{t}(u)
\end{equation}
valid for arbitrary $u,v$. $\hat{t}(u)$ is a polynomial of degree $N$, with operator-valued coefficients, operators on $\mathcal{H}_N$. On abstract grounds, the relation \eqref{11.14} ensures that the Taylor coefficients of $\hat{t}(u)$ commute with each other. 

In fact, the  family of operators can be worked out more concretely. Let us define 
\begin{equation}\label{11.15} 
F_N(u) =  \sum_{\{k+2l = N\}} (u-p_{i_1})(u-p_{i_2})\cdots (u- p_{i_k})(-X_{j_1})\cdots (-X_{j_l}),
\end{equation}
where the string of indices satisfies the conditions below \eqref{11.4}. We split as
\begin{equation}\label{11.16} 
F_N(u) =  F_N^{\diamond}(u) + F_N^{\diamond\diamond}(u),
\end{equation}
where $F_N^{\diamond}(u)$ collects all summands containing the factor $\mathrm{e}^{x_N}$.
Then, by an induction argument, 
\begin{equation}\label{11.17} 
T(u) = T_N(u) =
\begin{pmatrix}
 F_N^{\diamond}(u) & \mathrm{e}^{x_N} F_{N-1}^{\diamond}(u)\\
\mathrm{e}^{-x_{N+1}}F_{N+1}^{\diamond\diamond}(u)& F_N^{\diamond\diamond}(u)
 \end{pmatrix}.
\end{equation}
In particular
\begin{equation}\label{11.18} 
 \hat{t}(u) = F_N(u) = u^N \sum_{m =0}^N (-1)^m u^{-m} \hat{I}_m,
\end{equation}
where $\hat{I}_m$ is the quantization of $I_m$. Since $\{\hat{I}_1,...,\hat{I}_N\}$ is  a  family of commuting operators, one can still expand  $\log(u^{-N}\hat{t}(u))$ at $u = \infty$ as 
\begin{equation}\label{11.19} 
-\log\big(u^{-N} \hat{t}(u) \big) = u^{-1} \hat{Q}^{[1],N} + \tfrac{1}{2}u^{-2} \hat{Q}^{[2],N}... + \tfrac{1}{N}u^{-N} \hat{Q}^{[N],N} + \mathcal{O}\big(u^{-(N+1)}\big).
\end{equation}
By construction, $\hat{Q}^{[m],N}$ is a polynomial in $\hat{I}_1,...,\hat{I}_m$ of maximal order $m$, which coincides with the corresponding 
classical identity.  Thus also $\{ \hat{Q}^{[1],N},...,\hat{Q}^{[N],N}\}$ is a  family of commuting operators, also commuting with the
$\hat{I}_1,...,\hat{I}_N$. For the hamiltonian
\begin{equation}\label{11.20} 
H_{\mathrm{qt},N} =  \tfrac{1}{2} \hat{Q}^{[2],N} = \tfrac{1}{2}\hat{I}_1^2 - \hat{I}_2,
 \end{equation}
 which implies that  the $\hat{I}_m$'s and $\hat{Q}^{[m],N}$'s are conserved.  
 For the particular case $N=4$ the result is 
\begin{eqnarray}\label{11.21} 
&&\hspace{-30pt}\hat{Q}^{[1],4} =  \hat{I}_1,\quad  \hat{Q}^{[2],4} = \hat{I}_1^2 - 2  \hat{I}_2, \quad  \hat{Q}^{[3],4} =   \hat{I}_1^3 - 3  \hat{I}_1\hat{I}_2 + 3  \hat{I}_3,\nonumber\\
&&\hspace{-30pt}\hat{Q}^{[4],4} =   \hat{I}_1^4 - 4 \hat{I}_1^2\hat{I}_2 + 2 \hat{I}_2^2 + 4  \hat{I}_1 \hat{I}_3 - 4 \hat{I}_4.
\end{eqnarray}

The nonlocal H\'{e}non invariants, $ \hat{I}_m$, are converted to the Flaschka invariants, $\hat{Q}^{[m],N}$. For the classical chain,   because of \eqref{11.6}, the Flaschka invariants are 
indeed local. The claim is that the cancellation mechanism still holds for the quantum chain, where  $\hat{Q}^{[m],N}$ are operators whose local form has yet to be determined. In our standard example, $N=4$, one obtains
\begin{eqnarray}\label{11.22} 
&&\hspace{-30pt}\hat{Q}^{[1],4} = \sum_{j=1}^4 p_j,  \quad \hat{Q}^{[2],4} = \sum_{j=1}^4\big( p_j^2 + 2X_j\big), \quad
\hat{Q}^{[3],4}=  \sum_{j=1}^4\big( p_j^3 + 3(p_j+p_{j+1})X_j\big),\\
&&\hspace{-30pt}\hat{Q}^{[4],4} =  \sum_{j=1}^4\big( p_j^4+ 2(p_j^2 +p_jp_{j+1} +p_{j+1}^2)X_j + 2X_j (p_j^2 +p_jp_{j+1} +p_{j+1}^2) + 2 X_j^2 +4X_jX_{j+1}\big),
\nonumber
\end{eqnarray}
which are local, as anticipated. In fact we have obtained the minimal version of the densities, compare with the discussion above \eqref{2.24}.
Since $[p_j + p_{j+1},X_j] =0$, see \eqref{11.11}, the only non-obvious part is the particular operator ordering for $\hat{Q}^{[4],4}$. 

Actually, the precise functional form of the higher charges is fairly irrelevant for us. As crucial information, the density of the $n$-th charge apparently depends on local operators in a block of size at most $(n/2) +1$.
Since the hamiltonian is nearest neighbor this property implies that the $n$-th current operator density depends on a block of size at most $(n/2) +2$.
\subsection{Spectral properties}
\label{sec11.2}
Since the interaction depends only on relative positions, an eigenfunction of  $H_{\mathrm{qt},N}$ must be of the form
\begin{equation}\label{11.23} 
\psi_{E}(x_1,...,x_N) =\tilde{\psi}_{E_\perp}(x_2-x_1,..., x_N - x_{N-1}) \mathrm{e}^{\mathrm{i}\hbar^{-1}E_1N^{-1}(x_1 +...+x_N)},
\end{equation}
$E = (E_1,...,E_N) = (E_1,E_\perp)$, implying $\hat{I}_1 \psi_{E} = E_1 \psi_{E}$. More abstractly one can think of a fiber decomposition of $H_{\mathrm{qt},N}$ with respect to the total momentum. $E_1$ would be then the corresponding fiber parameter. In terms of the hamiltonian one explicitly splits into kinetic energy of the  center of mass plus relative internal motion,
\begin{equation}\label{11.24} 
H_{\mathrm{qt},N} = H_\mathrm{cm} + H_\mathrm{rel}.
\end{equation}
Thus $\tilde{\psi}_{E_\perp} \in L^2(\mathbb{R})^{\otimes {N-1}}$ is an eigenfunction of $H_\mathrm{rel}$.
In fact, working out $H_\mathrm{rel}$, one finds an interaction potential increasing exponentially in all directions. The eigenvalues of $H_\mathrm{rel}$ are isolated,
even nondegenerate, and the thermal operator $\exp(- H_\mathrm{rel})$ is trace class. 

It is of great advantage to consider simultaneously all eigenvalues of the H\'{e}non invariants, to say
\begin{equation}\label{11.25} 
\hat{I}_m \psi_{E} = E_m \psi_{E}, 
\end{equation}
$m = 1,...,N$,
with $E_1$ as in \eqref{11.23}. There is an explicit formula for the joint eigenfunctions and it is known that they are complete in the sense that linear combinations of the set $\{\tilde{\psi}_{E_\perp} \}$ span the entire
Hilbert space $L^2(\mathbb{R}^{N-1})$.  Then $\psi_{E}$ is also an eigenfunction of the transfer matrix $\hat{t}(u)$, 
\begin{equation}\label{11.26} 
\hat{t}(u)\psi_{E} = \tau(u,E)\psi_{E},
\end{equation}
where 
\begin{equation}\label{11.27} 
\tau(u,E) = u^N \Big(1 +  \sum_{m =1}^N (-1)^m u^{-m} E_m\Big)
\end{equation}
The $m$-th Flaschka invariant has eigenvalue $q_m$, 
\begin{equation}\label{11.28} 
\hat{Q}^{[m],N}\psi_{E}  = q_m\psi_{E}. 
\end{equation}
Since the operators commute, the $q_m$'s are determined by 
\begin{equation}\label{11.29} 
-\log\big(u^{-N} \tau(u,E) \big) = u^{-1} q_1 + \tfrac{1}{2}u^{-2} q_2 + ... + \tfrac{1}{N}u^{-N} q_N+ \mathcal{O}\big(u^{-(N+1)}\big).
\end{equation}
We now write
\begin{equation}\label{11.30} 
\tau(u) = \prod_{j=1}^N(u- \tau_j).
\end{equation}
Then 
\begin{equation}\label{11.31} 
-\log\big(u^{-N} \tau(u)\big) = -\sum_{j=1}^N \log\big(1- u^{-1}\tau_j\big) = \sum_{m=1}^N u^{-m} \sum_{j=1}^N\tfrac{1}{m} (\tau_j)^m + \mathcal{O}\big(u^{-(N+1)}\big)
\end{equation}
and
\begin{equation}\label{11.32} 
q_m = \sum_{j=1}^N (\tau_j)^m,
\end{equation}
$ m = 1,...,N$.
As already familiar, thereby one introduces an empirical density through
\begin{equation}\label{11.33} 
\rho_\mathrm{qt}(w)  = \frac{1}{N} \sum_{j=1}^N \delta(\tau_j - w)
\end{equation}
with the property that 
\begin{equation}\label{11.34} 
N^{-1} q_m = \int_\mathbb{R} \mathrm{d}w \rho_\mathrm{qt}(w) w^m.
\end{equation}

Before continuing, we have to recall the quantum mechanical analog of the scattering shift studied in Section \ref{sec2.3}. The relative motion of two Toda particles is governed by
the Schr\"{o}dinger equation
\begin{equation}\label{11.35} 
\mathrm{i}\partial_t \psi_t(x) = \big(-\hbar^2\partial_x^2  + \mathrm{e}^{-x}\big)\psi_t(x).
\end{equation}
The particle representing the relative motion travels inwards from the far right with momentum $\hbar k_\mathrm{in}$, gets reflected at the potential barrier, and moves outwards with momentum
$\hbar k_\mathrm{out} = - \hbar k_\mathrm{in}$.  The phase shift accumulated during the scattering process is given by 
\begin{equation}\label{11.36} 
\theta(k) = k \log \hbar^2 -\mathrm{i} \log \frac{\Gamma(1 +\mathrm{i}k)}{\Gamma(1 -\mathrm{i}k)}.
\end{equation}
Surprisingly, the phase shift is independent of $\hbar$, except for the additive constant $k \log \hbar^2$.
The scattering shift is then
\begin{equation}\label{11.37} 
\theta'(k) =  \log \hbar^2 + \psi_\mathrm{di}(1 +\mathrm{i}k) +\psi_\mathrm{di}(1 -\mathrm{i}k) 
\end{equation}
with the Digamma function $\psi_\mathrm{di} = \Gamma'/\Gamma$. From the convergent power series,
\begin{equation}\label{11.38} 
\psi_\mathrm{di}(1 +\mathrm{i}k) +\psi_\mathrm{di}(1 -\mathrm{i}k) = -\gamma_\mathrm{E} + \sum_{n = 1}^\infty \frac{k^2}{n(n^2 +k^2)},
\end{equation}
$\gamma_\mathrm{E} = 0.577...$ the Euler-Mascheroni number, one concludes that $\theta'$ has a positive curvature at $k=0$. The large $k$ behavior can be inferred from the expansion
\begin{equation}\label{11.39} 
\psi_\mathrm{di}(1 +\mathrm{i}k) +\psi_\mathrm{di}(1 -\mathrm{i}k) = \log(1+k^2) - \frac{1}{1+k^2} +     \mathcal{O}(k^{-4}).     
\end{equation}
For the semiclassical limit we set $\hbar k = w$ and
\begin{equation}\label{11.40} 
\lim_{\hbar \to 0} \theta'(\hbar^{-1}w )   = \log w^2,  
\end{equation}
as it should be.

In 2010 K.K. Kozlowski and J. Teschner obtained a remarkable result on the spectral properties of the chain, which comes handy now.
They fix the eigenvalue $E_1$ and establish that for each vector $(\tau_1,...,\tau_N)$ there is
another vector $(\delta_1,...,\delta_N)$ such that 
\begin{equation}\label{11.41} 
 \sum_{j=1}^N (\tau_j)^m =  \sum_{j=1}^N (\hbar \delta_j)^m  + \mathcal{O}(\mathrm{e}^{-N})
\end{equation}
for $m = 1,...,N$. In addition, for some $\{n_j \in \mathbb{Z}, j = 1,...,N\}$, the roots  $(\delta_1,...,\delta_N)$ satisfy the identity
\begin{equation}\label{11.42} 
2 \pi n_j = N\delta_j\log \hbar^2 +\mathrm{i} \log \zeta - \sum_{i=1}^N   \mathrm{i} \log \frac{\Gamma(1 +\mathrm{i}(\delta_j - \delta_i))}{\Gamma(1 -\mathrm{i}(\delta_j - \delta_i))}
+ \mathcal{O}(\mathrm{e}^{-N}).
\end{equation}
This expression has already the same flavor of as \eqref{10.12}  for the $\delta$-Bose gas, except for the model dependent phase shift, of course.
For the $\delta$-Bose gas \eqref{11.41} and \eqref {11.42} are a strict identities, no error term, whereas for the quantum Toda lattice,
they hold approximately for large $N$. 

Actually, for both error terms there is an explicit identity. So far, the exponential bound $\mathcal{O}(\mathrm{e}^{-N})$  has been established only for particular cases.  
\subsection{GGE and hydrodynamics}
\label{sec11.3}
Given the input from \eqref{11.42} and assuming $n_1 <...<n_N$, with small modifications one can follow the Yang-Yang scheme from the $\delta$-Bose gas.
By definition the parameter $\zeta$ has unit length, $|\zeta| = 1$, and hence drops out as $N \to \infty$.
The empirical density \eqref{11.33} satisfies the constraints
\begin{equation}\label{11.43} 
\rho_\mathrm{qt}(w) \geq 0,\quad \int_\mathbb{R}\mathrm{d}w\rho_\mathrm{qt}(w) =1, \quad \int_\mathbb{R}\mathrm{d}w\rho_\mathrm{qt}(w)w = N^{-1} E_1 =e_1.
\end{equation}
Omitting both correction terms, Eq. \eqref{11.42} becomes
\begin{equation}\label{11.44} 
\frac{1}{N} 2\pi n_j =  \frac{1}{N}\sum_{i=1}^N \theta(\delta_j - \delta_i) +e_1.
\end{equation}

The GGE is defined through the weight
\begin{equation}\label{11.45} 
\exp\big(- \sum_{j=1}^N\tilde{V} (\delta_j)\big), \quad \tilde{V}(w) = \sum_{n=2}^\infty\mu_n w^n.
\end{equation}
Following the steps in the analysis of the $\delta$-Bose gas, the free energy functional for quantum Toda reads
\begin{equation}\label{11.46}
\mathcal{F}_\mathrm{qt}(\rho) = \int_\mathbb{R}\mathrm{d}w \big( \rho \tilde{V} +\rho \log \rho +  \rho_\mathsf{h} \log \rho_\mathsf{h} - 
(\rho +  \rho_\mathsf{h})\log (\rho +  \rho_\mathsf{h})\big).
\end{equation}
This functional has to be minimized under the constraints
 \begin{equation}\label{11.47}
 \rho \geq 0,\quad \int_\mathbb{R}\mathrm{d}w\rho(w) =1,\quad \int_\mathbb{R}\mathrm{d}w\rho(w)w = e_1.
\end{equation}
As before, defining the operator $T$ through the convolution with the Toda scattering shift, 
\begin{equation}\label{11.48} 
Tf(w) =  \frac{1}{2\pi}  \int_\mathbb{R}\mathrm{d}w' \theta'(w - w')f(w'),
\end{equation}
the additional constraint \eqref{11.44} can be written as
\begin{equation}\label{11.49} 
\rho + \rho_\mathsf{h} =  T\rho. 
\end{equation} 
In contrast to the $\delta$-Bose gas,  entropy, free energy, and total momentum  are now per lattice site.
The constraint on the first moment can be absorbed into a chemical potential $\mu_1$, turning the confining potential 
$\tilde{V}$ into $V$. 
Denoting the minimizer of \eqref{11.46} by $\rho_\mathsf{p}$, the GGE average of the conserved charges in the limit $N \to \infty$
is given by 
\begin{equation}\label{11.50} 
\langle Q^{[n]}_0\rangle_{V} = \langle \rho_\mathsf{p} \varsigma_n \rangle. 
\end{equation}

The variational problem appears to be identical to the one of the $\delta$-Bose gas with volume $L=0$. This should come as no surprise,
since we used periodic boundary conditions. In this case the stretch vanishes by construction. As discussed already,  in the classical Toda lattice 
for this particular choice, the positions have the statistics of a random walk with step size of order $1$ and exactly zero drift, hence a fluctuating volume of order $\sqrt{N}$.

For the classical chain, on top of the Flaschka invariants based on the Lax matrix, also the stretch, resp. the particle density in the fluid picture, is conserved. But so far we have not identified a corresponding control parameter in the quantum model. 
Still, physically the stretch must be included in a proper hydrodynamic description. In the classical model, a stretch is  forced through the tilt boundary condition $q_{N+1} = q_1 + \nu N$ with $\nu \in \mathbb{R}$, compare with \eqref{3.1}.  No obvious  quantum analogue
seems to be readily available.  
Some time ago, E.K. Sklyanin worked out a concrete proposal, which somehow seems to have been completely overlooked
in the literature. Let me first explain his idea for the classical Toda lattice in thermal equilibrium, setting the inverse temperature to $1$.  Then of relevance is only the potential part, $\sum_{j=1}^N X_j$, which is invariant under the global translations, $x_j \rightsquigarrow x_j + a$ for all $j$. Since we are only interested 
in the relative coordinates $r_j = x_{j+1} - x_j$,  these translations can be subtracted by setting $x_1 = 0$. The Boltzmann weight 
for the relative coordinates is then given by
\begin{equation}\label{11.51}
 \prod_{j=1}^{N}  \mathrm{d}r_j \exp( - \mathrm{e}^{-r_j}) \delta(r_1+...+r_N).
\end{equation} 
Now let us change the strength of $X_N$, thereby resulting in the modified potential  energy
\begin{equation}\label{11.52} 
\sum_{j=1}^{N-1} X_j + \mathrm{e}^{-\nu N}X_N.
\end{equation} 
Following the same computation as before, we end up with the Boltzmann weight 
\begin{equation}\label{11.53} 
 \prod_{j=1}^{N}  \mathrm{d}r_j \exp( - \mathrm{e}^{-r_j}) \delta(r_1+...+r_N - \nu N),
\end{equation} 
which is precisely of the desired form.

The idea of Sklyanin is to copy \eqref{11.52} for the quantum model and thus to consider the modified hamiltonian
\begin{equation}\label{11.54} 
H_{{\nu},N} = -\sum_{j=1}^{N} \tfrac{1}{2} \hbar^2 \partial_{x_j}^2 +   \sum_{j=1}^{N-1} X_j + \mathrm{e}^{-\nu N}   X_N. 
\end{equation}
Note that $[H_{\mathrm{qt},N}, H_{{\nu},N}] \neq 0$ for $\nu \neq 0$. Hence the thermal state $\exp(- H_{{\nu},N})$
is not time-invariant, in sharp contrast to pressure ensemble of the classical Toda lattice. 
Without further additional  precautions $\exp(- H_{{\nu},N})$ cannot serve as a candidate for local equilibrium.
But here $H_{{\nu},N}$ is introduced only for the purpose of computing the GGE free energy. In fact, introducing the unitary 
\begin{equation}\label{11.55} 
U\psi(x_1,...,x_N) = \psi(x_1-\nu,...,x_N -N\nu),
\end{equation} 
one obtains
\begin{equation}\label{11.56} 
U^*p_jU = p_j, \qquad U^*X_jU = \mathrm{e}^{-\nu}X_j,\qquad U^*X_NU = \mathrm{e}^{\nu (N-1)}X_N,
\end{equation} 
and 
\begin{equation}\label{11.57}
U^*H_{\nu,N}U = H_{\mathrm{hom},\nu,N} = -\sum_{j=1}^{N} \tfrac{1}{2} \hbar^2 \partial_{x_j}^2 +  \sum_{j=1}^{N} \mathrm{e}^{-\nu}  X_j.
\end{equation}
The boundary coupling can be spread homogeneously over the entire lattice. 
Denoting by $\langle \cdot \rangle_{\nu,N}$
the average with respect to $\exp(- H_{\nu,N})$ and using unitary equivalence, one concludes
\begin{equation}\label{11.58} 
\langle (x_{j+1} - x_j) \rangle_{N,\nu} = \nu, \quad j = 1,...,N-1, \quad \langle (x_{1} - x_N) \rangle_{N,\nu} = -(N-1)\nu.
\end{equation}
At the expense of a single huge jump one indeed has induced a non-zero, constant stretch.

While convincing, one has to monitor how the higher invariants are affected by such a modification. 
We set 
\begin{equation}\label{11.59}
M_\nu = 
\begin{pmatrix}
1 & 0\\
0 & \mathrm{e}^{-\nu N}
 \end{pmatrix}
\end{equation}
Then the modified monodromy matrix reads 
\begin{equation}\label{11.60} 
T_\nu = M_\nu T_N(u)
\end{equation} 
and
\begin{equation}\label{11.61} 
\hat{t}_\nu (u) = F_N^{\diamond}(u) +  \mathrm{e}^{-\nu N}
F_N^{\diamond\diamond}(u).
\end{equation} 
Thus the H\'{e}non invariants of the modified hamiltonian are still given by \eqref{11.4}, compare with \eqref{11.18}. Only, everywhere
$X_N$ is substituted by  $\mathrm{e}^{-\nu N}X_N$.
The same rule transcribes to the Flaschka invariants. Denoting the Flaschka invariants of   $H_{\nu,N}$ by $\hat{Q}^{[m],N}_\nu$ and those of  
$H_{\mathrm{hom}, \nu,N}$ by $\hat{Q}^{[m],N}_{\mathrm{hom},\nu}$, one arrives at 
\begin{equation}\label{11.62} 
U^*\hat{Q}^{[m],N}_\nu U = \hat{Q}^{[m],N}_{\mathrm{hom},\nu}.
\end{equation} 

This equation teaches us that also under a generalized Gibbs ensemble for $H_{\nu,N}$, the average stretch equals $\nu$, except for the jump of $-\nu(N-1)$ at the bond $(N,1)$. Secondly, the generalized free energy can be computed from the Flaschka invariants of the lattice with coupling strength $\mathrm{e}^{-\nu}$. By scaling this means that in \eqref{11.36} $\log \hbar^2$ is replaced by 
$\log (\hbar^2 \mathrm{e}^{\nu})$. Thus, combining the terms as leading to \eqref{11.44}, the stretch modifies this relation to 
\begin{equation}\label{11.63} 
 2\pi n_j =  \nu N\delta_j  + \sum_{i=1}^N \theta(\delta_j - \delta_i) +N e_1.
\end{equation}
Up to the additional term coming from the fixed total momentum, we have obtained perfect analogy to the $\delta$-Bose gas,
see Eq. \eqref{10.12}.
However, in contrast to the $\delta$-Bose gas, the stretch parameter $\nu N$ can be negative, just as for the classical chain.

In our argument for the average currents of the $\delta$-Bose gas we never used the specific form of the scattering shift. For the quantum Toda lattice
the density is conserved and its current is the momentum, which itself is conserved. No extra work is required. To conclude, in the fluid picture the hydrodynamic equations of the quantum Toda lattice read 
\begin{equation}
\label{11.64} 
\partial_t\rho_\mathsf{p}(x,t;v) + \partial_x\big(v^\mathrm{eff}(x,t;v)\rho_\mathsf{p}(x,t;v)\big) = 0.
\end{equation}
The effective velocity is determined as in \eqref{10.39}, resp. \eqref{10.45}. Of course the appropriate $T$ operator is the one of \eqref{11.48}.  \bigskip\\
\textbf{\large{Notes and references}}\bigskip\\
 \textbf{ad 11.0}: B. Sutherland \cite{S78} first discussed the integrability of the quantum Toda lattice, using a low density
 approximation of the integrable quantum fluid with the $1/\sinh^2$ interaction potential. He developed the method of asymptotic Bethe ansatz, which uses properties of the eigenfunctions when particles are far apart \cite{S95}. Finite volume has been introduced ad hoc 
 by imposing periodic boundary conditions on the wave numbers. Sutherland studied TBA for the ground state. In a series of papers Mertens and his group  \cite{TM83,M84,O85,HM86,GM88}, see also \cite{T84a,TI86}, investigated TBA at non-zero temperatures and also the semi-classical classical limit. At the time, only thermal equilibrium was in focus. A most useful overview is  \cite{SS97}, see also \cite{SY10}.
 An early summary on the more mathematical activities in the 
 quantum integrable camp is the review \cite{OP83}. Highly recommended is the instructive  book by Sutherland \cite{S04}. 
    
 Gutzwiller \cite{G80} studied the spectrum for few, $N=3,4$, particles, later being extended to an arbitrary number  \cite{GP92}. 
 E.K. Sklyanin \cite{S85}
 achieved a crucial advance by linking the Toda lattice to the flourishing field of quantum integrable systems, with notions as Yang-Baxter equations and monodromy matrix. He developed the method of separation of variables, introducing the Baxter equation  in this context. 
 He also discussed the spectrum of eigenvalues in the limit of large $N$ and supported the more heuristic results in \cite{S78}.  
 Following a proposal by Nekrasov and Shatashvili \cite{SN10}, Kozlowski and Teschner \cite{KT10}  substantially advanced such methods. \medskip\\
 \textbf{ad 11.1}: Our discussion is based on \cite{S85} and \cite{SS97}. Identities as \eqref{11.6} are proved in \cite{RS79},
Section 17.\medskip\\
 \textbf{ad 11.2}:  A fairly explicit formula  for the eigenfunctions of the Toda lattice have been obtained by Kharchev and Lebedev \cite{KL99,KL01}. Their completeness is proved
 in \cite{A09}.  
 The results \eqref{11.40} and \eqref{11.41} are quoted from \cite{KT10}. The reader is invited to look up the precise statements. 
  \medskip\\
  \textbf{ad 11.3}: The kernel of the $T$ operator of the $\delta$-Bose gas decays to $0$ for $|w| \to \infty$, while for the Toda lattice the kernel diverges logarithmically. Hence one has to check whether the techniques from \cite{DLP93} still carry through.
Sklyanin introduced already the parameter, which we call $\nu$, and established that it reappears in the   ground state TBA as the chemical potential dual to the density. The analysis in \cite{KT10} starts with the hamiltonian $H_{\nu,N}$  from the outset.
 The parameter $\nu$ then naturally appears  at the correct location. The short-cut discussed in the main text seems to be new. 
 
 Comparing \eqref{11.63} and \eqref{10.12} complete correspondence is noted. However, for the $\delta$-Bose gas it is has been proved that for given $I_1 < ...< I_N$ there is a unique solution of \eqref{10.12} \cite{D93}.  Apparently no such information is available for
 the Toda lattice equation \eqref{11.63}.
  
 
 \section{Beyond the Euler time scale}
\label{sec12}
\setcounter{equation}{0} 
As a common experience, sound waves in air propagate ballistically. When the air is confined to a thin tube, an initial perturbation maintains its shape and travels with the speed of sound. There could be dispersive effects due to a nonlinear energy-momentum relation.  The respective dynamics would be still time-reversible, no entropy is produced.  But on top, one observes irreversible sound damping, a phenomenon which can be traced back to molecular disorder.    
Very commonly, damping is modelled by a multiplicative diffusive factor as $\exp[-Dk^2|t|]$ in wave number space,
$D$ the diffusion coefficient. On the level of hydrodynamic equations, dissipation  amounts to the Navier-Stokes correction of Euler equations.

It is not immediately obvious whether such a conventional picture extends to integrable systems. As already at other occasions, the hard rod fluid serves as a convenient guiding example. Restricting to ballistic spacetime scales, a first step is to consider the motion of a tracer quasiparticle when the fluid 
is initially in some GGE.  To lowest order the tracer  acquires an effective velocity through collisions with fluid quasiparticles. 
Because of the randomness in the initial state these collisions are statistically independent and the effective velocity is the result 
of summing  over many collisions, said differently a law of large numbers.
As well known from the theory of independent random variables, a more precise description would be  a central limit theorem
\begin{equation}\label{12.1} 
q_\mathrm{tr}(t) \simeq  q_\mathrm{tr}(0) +v^\mathrm{eff} t + D_\mathrm{tr} b(t),
 \end{equation}
where $q_\mathrm{tr}(t)$ is the position of the tracer quasiparticle, $b(t)$ a standard Brownian motion, and  $D_\mathrm{tr}$  the diffusion constant,  which depends on the particular GGE.
When translating to the motion of conserved fields,  one should expect a Navier-Stokes correction, second order in $\partial_x$, whose analytical form has still to be worked out.
For this purpose we follow a standard strategy. 
The initial goal is to find out diffusive corrections for the propagation of a small perturbation away from the spatially homogeneous   GGE. 
This will be soft mathematics, providing merely a general framework and  not yet distinguishing between integrable and nonintegrable systems. 
The result is then written in a form, for which the step from global GGE to local GGE  will be compelling. 

\subsection{General framework}
\label{sec12.1}

As discussed in Section \ref{sec7}, the basic field correlator is
\begin{equation}\label{12.2} 
S_{m,n} (j,t) = \langle Q^{[m]}_j(t)Q^{[n]}_0(0) \rangle_\mathrm{gg}^\mathrm{c}.
\end{equation}
Here $\langle\cdot \rangle_\mathrm{gg}$ stands for the expectation in some GGE, which remains fixed throughout our discussion. 
By Lieb-Robinson type bounds, for fixed $t$, the correlator $S_{m,n}(j,t) $ decays exponentially outside the sound cone.
So $j$-summations are under control, while the time integration will turn out to be more delicate. To exploit spacetime symmetry,
we redefine 
\begin{equation}\label{12.3} 
Q^{[0]}_j =    \tfrac{1}{2}(r_{j-1} + r_j).
\end{equation}
So to avoid too many extra symbols, for this section only, the basic correlator is still denoted by $S$, taking \eqref{12.3} into account. 
\bigskip\\
$\blackdiamond\hspace{-1pt}\blackdiamond$~\textit{Spacetime inversion}.\hspace{1pt}
By spacetime stationarity one concludes
\begin{equation}\label{12.4} 
S_{m,n} (j,t) = S_{n,m} (-j,-t).
\end{equation}
In fact, the stronger property 
\begin{equation}\label{12.5} 
S_{m,n} (j,t) = S_{m,n} (-j,-t)
\end{equation}
holds by using invariance under spacetime inversion.

 We define time reversal through
\begin{equation}\label{12.6}
\mathcal{R}_\mathrm{tr}: (r, p) \mapsto (r,-p)
\end{equation}
and space inversion by
\begin{equation}\label{12.7}
\mathcal{R}_\mathrm{si}: (r, p) \mapsto (\tilde{r},\tilde{p}), \quad \tilde{r}_j = r_{-j-1}, \,\,\tilde{p}_j = - p_{-j}.
\end{equation}
Since $r_j$ should be viewed as bond variable, the inversion is relative to the origin. The equations of motion \eqref{2.5}, generate a flow on phase space denoted by $T_t$,  $T_t: (r,p) \mapsto (r(t), p(t))$. Using the flow equations, one concludes that
\begin{equation}\label{12.8}
\mathcal{R}_\mathrm{tr}\circ T_t =  T_{-t} \circ\mathcal{R}_\mathrm{tr}
\end{equation}
and 
\begin{equation}\label{12.9}
 \mathcal{R}_\mathrm{si}\circ T_t = T_{t} \circ\mathcal{R}_\mathrm{si} .
\end{equation}
Spacetime inversion is then defined through
\begin{equation}\label{12.10}
\mathcal{R}_\mathrm{si}\circ\mathcal{R}_\mathrm{tr} = \mathcal{R}: (r, p) \mapsto (\tilde{r},\tilde{p}), \quad \tilde{r}_j = r_{-j-1}, \,\,\tilde{p}_j =  p_{-j}.
\end{equation}
and hence
\begin{equation}\label{12.11}
\mathcal{R} \circ T_t =  T_{-t} \circ \mathcal{R}.
\end{equation}
In the random walk summation for $Q^{[n]}_j$, see  \eqref{2.14}, to each path from $j$ to $j$ in $n$ steps
there is a at level $j$ reflected path. Hence
\begin{equation}\label{12.12}
Q^{[n]}_j \circ\mathcal{R} = Q^{[n]}_{-j}.
\end{equation}
In particular, the GGE is invariant under $\mathcal{R}$. Combining \eqref{12.11} and \eqref{12.12}, the claim  \eqref{12.5} follows.

For the currents, the spacetime inversion is slightly more complicated. We introduce the down- and up-currents
\begin{equation}\label{12.13}
J^{[n]}_{j} = J^{[n]\scriptscriptstyle \downarrow}_{j} = \big(L^n L^{\scriptscriptstyle \downarrow}\big)_{j,j}, \qquad
J^{[n]\scriptscriptstyle \uparrow}_{j} = \big(L^n L^{\scriptscriptstyle \uparrow}\big)_{j,j}.
\end{equation}
Then
\begin{equation}\label{12.14}
J^{[n]\scriptscriptstyle \downarrow}_{j} \circ\mathcal{R}  = J^{[n]\scriptscriptstyle \uparrow}_{-j}.
\end{equation}
Correspondingly there are two current-current correlators
\begin{equation}\label{12.15}
\Gamma^{\scriptscriptstyle \downarrow(\scriptscriptstyle \uparrow)}_{m,n}  (j,t) =   
\langle J^{[m]\scriptscriptstyle \downarrow(\scriptscriptstyle \uparrow)}_j( t) J^{[n]\scriptscriptstyle \downarrow(\scriptscriptstyle \uparrow)}_0 (0) \rangle_\mathrm{gg}^\mathrm{c}
\end{equation} 
and hence spacetime inversion implies
\begin{equation}\label{12.16} 
\Gamma^{\scriptscriptstyle \downarrow}_{m,n} (j,t)= \Gamma^{\scriptscriptstyle \uparrow}_{m,n}  (-j,-t). 
\end{equation}
The down- and up-currents seem to be a special feature of the Toda lattice. \hfill $\blackdiamond\hspace{-1pt}\blackdiamond$\bigskip\\

Besides invariance under spacetime reversal, the correlator also satisfies
the conservation law
\begin{equation}\label{12.17} 
\partial_t S_{m,n} (j,t) = -\partial_j B_{m,n} (j,t),
\end{equation}
employing the difference operator  $\partial_jf(j) = f(j+1) - f(j)$. $B(j,t)$ is the charge-current time correlation 
\begin{equation}\label{12.18} 
B_{m,n}(j,t) = \langle J^{[m]}_j(t)Q^{[n]}_0(0) \rangle_\mathrm{gg}^\mathrm{c},
\end{equation}
whose static version has been encountered before, see \eqref{7.9}. Using both properties, one derives identities, known as sum rules, 
for the lowest moments of $S(t)$.

For the zeroth moment one obtains  
\begin{equation}\label{12.19}
\sum_{j \in \mathbb{Z}} S_{m,n} (j,t) =  \sum_{j \in \mathbb{Z}} C_{m,n}(j) = C_{m,n},
\end{equation}
where we used the conservation law \eqref{12.17} and the decay of $S_{m,n}(j,t)$ ensuring boundary terms to vanish.
 $C_{m,n}(j)$ is the static correlator,
\begin{equation}\label{12.20} 
C_{m,n}(j) = S_{m,n} (j,0),
\end{equation}
and  the spatially summed $C_{m,n}$ is the matrix of static susceptibilities, $C_{m,n} = C_{n,m}$.\bigskip\\
\textit{First moment}. By \eqref{12.5}, 
$C_{m,n}(j) = C_{m,n}(-j)$ and the sum rule
\begin{equation}\label{12.21} 
  \sum_{j \in \mathbb{Z}}j C_{m,n}(j) = 0
\end{equation}
follows. In addition, multiplying \eqref{12.17} with $j$ and summing over $j$, one obtains
\begin{equation}\label{12.22} 
\partial_t  \sum_{j \in \mathbb{Z}} jS_{m,n} (j,t) =  \sum_{j \in \mathbb{Z}} B_{m,n} (j,t) = B_{m,n} = (AC)_{m,n} 
\end{equation}
and hence 
\begin{equation}\label{12.23} 
 \sum_{j \in \mathbb{Z}} jS_{m,n} (j,t) = (AC)_{m,n} t 
\end{equation}
with the linarization matrix $A$, as discussed in Section \ref{sec7}. The $B$ matrix is symmetric, equivalently $AC = CA^\mathrm{T}$.\bigskip\\
\textit{Second moment}.  We pick some rapidly decaying test function $f$. Then 
\begin{equation}\label{12.24} 
  \sum_{j \in \mathbb{Z}} f_j\big( Q^{[n]}_j(t) - Q^{[n]}_j(0)\big)  = 
 \int_0^t\mathrm{d}s   \sum_{j \in \mathbb{Z}} ( \partial_j f_j)J^{[n]}_j(s).
\end{equation}
Squaring and using translation invariance,
\begin{eqnarray}\label{12.25} 
&&\hspace{0pt}   \sum_{j \in \mathbb{Z}} f_j   \sum_{j' \in \mathbb{Z}}\tilde{f}_{j'}\langle\big( Q^{[m]}_j(t) - Q^{[m]}_j(0)\big)
\big( Q^{[n]}_{j'}(t) - Q^{[n]}_{j'}(0)\big) \rangle_\mathrm{gg}^\mathrm{c} \nonumber\\
&&\hspace{0pt} =  -\int_0^t\mathrm{d}s  \int_0^t\mathrm{d}s' 
 \sum_{j \in \mathbb{Z}}  \sum_{j' \in \mathbb{Z}} (\partial_j^\mathrm{T}\partial_jf_j)\tilde{f}_{j'}\langle J^{[m]}_j(s)
J^{[n]}_{j'}(s')  \rangle_\mathrm{gg}^\mathrm{c}.
\end{eqnarray}
The current-current correlator is the central object for controlling diffusive behavior. We will use here the notation
\begin{equation}\label{12.26}
\Gamma_{m,n}  (j,t) =   \langle J^{[m]}_j(t)  
J^{[n]}_{0}(0)  \rangle_\mathrm{gg}^\mathrm{c} =\Gamma^{\scriptscriptstyle \downarrow}_{m,n}  (j,t)
\end{equation} 
and for the total current-current correlation
\begin{equation}\label{12.27}
 \sum_{j \in \mathbb{Z}} \Gamma_{m,n}  (j,t) = \Gamma_{m,n}  (t).
\end{equation} 
Note that by \eqref{12.16}, $\Gamma_{m,n}  (t) = 
\Gamma^{\scriptscriptstyle \downarrow}_{m,n} (t)= \Gamma^{\scriptscriptstyle \uparrow}_{m,n}  (-t) = \Gamma_{m,n}  (-t)
$ and thus $\Gamma_{m,n}  (\infty) = 
\Gamma_{m,n}  (-\infty)$. 
Setting $f_j = \tfrac{1}{2} j^2$, $\tilde{f}_j = \delta_{0,j}$ and using spacetime inversion \eqref{12.5}, one concludes that 
\begin{equation}\label{12.28} 
  \sum_{j \in \mathbb{Z}} j^2 \big(S_{m,n} (j,t) - S_{m,n} (j,0)\big)  = \Gamma_{m,n}(\infty) t^2 + \int_0^t\mathrm{d}s  \int_0^t\mathrm{d}s' \big(\Gamma_{m,n}(s - s') -  \Gamma_{m,n}(\infty)\big) .
\end{equation}
$\Gamma_{m,n}(\infty)$ is the \textit{Drude weight}. If $\Gamma_{m,n}(t) -  \Gamma_{m,n}(\infty)$ is integrable, one can define the \textit{Onsager matrix} 
\begin{equation}\label{12.29}
L_{m,n} =  \int_\mathbb{R}\mathrm{d}t \big(\Gamma_{m,n}(t) -  \Gamma_{m,n}(\infty)\big).
\end{equation}
$L$ is a symmetric  matrix with $L \geq 0$ as covariance matrix. Hence for large times 
\begin{equation}\label{12.30} 
 \sum_{j \in \mathbb{Z}} j^2 S_{m,n} (j,t)  \simeq \Gamma_{m,n}(\infty) t^2 + L_{m,n} |t|.
 \end{equation}
(We used the symbol $L$ already for the infinite volume Lax matrix. For the remainder of this section $L$ denotes 
the Onsager matrix).\\

The reader may worry that we have lost our way, lots of definitions, except for the diffusion matrix of interest.
To appreciate our preparations, a useful example is the shear viscosity of a fluid. A standard scheme to experimentally define
the shear viscosity, $\nu$,  is the Couette  flow. One plate is fixed and another plate, distance $y$ and in parallel, is moved with constant  
velocity $u$. For small $u$ one finds the relation
\begin{equation}\label{12.31}
 \frac{F}{A} = \nu \frac{u}{y}.
\end{equation}
Here $F/A$ is the pushing force per area acting on the moving plate. The shear viscosity reappears on a more sophisticated level in the Navier-Stokes equations. For example, solving these equations for the Couette flow confirms \eqref{12.31}. But having the full equations, one can think of different setups to yield $\nu$. In our context one example would be the propagation of a small perturbation of an equilibrium state. Of course, experimental accessibility may vary. But, independently of the method, measured is always the same shear viscosity.

Integrable systems are not readily available in the lab.  Diffusion is governed by a high-dimensional matrix and a priori 
it is not so obvious whether and how familiar methods would apply. For the Euler time scale, we already identified   
the dynamic correlator as
\begin{equation}\label{12.32}
\hat{S}_{m,n}(k,t) \simeq \big(\mathrm{e}^{-\mathrm{i}kAt}C\big)_{m,n}
\end{equation}
in wave number space,  valid for small $k$ and large $t$, $kt$ fixed, compare with \eqref{7.20} and  \eqref{7.29}. Most commonly, diffusive corrections are added through
\begin{equation}\label{12.33}
\hat{S}_{m,n}(k,t) \simeq \big(\mathrm{e}^{-\mathrm{i}kAt - k^2 D |t|}C\big)_{m,n}.
\end{equation}
By definition $D$ is the infinite-dimensional \textit{diffusion matrix}. In general $D$ is not symmetric. However, to describe dissipation,
$D$ is  required to have nonnegative eigenvalues. In generic examples, $D$ has zero eigenvalues and hence there is no diffusive decay along some particular directions. But $[A,D] \neq 0$ and $A$ mixes the modes such that $\hat{S}(k,t)$ still decays to $0$ in the long time limit. 

The first moment sum rule is obviously satisfied. Computing the second moment in the position space version of \eqref{12.33} and comparing with the sum rule 
\eqref{12.30}, one obtains 
\begin{equation}\label{12.34}
A^2Ct^2 + DC |t| = \Gamma(\infty) t^2 + L|t|.
\end{equation}
Hence the Drude weight matrix is given by
\begin{equation}\label{12.35}
\Gamma(\infty) = A^2C = A CA^\mathrm{T} = B\frac{1}{C}B.
\end{equation}
The latter expression is a natural symmetric form, which emphasizes that the Drude weight is a property
determined by  static correlations only. In addition,  the diffusion matrix is obtained as
\begin{equation}\label{12.36}
DC = CD^{\mathrm{T}} = L, \quad D = LC^{-1},
\end{equation}
which is the generalization of the \textit{Onsager relation} to integrable systems. Since $L \geq 0$ and $C >0$, $D$ is a similarity transform of $L$ and has thus only nonnegative eigenvalues.  
\bigskip\\
$\blackdiamond\hspace{-1pt}\blackdiamond$~\textit{Drude weight}.\hspace{1pt} In the context of condensed matter physics, it is fairly common to equate a non-zero Drude weight with infinite conductivity, implicitly suggesting that no further properties are to be investigated.   In fact, according to our discussion, a non-zero Drude weight simply indicates a ballistic component of the dynamic correlator. For nonintegrable systems in one dimension,  the ballistic component consists of sharp $\delta$-peaks moving with constant velocity on the Euler scale. Diffusion is then easily detected through the broadening of the peaks.
 For integrable systems the ballistic component is extended and of the generic form $t^{-1}g(t^{-1}x)$ in position space with some smooth scaling function $g$. Now diffusive corrections become harder to detect,
 unless special care is taken to suppress the broad ballistic background. An example is the domain wall discussed in Section \ref{sec8}. At the contact line the profile jumps from $\mathsf{n}_{-}$  to $\mathsf{n}_{+}$. Diffusion will broaden the step to an error function. Another example is the 
 XXZ model at zero magnetization (half filling). At this specific point the spin Drude weight vanishes and diffusive effects can be observed. 
 \hfill$\blackdiamond\hspace{-1pt}\blackdiamond$


\subsection{Nonintegrable chains}
\label{sec12.2}
Our preparations have been accomplished already in Section \ref{sec7.1} and, in essence, we are left with copying from the previous
 subsection, recalling that now there are only three fields, hence $n=0,1,2$.  It will turn out to be instructive to dwell on further details. 
 In Section \ref{sec7.1}  serif letters were used for matrices, so to distinguish from the integrable case. The same convention is followed here. Also in this subsection, to ease the comparison with the literature,  we switch to the physical pressure, still denoted by $P$, which means that in Eq. 
 \eqref{7.5} $P$ has to be substituted by $\beta P$. It is convenient to set $u=0$. Then the Gibbs state is invariant under time-reversal, which provides additional symmetries.
 The average stretch is $\nu = \langle r_0\rangle_{P,\beta}$ and average energy $\mathsf{e} =  \langle e_0\rangle_{P,\beta} $, where we dropped the mean velocity as parameter.
By convexity, one can view $P$ as a function of $\nu,\mathsf{e}$, $P = P(\nu,\mathsf{e})$. The linearization matrix reads
\begin{equation}\label{12.37} 
\mathsfit{A} = 
\begin{pmatrix}
0 & -1&0 \\
\partial_\nu P& 0 &\partial_\mathsf{e} P \\
0& P& 0\\
\end{pmatrix}. 
\end{equation}
The static correlator is given by
\begin{equation}\label{12.38} 
\mathsfit{C} = 
\begin{pmatrix}
 \langle r_0r_0\rangle_{P,\beta}^\mathrm{c} & 0&\langle r_0V_{\mathrm{ch,0}}\rangle_{P,\beta}^\mathrm{c} \\
0&\beta^{-1}&0\\
\langle r_0V_{\mathrm{ch,0}}\rangle_{P,\beta}^\mathrm{c}& 0& \langle e_0e_0\rangle_{P,\beta}^\mathrm{c}\\
\end{pmatrix} 
\end{equation}
and the current-charge cross correlation by  
\begin{equation}\label{12.39} 
\mathsfit{B} = \beta^{-1}
\begin{pmatrix}
0 & -1&0 \\
-1& 0&P \\
0& P& 0\\
\end{pmatrix}. 
\end{equation}
With this information one obtains the Drude weight
\begin{equation}\label{12.40} 
\Gamma({\infty}) = \beta^{-1}
\begin{pmatrix}
 1 & 0&-P \\
0&c^2&0\\
-P& 0& P^2\\
\end{pmatrix}, 
\end{equation}
where $c$ is the isentropic speed of sound,
\begin{equation}\label{12.41}
c^2 = \frac{\beta \langle (e_0 + Pr_0)(e_0+Pr_0)\rangle_{P,\beta}^\mathrm{c} }
{\langle r_0r_0\rangle_{P,\beta}^\mathrm{c}\langle e_0e_0 \rangle_{P,\beta}^\mathrm{c} -
(\langle r_0 e_0\rangle_{P,\beta}^\mathrm{c})^2}\,.
\end{equation}

Finally we have to list the two dynamic characteristics
introduced before. The Onsager matrix turns out as
\begin{equation}\label{12.42} 
\mathsfit{L} = 
\begin{pmatrix}
 0 & 0&0 \\
0&\sigma_\mathsf{p}^2&0\\
0& 0& \sigma_\mathsf{e}^2\\
\end{pmatrix}. 
\end{equation}
The only nonvanishing matrix elements are momentum-momentum and energy-energy current. The notation is supposed to indicate that they are related to the noise strength in a Ginzburg-Landau fluctuation theory.
But the defining time integral \eqref{12.29} cannot be expected to yield an explicit expression. The $0$'s at the upper left borders of the matrix result from
the stretch current being conserved. The $1,2$ matrix element vanishes, since  momentum and energy have opposite signs under time reversal. The resulting diffusion matrix is given by
\begin{equation}\label{12.43} 
\mathsfit{D} = 
\begin{pmatrix}
 0 & 0&0 \\
0&\sigma_\mathsf{p}^2&0\\
 \alpha\sigma_\mathsf{e}^2& 0& \sigma_\mathsf{e}^2\\
\end{pmatrix}, 
\end{equation}
where $\alpha = -\langle r_0 V_{\mathrm{ch,0}}\rangle_{P,\beta}^\mathrm{c}\big/\langle r_0r_0 \rangle_{P,\beta}^\mathrm{c}$. Note that $\mathsfit{L}$, and hence $\mathsfit{D}$, has a zero eigenvalue.  Since $[\mathsfit{A},\mathsfit{D}] \neq 0$,  still  $\hat{\mathsfit{S}}^\diamond (k,t)\to 0$ as $t \to \infty$,
except for $k=0$. If one reinstalls a nonzero mean velocity $u$, the Gibbs state is no longer invariant under time reversal. By a Galilean transformation,
one can still figure out the various matrices. As a particular consequence,  while $\mathsfit{L}_{0,n} = 0$ for $n=0,1,2$ because of  stretch current conservation, the cross term 
$\mathsfit{L}_{1,2} \neq 0$. The diffusion matrix also picks up further non-vanishing entries. 

In position space, on the Euler scale $\mathsfit{S}^\diamond(x,t) $  consists of three $\delta$-peaks, the heat peak at rest and the two sound peaks traveling with velocity $\pm c$. The diffusion term broadens the peaks as $\sqrt{t}$. While this prediction looks innocent,  it is completely off the track as noted
already in the mid-1970ies. We quietly assumed that $\Gamma_{1,1}(t)$ and  
$\Gamma_{2,2}(t)$ are integrable. A more refined theory arrives at the conclusion that both current correlations 
decay generically as $|t|^{-2/3}$, which is well confirmed by molecular dynamics simulations. Also the shape functions differ from a Gaussian.
The heat peak broadens as  $t^{3/5}$ with the symmetric Levy-$\tfrac{5}{3}$ function as shape and the sound peaks broaden
as  $t^{2/3}$. Their shape function equals the spacetime covariance of the stationary stochastic Burgers equation. 
In fact, there are three universality classes. For one of them, $\Gamma_{1,1}(t)$ and $\Gamma_{2,2}(t)$ have an integrable  
decay and then the claims from the linear theory are fully confirmed.
 
\subsection{Navier-Stokes equations for the Toda fluid}    
\label{sec12.3}
For the Toda lattice there is currently no method to compute the Onsager matrix directly on the basis of the microscopic model. 
However, using form factor expansions, for the $\delta$-Bose gas a closed formula for the $L$-matrix has become available. By itself this is a surprising result. To explicitly compute transport coefficients for many-body systems  is a rare exception.  
By the much emphasized analogy between integrable systems, one arrives at a firm prediction for the Toda fluid. We state here merely the result and discuss some of its consequences. The notations introduced Section \ref{sec9.2} will be used.

The matrix $L_{m,n}$ uses the basis consisting of monomials  $\varsigma_n$. Structurally more transparent are matrix elements computed for general functions over $\mathbb{R}$. We first introduce the kernel

\begin{equation}\label{12.44} 
K(w_1,w_2) = \rho_\mathsf{p}(w_1)\rho_\mathsf{p}(w_2) |v^\mathrm{eff}(w_1) - v^\mathrm{eff}(w_2) | |T^\mathrm{dr}(w_1,w_2) |^2,
\end{equation}
the integral operator $T^\mathrm{dr}$ being defined through 
\begin{equation}\label{12.45} 
T^\mathrm{dr} = (1 - T\rho_\mathsf{n} )^{-1}T,
\end{equation}
and its action on the constant function,
\begin{equation}\label{12.46} 
\kappa(w_1) = \int_\mathbb{R}\mathrm{d}w_2 K(w_1,w_2).
 \end{equation}
Clearly, $T^\mathrm{dr}$ is a symmetric operator and thus also $K$. Then, for general functions $f,g$ on $\mathbb{R}$, 
\begin{equation}\label{12.47} 
\langle f, L g\rangle = \frac{1}{2} \int_{\mathbb{R}^2} \mathrm{d} w_1  \mathrm{d} w_2 K(w_1,w_2)\Big(\frac{f^\mathrm{dr}(w_2)}{\rho_\mathsf{s}(w_2)} - \frac{f^\mathrm{dr}(w_1)}{\rho_\mathsf{s}(w_1)}\Big)
\Big(\frac{g^\mathrm{dr}(w_2)}{\rho_\mathsf{s}(w_2)} - \frac{g^\mathrm{dr}(w_1)}{\rho_\mathsf{s}(w_1)}\Big).
\end{equation} 
As it should be, $L$ is symmetric and $L \geq 0$. To determine possible zero eigenvalues, we note that $K(w_1,w_2) >0$. Hence 
$Lf = 0$ implies
\begin{equation}\label{12.48} 
f^\mathrm{dr}(w)  = c \rho_\mathsf{s}(w)
\end{equation}
with arbitrary $c \in \mathbb{R}$, which has $f = c  \varsigma_0$ is only solution. The zero subspace of $L$ corresponds to the 
projector $|\varsigma_0\rangle \langle \varsigma_0|$. 

The diffusion matrix is obtained from the Onsager matrix as $D = LC^{-1}$. The $C$ matrix for the Toda lattice has been stated in \eqref{7.22}.
For the Toda fluid the comoving terms $q_n \varsigma_n$ and the factors of $\nu$ are dropped, with the result 
\begin{equation}\label{12.49} 
C = (1 -   \rho_\mathsf{n}T)^{-1} \rho_\mathsf{p} (1 - T \rho_\mathsf{n} )^{-1}. 
\end{equation}
Hence 
\begin{equation}\label{12.50} 
C^{-1} = (1 - T \rho_\mathsf{n} ) \rho_\mathsf{p}^{-1} (1 -   \rho_\mathsf{n}T)
\end{equation}
and 
\begin{equation}\label{12.51} 
(C^{-1}g)^\mathrm{dr} =  \rho_\mathsf{p}^{-1}(1 -  \rho_\mathsf{n}T) g.
\end{equation}
Note that the zero eigenvalue of $D$ has the left eigenvector $\varsigma_0$ and the right eigenvector  $
(1 - \rho_\mathsf{n} T)^{-1}\rho_\mathsf{p} \rho_\mathsf{s}$.

For applications it will be more convenient to have  also the integral kernel of $D$ available. To separate into a diagonal multiplication operator 
and smooth off-diagonal integral kernel, we set
\begin{equation}\label{12.52} 
Uf(w) = \rho_\mathsf{s}(w)^{-1}( (1 - T\rho_\mathsf{n})^{-1}f)(w)
\end{equation}
and 
\begin{equation}\label{12.53} 
UC^{-1}g(w)  = ( \rho_\mathsf{s}(w) \rho_\mathsf{p}(w))^{-1}((1- \rho_\mathsf{n}T)g)(w).
\end{equation}
Using the symmetry of $K$, one arrives at 
\begin{equation}\label{12.54} 
 \langle f, D g\rangle =   \int_{\mathbb{R}^2} \mathrm{d}w_1 \mathrm{d} w_2 K(w_1,w_2) \big(Uf(w_1) UC^{-1}g(w_1).
 - Uf(w_1) UC^{-1}g(w_2)\big)
\end{equation}
Setting $\bar{K}  =   U^\mathrm{T}KUC^{-1} $, it follows
\begin{equation}\label{12.55} 
D(w_1,w_2) = -\bar{K}(w_1,w_2) + \delta(w_1 - w_2)U^\mathrm{T}\kappa UC^{-1} (w_1,w_1),
\end{equation}
where $\kappa$ is regarded as multiplication operator.
Since the left eigenvector of $D$ is $\varsigma_0$,
if $\bar{K}(w_1,w_2) \geq 0$ and also  $U^\mathrm{T}\kappa UC^{-1} (w_1,w_1) \geq 0$, then the operator 
$-D$ has the structure of the generator of a time-continuous Markov jump process.

A control check of \eqref{12.47} is easily performed using the model of hard rods. Firstly, the various kernels and functions appearing in  \eqref{12.47} and \eqref{12.54} are explicit. Secondly,
there is an exact expression for the structure function, here denoted by $\hat{\mathsfit{S}}_{m,n}(k,t)$, and the asymptotic behavior 
\eqref{12.33} is confirmed. Using a sum rule as in \eqref{12.28}, the total current correlator $\Gamma_{m,n}(t)$ can be worked out, with the result being proportional to $\delta(t)$. Thereby the time integral trivializes and the Onsager matrix is obtained as
\begin{equation}\label{12.56}
L(w_1,w_2) = (\mathsfit{a}\rho)^2 (1-\mathsfit{a}\rho)^{-1}\big(\delta(w_1-w_2) r(w_1) - |w_1- w_2|h(w_1)h(w_2)\big),
\end{equation}
where $r(w) = \int_\mathbb{R}\mathrm{d}w' h(w')|w - w'| $, and the diffusion matrix as
\begin{equation}\label{12.57}
D(w_1,w_2) = \mathsfit{a}(\mathsfit{a}\rho) (1-\mathsfit{a}\rho)^{-1}\big(\delta(w_1-w_2) r(w_1) - h(w_1)|w_1-w_2|\big).
\end{equation}
Concluding, for hard rods \eqref{12.47}, \eqref{12.54} are in agreement with \eqref{12.56}, \eqref{12.57}.

Based on our input, in analogy to classical fluids, the Navier-Stokes type equation of the Toda fluid is given by
\begin{equation}\label{12.58} 
\partial_t\rho_\mathsf{p}(x,t) + \partial_x\big(v^\mathrm{eff}(x,t)\rho_\mathsf{p}(x,t)\big) = \partial_xD\partial_x \rho_\mathsf{p}(x,t)
\end{equation}
with diffusion matrix $D$ of \eqref{12.55}. The transformation to quasilinear form can still be carried out and yields
\begin{equation}\label{12.59} 
\partial_t\rho_\mathsf{n}(x,t) + v^\mathrm{eff}(x,t)\partial_x\rho_\mathsf{n}(x,t) =
\rho_\mathsf{s}(x,t)^{-1}(1- \rho_\mathsf{n}T)\partial_xD\partial_x \rho_\mathsf{p}(x,t).
\end{equation}

There is one fundamental physics property which can be checked directly. The balance equation for the local entropy will have a flow term 
and a production term. The latter should be positive, thereby reflecting the second law of thermodynamics. The entropy is defined by
minus the second term of
 \eqref{9.28}, upon setting $\rho = \rho_\mathsf{p}$.  Hence the time derivative reads
\begin{eqnarray}\label{12.60} 
&&\hspace{0pt}\partial_t s = -\partial_t \langle  \rho_\mathsf{s}\rho_\mathsf{n}\log \rho_\mathsf{n} \rangle = 
- \langle (\log \rho_\mathsf{n}) \partial_t \rho_\mathsf{p}\rangle - \langle  \rho_\mathsf{s} \partial_t \rho_\mathsf{n} \rangle\nonumber\\
&&\hspace{0pt} =  \langle( \log \rho_\mathsf{n}) \partial_x (v^\mathrm{eff}\rho_\mathsf{p})\rangle
+\langle  \rho_\mathsf{s} v^\mathrm{eff}\partial_x\rho_\mathsf{n} \rangle -\langle  \big(\log\rho_\mathsf{n} 
+(1- \rho_\mathsf{n}T)\big) \partial_x(D \partial_x \rho_\mathsf{p})\rangle,
\end{eqnarray}
where \eqref{12.58} and \eqref{12.59} have been inserted. The first order derivative terms  contribute to the flow term,  where the second summand is cancelled by writing
\begin{eqnarray}\label{12.61} 
&&\hspace{-40pt}\langle (\log \rho_\mathsf{n}) \partial_x (v^\mathrm{eff}\rho_\mathsf{p})\rangle 
+\langle  \rho_\mathsf{s} v^\mathrm{eff}\partial_x\rho_\mathsf{n} \rangle \nonumber\\
 &&\hspace{-20pt}
= \partial_x \langle (\log \rho_\mathsf{n})  v^\mathrm{eff}\rho_\mathsf{p}\rangle 
- \langle (\partial_x \log \rho_\mathsf{n})  v^\mathrm{eff}\rho_\mathsf{p}\rangle + \langle  \rho_\mathsf{s} v^\mathrm{eff}\partial_x\rho_\mathsf{n} \rangle= \partial_x \langle (\log \rho_\mathsf{n})  v^\mathrm{eff}\rho_\mathsf{p}\rangle.
\end{eqnarray}
For the third summand we move the leftmost $\partial_x$ in front, as before. This yields a second contribution to the flow term as  
\begin{equation}\label{12.62} 
-\partial_x \langle \big(\log \rho_\mathsf{n} +(1 - T\rho_\mathsf{n})\big) D \partial_x \rho_\mathsf{p}\rangle
\end{equation}
and the dissipative term
\begin{equation}\label{12.63} 
 \langle\big(\rho_\mathsf{n}^{-1}\partial_x\rho_\mathsf{n} - T\partial_x\rho_\mathsf{n}\big) D \partial_x \rho_\mathsf{p}\rangle.
\end{equation}
Using $\partial_x\rho_\mathsf{s} = T \partial_x \rho_\mathsf{p}$, one arrives at the identity
\begin{equation}\label{12.64} 
\rho_\mathsf{n}^{-1}\partial_x \rho_\mathsf{n} = \rho_\mathsf{p}^{-1}(1 - \rho_\mathsf{n}T)\partial_x\rho_\mathsf{p}.
\end{equation}
We now substitute $D = LC^{-1}$ and note that 
 \begin{equation}\label{12.65} 
 C^{-1} \partial_x  \rho_\mathsf{p} = (1 - T\rho_\mathsf{n})\rho_\mathsf{p}^{-1}(1 - \rho_\mathsf{n}T)\partial_x\rho_\mathsf{p} = (1- T \rho_\mathsf{n})\rho_\mathsf{n}^{-1}\partial_x\rho_\mathsf{n}
\end{equation}
implying the dissipative term
\begin{equation}\label{12.66} 
\langle (\partial_x \rho_\mathsf{p}),C^{-1}LC^{-1}(\partial_x \rho_\mathsf{p})\rangle \geq 0.
\end{equation}

Altogether the entropy balance for the Toda fluid reads
\begin{equation}
\label{12.67} 
\partial_t s +\partial_x \mathfrak{j}_\mathrm{s} = \sigma.
\end{equation}
The entropy current is determined to 
\begin{equation}\label{12.68} 
 \mathfrak{j}_\mathrm{s} = - \langle (\log \rho_\mathsf{n} )v^\mathrm{eff}\rho_\mathsf{p}\rangle +\langle \big(\log \rho_\mathsf{n} + (1 - T\rho_\mathsf{n})\big)D \partial_x \rho_\mathsf{p} \rangle.
\end{equation}
For the entropy production the very concise formula
\begin{equation}\label{12.69} 
\sigma = \langle (\partial_x \rho_\mathsf{p}),C^{-1}LC^{-1}(\partial_x \rho_\mathsf{p})\rangle.
\end{equation}
is unveiled.
\bigskip\\
\textbf{\large{Notes and references}}\bigskip\\
 \textbf{ad 12.1}: For general anharmonic chains the sum rules are discussed in \cite{MS16}. While time reversal is 
 a standard item, the use of spacetime reversal  I learned from \cite{DBD18,DBD19}. The exact formula in \eqref{12.32}
 is derived in \cite{LPS68} and the respective current correlations can be found in \cite{S82}. The Navier-Stokes correction
 for hard rods is proved in \cite{BS97}, with a result in complete agreement with \eqref{12.58}.
 
 For simple fluids the necessity to subtract the Drude weight has been recognized in the pioneering paper \cite{G54},
 see \cite{S91} for a textbook discussion. In condensed matter physics mostly one refers back to the Mazur bound \cite{M69},
 which in our context means to sum in \eqref{12.35} only over a restricted number of conserved fields. The general formula is stated in 
 \cite{DS17}.
 Applying  \eqref{12.35} naively to the XXZ chain, one would conclude  that the spin Drude weight vanishes. In actual fact, the Drude weight vanishes only for $\Delta >1$ in the standard units, while it is non-zero and nowhere continuous for $0 \leq \Delta \leq 1$. The resolution 
 can be traced back on our insistence on strictly local conservations. The XXZ chain has additional conservation laws which have exponential tails and thus contribute to hydrodynamics, in particular to the Drude weight. We refer to \cite{MPP15,ID17a} for more details.   
 \medskip\\
\textbf{ad 12.2}: Nonintegrable anharmonic chains are studied at length in \cite{S14}, where also the connection to the three component stochastic Burgers equation, alias KPZ equation, is explained. The hydrodynamic limit of anharmonic chains is studied in
\cite{BO20}. Interesting molecular dynamics simulations are reported in \cite{GCD20}
\medskip\\
\textbf{ad 12.3}: The section is entirely based on \cite{DBD18,DBD19}, where the Navier-Stokes corrections are obtained through form factor expansions. The simple looking formula for the entropy production is new.

\end{document}